\documentclass[pdflatex,sn-nature,referee]{sn-jnl}

\usepackage{graphicx}%
\usepackage{multirow}%
\usepackage{amsmath,amssymb,amsfonts}%
\usepackage{amsthm}%
\usepackage{mathrsfs}%
\usepackage[title]{appendix}%
\usepackage{xcolor}%
\usepackage{textcomp}%
\usepackage{manyfoot}%
\usepackage{booktabs}%
\usepackage{algorithm}%
\usepackage{algorithmicx}%
\usepackage{algpseudocode}%
\usepackage{listings}%
\usepackage{url}%
\usepackage{makecell}%

\begin{document}

\title[ArticleTitle]{Lattice-renormalized geometric frustration drives fast ionic transport}


\author[1,2]{\fnm{Jianmin} \sur{Yang}}\email{e0546047@u.nus.edu}

\author*[1]{\fnm{Lin} \sur{Xie}}\email{xielin@gbu.edu.cn}

\affil*[1]{\orgdiv{Dongguan Key Laboratory of Solid-State Battery Materials, School of Physical Sciences}, \orgname{Great Bay University}, \city{Dongguan}, \postcode{523000}, \state{Guangdong}, \country{China}}

\affil[2]{\orgdiv{Department of Materials Science and Engineering}, \orgname{National University of Singapore}, \postcode{117575}, \country{Singapore}}


\abstract{Fast ionic transport is commonly understood from two complementary perspectives: soft lattices that facilitate ion migration, and frustrated ionic sublattices that host multiple nearly degenerate configurations. Here, we show that the host lattice and ionic geometric frustration are fundamentally coupled, and together shape a lattice-renormalized free-energy landscape that unifies these two perspectives. Starting from a coupled ion–lattice Hamiltonian and eliminating the adiabatic lattice response, we derive a configuration-dependent free-energy renormalization that takes a quadratic form of the ionic configurational forces weighted by the inverse stiffness of the host lattice. In a coarse-grained representation, this renormalization decomposes into local lattice relaxation and non-local interference between lattice-response fields. These two contributions reshape local ionic configurations and their correlations, thereby redistributing the configurational statistical weights underlying collective ionic transport. Atomistic simulations of cubic Li$_7$La$_3$Zr$_2$O$_{12}$ and AgCrSe$_2$ substantiate this physical picture, from microscopic configurational rearrangements to macroscopic diffusion dynamics. This framework therefore recasts fast ionic transport as a lattice-renormalized geometric frustration problem, in which lattice softness, ionic frustration, and collective diffusion all arise from the same underlying landscape.}




\maketitle


Fast ionic transport constitutes a physical process in which ions diffuse with liquid-like mobility in a structurally ordered crystalline host lattice \cite{Boyce1979, Hull2004}. The microscopic origin of this macroscopic behavior is commonly understood from two complementary perspectives. The first focuses on the lattice: soft phonon modes \cite{Wakamura1997, Muy2021, Jun2024}, large polarizability \cite{Wang2015, Kraft2017}, strong anharmonicity \cite{Gupta2021, Brenner2022}, and facile anion motion \cite{Zhang2022, Liu2024} promote fast ionic transport by reshaping local transport pathways and reducing migration barriers \cite{Muy2021, Jun2024}. The second perspective emphasizes the ionic subsystem itself. In many prototypical fast ionic conductors, mobile ions explore a complex energy landscape containing multiple nearly degenerate configurations \cite{Wood2021, DiStefano2019, Duvel2017, Adelstein2016, Kweon2017}, where local preferences cannot be simultaneously satisfied under the constraints of the host \cite{Ling2023, Yang2026}. This configurational disorder, also referred to as geometric frustration \cite{Keen2015, Irvine2022}, is suggested to account for low activation energies \cite{Wood2021, DiStefano2019}, dynamic ionic correlations \cite{Yang2026}, and concerted ionic migration \cite{Irvine2022}. While both views are supported by experimental and computational evidence, they are typically treated independently, rather than as intrinsically coupled degrees of freedom.

Microscopically, as shown in Fig. \ref{fig1}a, geometrically frustrated mobile ions continuously deform the surrounding host lattice, while the deformed lattice in turn dynamically feeds back to the ions. This mutual coupling generates an emergent free-energy landscape from which fast ionic transport arises thermodynamically. However, a microscopic theory that treats both ionic geometric frustration and host lattice response on the same footing to explicitly formulate how their coupling renormalizes the free-energy landscape is still lacking.

In this work, we construct such a microscopic framework that recasts fast ionic transport as a lattice-renormalized geometric frustration problem. Starting from a coupled ion-lattice Hamiltonian and eliminating the lattice degrees of freedom adiabatically, we derive an analytical expression for the renormalized free-energy landscape. Specifically, the lattice response induces a configuration-dependent renormalization expressed as a quadratic form of ionic configurational forces weighted by the host's inverse stiffness. At the microscopic level, this global renormalization can be coarse-grained into local lattice relaxation and non-local interference between lattice-response fields. This formulation, substantiated by atomistic simulations across garnet and layered superionic materials, reveals how cooperative ionic correlations and collective transport emerge from a single renormalized free-energy landscape.

\section*{Theory}\label{sec_theory}
\subsection*{Lattice-renormalized geometric frustration}\label{subsec_LMGF}
We consider mobile ions within a deformable host lattice. Let $\mathbf{r}_i$ denote the $i$-th mobile ion's coordinates, and $\mathbf{R}_{l\kappa} = \mathbf{R}_{l\kappa}^0 + \mathbf{u}_{l\kappa}$ the position of framework atom $\kappa$ in unit cell $l$ displaced from the force equilibrium position $\mathbf{R}_{l\kappa}^0$ by $\mathbf{u}_{l\kappa}$. The lattice potential acting on ions is decomposed into a static periodic potential $V_0(\{\mathbf r_i\})$, defined by the reference host lattice, and a residual coupling $\delta\mathcal{V}_{\text{lat-ion}}$ to lattice displacements. The total Hamiltonian is:
\begin{equation}
\begin{split}
    \mathcal{H} = & \underbrace{\sum_i \frac{\mathbf{p}_i^2}{2m_i} + V_0(\{\mathbf{r}_i\}) + U_{\text{ion}}^{\text{bare}}(\{\mathbf{r}_i\})}_{\displaystyle\mathcal{H}_{\text{GF}}} \\ & + \underbrace{\sum_{l,\kappa}\frac{\mathbf{p}_{l\kappa}^2}{2M_{l\kappa}} + \delta\mathcal{V}_{\text{lat-ion}}(\{\mathbf{r}_i\}, \{\mathbf{R}_{l\kappa}\}) + U_{\text{lat}}(\{\mathbf{R}_{l\kappa}\})}_{\displaystyle\mathcal{H}_{\text{lat}}}
\end{split} \tag{1} \label{eq:eq1}
\end{equation}
The first term, $\mathcal{H}_{\text{GF}}$, defines the ionic geometric frustration (GF) problem in a frozen lattice. It contains the periodic potential $V_0$ imposed by the host and the intrinsic inter-ionic interactions $U_{\text{ion}}^{\text{bare}}$, which favor specific ionic correlations. Geometric frustration arises when these preferred ionic correlations are incompatible with the underlying host periodicity \cite{Wood2021, DiStefano2019, Duvel2017, Yang2026, Irvine2022}. The second term, $\mathcal{H}_{\text{lat}}$, captures the lattice response induced by the residual ion-lattice coupling. The remaining question is therefore how this deformable lattice reshapes the frustrated ionic landscape.

From equation \eqref{eq:eq1}, we evaluate the corresponding partition function for the coupled ion-lattice system and construct a conditional free energy associated with a fixed ionic configuration $\{\mathbf{r}_i\}$ following standard statistical-mechanical theories \cite{Landau1980, Chandler1987} (see Supplementary Information S1.1 for full derivations). In collective normal-mode coordinates of lattice displacements $x$, this conditional free energy at finite temperature can be expanded around the reference equilibrium lattice structure ($x=0$) to second order in matrix form as
\begin{equation}
    \mathcal{F}(\{\mathbf{r}_i\},x,T) \approx \mathcal{F}_{\text{GF}}(\{\mathbf{r}_i\},T) + \mathcal{F}_{\text{lat}}^0(\{\mathbf{r}_i\},T) - \widetilde{F}^\dagger(\{\mathbf{r}_i\},T) x + \tfrac{1}{2}x^\dagger \widetilde{K}(\{\mathbf{r}_i\},T) x \tag{2} \label{eq:eq2}
\end{equation}
where $\mathcal{F}_{\text{GF}}$ is the conditional free energy of the frustrated ionic subsystem in the absence of lattice response, $\mathcal{F}_{\text{lat}}^0$ is the lattice vibrational free energy conditioned at the given $\{\mathbf{r}_i\}$, and $\dagger$ is the Hermitian conjugate. The last two terms capture the exact physical picture described earlier: mobile ions deform the host lattice with effective configuration-dependent generalized forces $\widetilde{F}(\{\mathbf{r}_i\},T)$, while the host responds to this deformation with an effective stiffness matrix $\widetilde{K}(\{\mathbf{r}_i\},T)$.

Under the adiabatic approximation, the host lattice relaxes to its conditional equilibrium state. The equilibrium condition, $\widetilde{F} = \widetilde{K}x_{\text{relax}}$, is obtained by minimizing equation \eqref{eq:eq2} with respect to the normal-mode coordinate $x$, leading to
\begin{equation}
    \mathcal{F}(\{\mathbf{r}_i\},x_{\text{relax}},T) = \mathcal{F}_{\text{GF}}(\{\mathbf{r}_i\},T) + \mathcal{F}_{\text{lat}}^0(\{\mathbf{r}_i\},T) \underbrace{- \tfrac{1}{2}\widetilde{F}^\dagger \widetilde{K}^{-1}\widetilde{F}}_{\displaystyle\equiv\mathcal{F}_{\text{LR}}(\{\mathbf{r}_i\},T)} \tag{3} \label{eq:eq3}
\end{equation}
where $\widetilde{K}^{-1}$ is the pseudo-inverse of the effective stiffness matrix $\widetilde{K}$ \cite{Penrose1955, Dove_1993, Gonze1994}. Equation \eqref{eq:eq3} establishes the central result of the present theory: the ``bare" ionic free energy $\mathcal{F}_{\text{GF}}$ is renormalized by the lattice vibrational contribution $\mathcal{F}_{\text{lat}}^0$ and a lattice-relaxation contribution $\mathcal{F}_{\text{LR}}$. The vibrational term $\mathcal{F}_{\text{lat}}^0$ mainly contributes a smooth background entropy and plays a secondary role in renormalization. Explicit calculations indicate that its variation with respect to ionic configurations is typically an order of magnitude smaller than that of $\mathcal{F}_{\text{LR}}$ (Supplementary Information S2). The primary contribution is therefore $\mathcal{F}_{\text{LR}}$, a quadratic form of the configurational forces weighted by the lattice softness. As schematically illustrated in Fig. \ref{fig1}a, this non-positive quadratic coupling reshapes the bare ionic free-energy landscape by selectively stabilizing configurations that couple strongly to the soft lattice modes. Consequently, the hierarchy of geometrically accessible ionic configurations and the transition pathways connecting them are reorganized. Macroscopic fast ionic transport emerges once this renormalized landscape varies weakly and establishes sufficient connectivity across diverse ionic configurations.

\subsection*{Microscopic structure of lattice-relaxation contribution $\mathcal{F}_{\text{LR}}$}\label{subsec_CG}
In order to explicitly show how $\mathcal{F}_{\text{LR}}$ reorganizes the ionic configurations microscopically, we introduce a minimal coarse-grained representation \cite{Pardee1975, Tomoyose1991}, in which discrete occupation variables $n_{l\mu}$ ($n_{l\mu}=1$ if site $\mu$ of unit cell $l$ is occupied by ions and $n_{l\mu}=0$ otherwise) are introduced as source fields driving the lattice:
\begin{equation}
    \widetilde{F}_{\mathbf{q}s}(\{n_{l\mu}\}, T) = \frac{1}{\sqrt{N_c}}\sum_{l,\mu} \lambda_{\mathbf{q}s}^{\mu}(T)\,n_{l\mu}\, e^{-i\mathbf{q}\cdot\mathbf{R}_l^0} \tag{4} \label{eq:eq4}
\end{equation}
Here $\lambda_{\mathbf{q}s}^{\mu}(T)$ is the amplitude with which $n_{l\mu}$ couples to lattice mode $(\mathbf{q},s)$, and $N_c$ is the number of unit cells. Retaining the leading order in a systematic expansion of $\widetilde{K}^{-1}$ (Supplementary Information S1.3), i.e. approximating $\widetilde{K}(\{\mathbf{r}_i\},T)$ by its configuration-independent part $\widetilde{K}_0(T)$, we recast $\mathcal{F}_{\text{LR}}$ using equation \eqref{eq:eq4} and the eigenfrequencies $\Omega_{\mathbf{q}s}(T)$ of $\widetilde{K}_0(T)$ as
\begin{equation}
    \mathcal{F}_{\text{LR}}(\{n_{l\mu}\}, T) = -\frac{1}{2N_c} \sum_{\mathbf{q},s} \frac{1}{\Omega_{\mathbf{q}s}^2(T)} \left|\sum_{l,\mu} \lambda_{\mathbf{q}s}^{\mu}(T)\, n_{l\mu}e^{-i\mathbf{q}\cdot\mathbf{R}_l^0}\right|^2 \tag{5} \label{eq:eq5}
\end{equation}
Equation \eqref{eq:eq5} reveals the intrinsic collective nature of $\mathcal{F}_{\text{LR}}$. Each lattice mode contributes through two complementary factors: the inverse squared mode frequency $\Omega_{\mathbf{q}s}^{-2}$ that highlights soft lattice modes, and the squared amplitude of a coherent configurational field generated collectively by the entire ionic ensemble.

Using the relation $n_{l\mu}^2 = n_{l\mu}$, the collective renormalization in equation \eqref{eq:eq5} can be decomposed into one-body and two-body terms as:
\begin{equation}
    \mathcal{F}_{\text{LR}}(\{n_{l\mu}\}, T) = \sum_{l,\mu} h_{\mu}(T)\, n_{l\mu} \;+\; \sum_{(l,\mu)<(l',\mu')} V_{l\mu,l'\mu'}(T)\, n_{l\mu}\, n_{l'\mu'} \tag{6a} \label{eq:eq6a}
\end{equation}
where $(l,\mu)<(l',\mu')$ indicates $l<l'$, or $\mu < \mu'$ with $l' = l$.

\begin{figure}[h]
    \centering
    \includegraphics[width=1.0\linewidth]{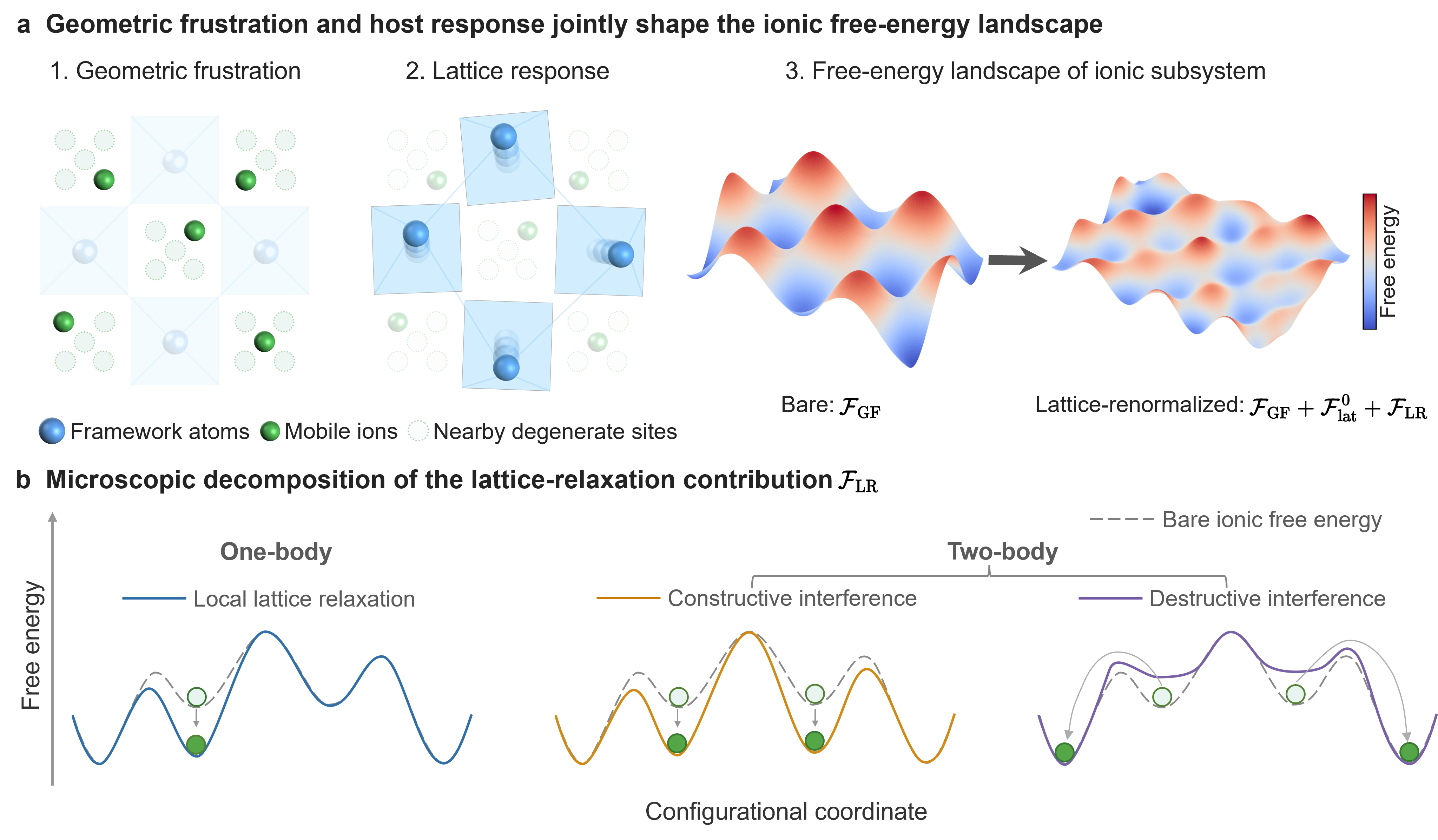}
    \caption{\textbf{Lattice-renormalized geometric frustration.} \textbf{a,} Multiple nearly degenerate ionic configurations generated by intrinsic geometric frustration and the host lattice response to ionic frustration. The bare conditional free-energy landscape of the ionic subsystem $\mathcal{F}_{\text{GF}}$ is renormalized by lattice vibration $\mathcal{F}_{\text{lat}}^0$ and lattice relaxation $\mathcal{F}_{\text{LR}}$ as  $\mathcal{F}_{\text{GF}}+\mathcal{F}_{\text{lat}}^0+\mathcal{F}_{\text{LR}}$ [equation \eqref{eq:eq3}]. \textbf{b,} Microscopic structure of lattice-relaxation contribution $\mathcal{F}_{\text{LR}}$ decomposed in a minimal coarse-grained representation [equation \eqref{eq:eq6a}]. The one-body term leads to local lattice relaxation that stabilizes ionic frustration. The constructive (destructive) two-body interference results in effective attraction (repulsion) of two ions and redistributes ionic correlations.}
    \label{fig1}
\end{figure}

The one-body contribution is
\begin{equation}
    h_{\mu}(T) = -\frac{1}{2N_c} \sum_{\mathbf{q},s} \frac{|\lambda_{\mathbf{q}s}^{\mu}(T)|^2}{\Omega_{\mathbf{q}s}^2(T)} \leqslant 0 \tag{6b} \label{eq:eq6b}
\end{equation}
It measures the degree to which occupying site $\mu$ lowers local free energy and thus acts as an effective potential that stabilizes arbitrary local frustrated ionic occupations through lattice relaxation (Fig. \ref{fig1}b). This local contribution is analogous to the self-trapping concept in polaron theory \cite{Klinger1985, Emin2012}, but here it arises as a configurational free-energy renormalization of ionic occupation rather than the dressing of an individual electron.

The two-body contribution
\begin{equation}
    V_{l\mu,l'\mu'}(T) = -\frac{1}{N_c}\sum_{\mathbf{q},s} \frac{\mathrm{Re}\left[\lambda_{\mathbf{q}s}^{\mu*}(T)\,\lambda_{\mathbf{q}s}^{\mu'}(T)e^{-i\mathbf{q}\cdot(\mathbf{R}_{l'}^0-\mathbf{R}_l^0)}\right]}{\Omega_{\mathbf{q}s}^2(T)} \tag{6c} \label{eq:eq6c}
\end{equation}
describes the non-local interference that extends beyond local lattice relaxation. Its sign and magnitude are determined by the constructive or destructive interference of the lattice-response fields induced by two occupied sites. In this picture, ionic correlations are not only determined by the bare ionic free-energy landscape, but also modulated by an effective attractive ($V_{l\mu,l'\mu'}<0$) or repulsive ($V_{l\mu,l'\mu'}>0$) potential generated by the constructive or destructive interference (Fig. \ref{fig1}b). The resulting ionic modulations explain the thermodynamic origin of concerted ionic migration \cite{He2017, Irvine2022}.

The decomposition above clarifies how $\mathcal{F}_{\text{LR}}$ gives rise to emergent reorganization of ionic configurations through local relaxation and non-local interference in real space. To identify the collective spatial rearrangements of ionic configurations across the crystal, it is natural to reformulate equation \eqref{eq:eq5} in reciprocal space. Fourier transforming the occupation variable $n_{l\mu}$ in equation \eqref{eq:eq5} yields (Supplementary Information S1.3)
\begin{equation}
    \mathcal{F}_{\text{LR}}(\{n_{\mathbf{q}\mu}\}, T) = -\frac{1}{2}\sum_{\mathbf{q}}\sum_{\mu\mu'} \mathcal{J}_{\mathbf{q}}^{\mu\mu'}(T)\, n_{\mathbf{q}\mu}^*\, n_{\mathbf{q}\mu'} \tag{7} \label{eq:eq7}
\end{equation}
where $n_{\mathbf{q}\mu}$ and $\mathcal{J}_{\mathbf{q}}^{\mu\mu'}(T) \equiv \sum_s \Omega_{\mathbf{q}s}^{-2}(T)\lambda_{\mathbf{q}s}^{\mu*}(T)\,\lambda_{\mathbf{q}s}^{\mu'}(T)$ represent the reciprocal-space occupation modulation and renormalization kernel, respectively. The dominant eigenmodes of $\mathcal{J}_{\mathbf{q}}^{\mu\mu'}$ identify the collective ionic correlations $n_{\mathbf{q}\mu}^* n_{\mathbf{q}\mu'}$ that are most strongly renormalized. This renormalization can amplify weak or latent correlations that are geometrically accessible within the bare frustrated ionic configuration space, rendering them observable as enhanced diffuse scattering in the static structure factor $S(\mathbf q)$ \cite{Hansen2006}. Equation \eqref{eq:eq7} therefore establishes a direct connection between free-energy landscape renormalization, ensemble-averaged ionic correlations $\langle n_{\mathbf{q}\mu}^*n_{\mathbf{q}\mu'}\rangle$, and the experimentally accessible partial static structure factor $S(\mathbf{q})\propto \sum_{\mu\mu'}\langle n_{\mathbf q\mu}^*n_{\mathbf q\mu'}\rangle$ of the ionic subsystem \cite{Keen2002}.

\section*{Multiscale signatures of lattice-renormalized geometric frustration}\label{sec_MD}
We now illustrate how lattice-renormalized geometric frustration manifests across multiple scales through atomistic simulations of cubic Li$_7$La$_3$Zr$_2$O$_{12}$ (c-LLZO) and AgCrSe$_2$. The dynamics driven by intrinsic geometric frustration, $\mathcal{H}_{\text{GF}}$, is separated numerically from the full ion-lattice coupled dynamics, $\mathcal{H}_{\text{GF}} + \mathcal{H}_{\text{lat}}$, by comparing individual frozen-lattice and full-lattice simulations (see Methods).

\begin{figure}[h]
\centering
\includegraphics[width=1.0\textwidth]{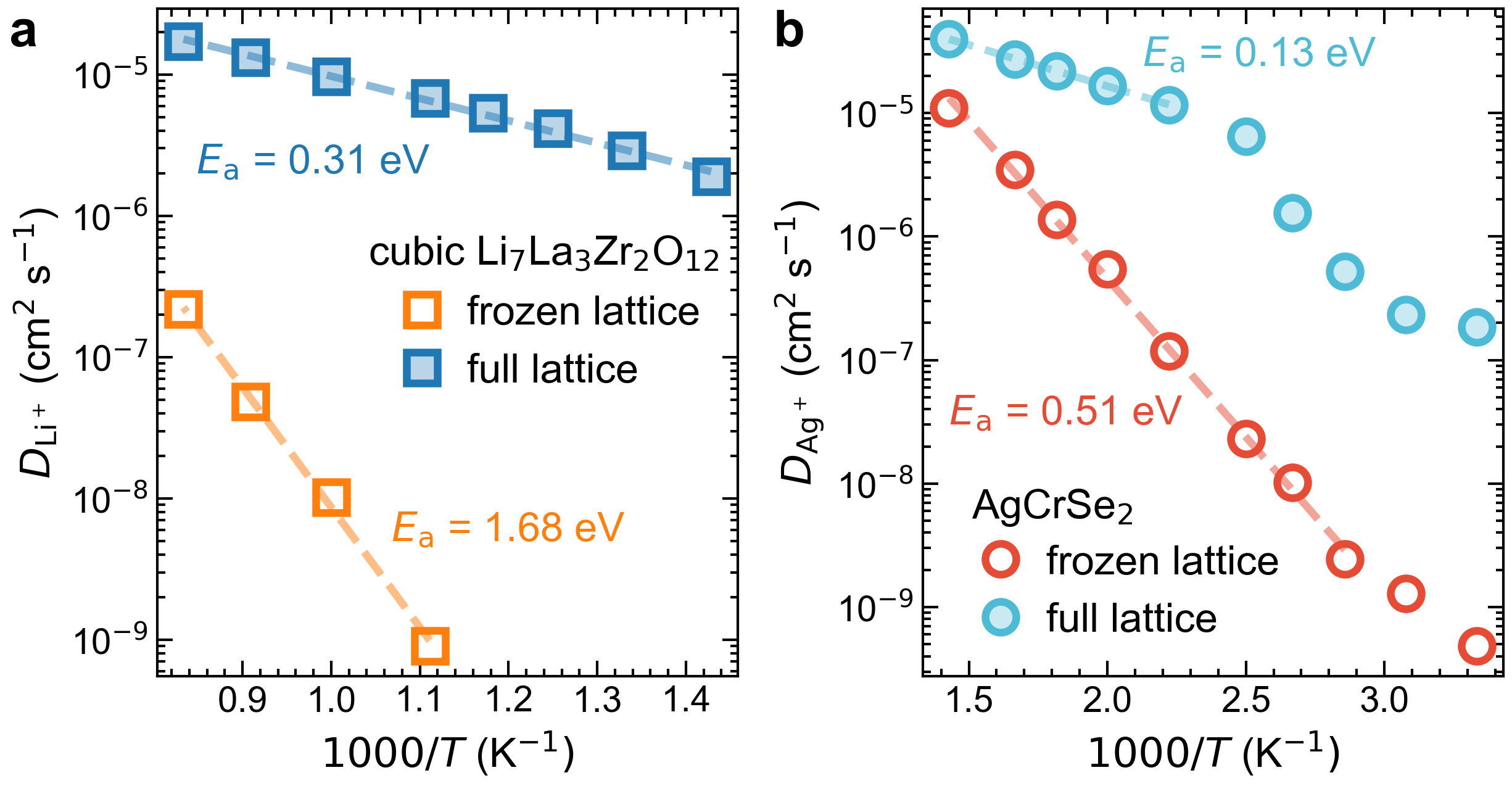}
\caption{\textbf{Lattice-renormalized ion diffusion.} \textbf{a,} Frozen-lattice and full-lattice diffusion coefficients (symbols) and Arrhenius fits (dashed lines) for cubic Li$_7$La$_3$Zr$_2$O$_{12}$ from 700 to 1200 K. Lattice renormalization reduces the activation energy from 1.68 eV to 0.31 eV. \textbf{b,} Frozen-lattice and full-lattice diffusion coefficients (symbols) \cite{Yang2026} and Arrhenius fits (dashed lines) for AgCrSe$_2$ from 300 to 700 K. The activation energy is reduced from 0.51 eV to 0.13 eV by lattice renormalization. The substantial reduction of activation energies in both materials provides a macroscopic signature of the free-energy renormalization described by equation \eqref{eq:eq3}.} \label{fig2}
\end{figure}

At the macroscopic scale, the immediate consequence of lattice-renormalized geometric frustration is the emergence of fast ionic transport. Specifically, the migration barrier on the landscape defined by equation \eqref{eq:eq3} is effectively ``flattened" through lattice response. In frozen-lattice c-LLZO, no Li$^+$ diffusion occurs below 900 K, whereas full lattice dynamics enhances diffusion coefficients by orders of magnitude (Fig. \ref{fig2}a). The corresponding activation energy, $E_{\text{a}}$, decreases from 1.68 eV in a frozen lattice to 0.31 eV in a fully responsive lattice. The parallel trend observed in AgCrSe$_2$ (Fig. \ref{fig2}b), characterized by a reduction of $E_{\text{a}}$ from 0.51 eV to 0.13 eV, indicates the same lattice-renormalization mechanism across chemically distinct materials. In comparison, the characteristic energy scales of $\mathcal{F}_{\text{LR}}$ are comparable to the changes in $E_{\text{a}}$, ranging from -0.41 to -0.25 eV per Li$^+$ ion for c-LLZO and from -0.34 to -0.10 eV per Ag$^+$ ion for AgCrSe$_2$ (see Supplementary Information S2). The quantitative discrepancies are likely due to an overestimation of the number of mobile ions that are actually involved in the diffusion process.

\begin{figure}[ht]
\centering
\includegraphics[width=1.0\textwidth]{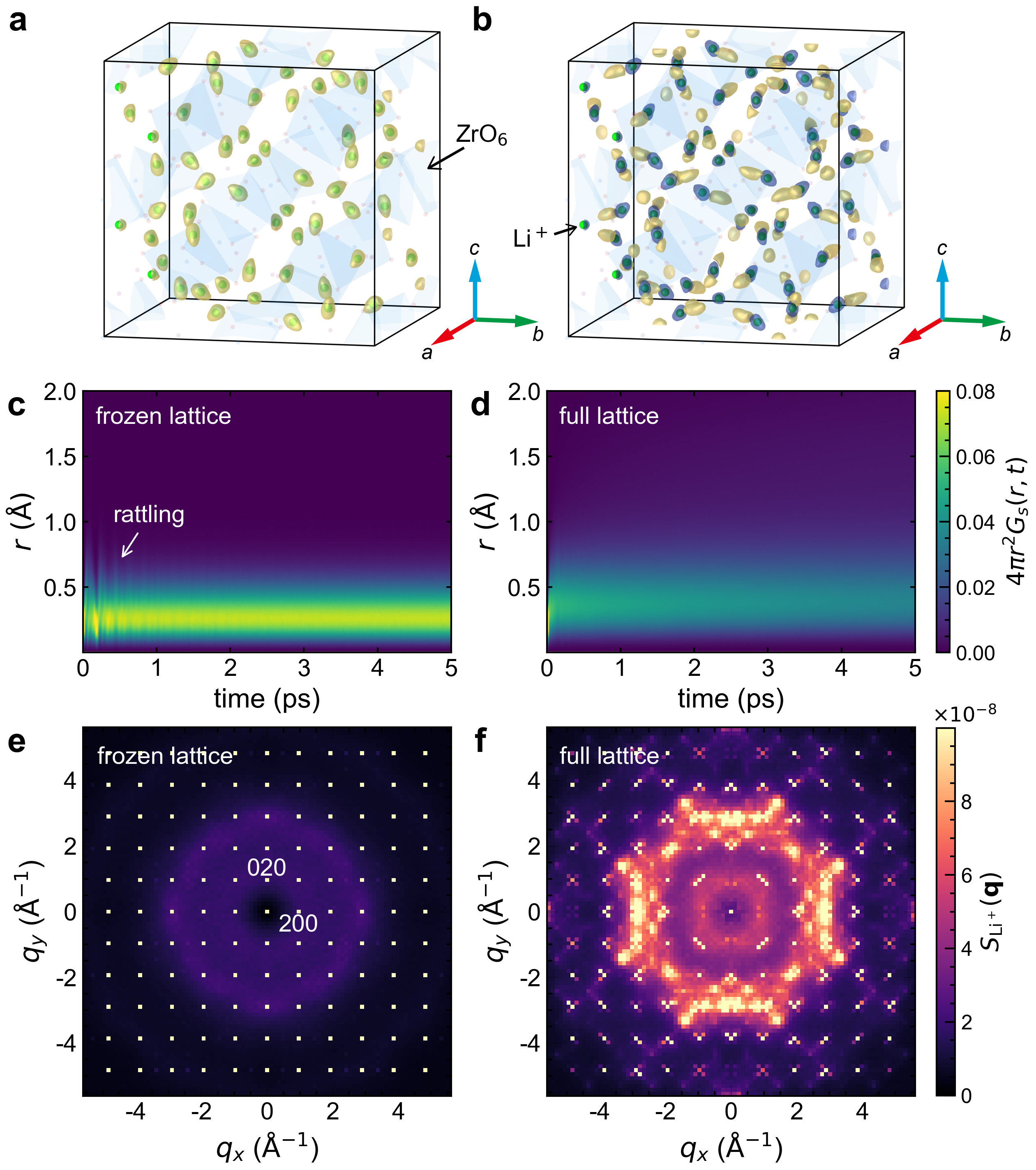}
\caption{\textbf{Lattice renormalization reshapes the frustrated ionic configuration through 
local lattice relaxation and interference.} Analysis of cubic Li$_7$La$_3$Zr$_2$O$_{12}$ at 700 K. \textbf{a,} Probability density distribution of Li$^+$ ions in the frozen-lattice limit, showing ions localized at regular tetrahedral and octahedral sites. \textbf{b,} Probability density difference $\Delta\rho_{\text{Li}^+}$ between the full-lattice density $\rho_{\text{Li}^+}^{\text{full}}$ and the frozen-lattice density $\rho_{\text{Li}^+}^{\text{GF}}$, revealing lattice-renormalized redistribution from regular sites (blue, negative) to nearby configurations (yellow, positive). The isosurface is $\pm$30$\rho_0$, where $\rho_0$ is the average density. \textbf{c,d,} Single Li$^+$ dynamics probed by the self-part of the van Hove correlation function $4\pi r^2G_s(r,t)$. In the frozen lattice (\textbf{c}), Li$^+$ remains dynamically 
localized despite substantial rattling, while the full lattice enables 
sustained migration (\textbf{d}). \textbf{e,f,} Simulated partial static structure factors $S_{\text{Li}^+}(\mathbf{q})$. Compared to the frozen-lattice result (\textbf{e}), the enhanced diffuse streaks around $\mathbf{q}$=\{600\}$^*$ in the full-lattice simulation (\textbf{f}) signify the emergence of collective short-range order, consistent with the collective correlations predicted by equation \eqref{eq:eq7}. Together, panels (\textbf{b}), (\textbf{d}) and (\textbf{f}) show how lattice renormalization progressively reshapes frustrated ionic configurations, from local real-space redistribution to collective reciprocal-space correlations.} \label{fig3}
\end{figure}

Microscopically, the local signature of lattice renormalization is the stabilization of local configurations predicted by equation \eqref{eq:eq6b}. This local renormalization appears structurally as a redistribution of ionic occupation probability. Figures \ref{fig3}a-b compare the frozen-lattice probability density distribution $\rho_{\text{Li}^+}^{\text{GF}}$ of Li$^+$ at 700 K and its redistribution under lattice response: $\Delta\rho = \rho_{\text{Li}^+}^{\text{full}} - \rho_{\text{Li}^+}^{\text{GF}}$ ($\rho_{\text{Li}^+}^{\text{full}}$: ion distribution in full lattice), in c-LLZO. The flexible lattice redistributes Li$^+$ ions from the conventional tetrahedral and octahedral occupations \cite{Xie2011} (blue negative isosurfaces) into neighboring configurations (yellow positive isosurfaces), consistent with the local relaxation described by equation \eqref{eq:eq6b}. The dynamic signature of this structural redistribution is evident in the self-part of the van Hove correlation function $G_s(r,t)$ \cite{Hansen2006, PhysRev.95.249}. In the frozen lattice, Li$^+$ ions remain dynamically localized despite intense rattling motion (Fig. \ref{fig3}c), whereas the full lattice enables sustained migration through a broader ensemble of accessible configurations (Fig. \ref{fig3}d).

Beyond local relaxation, the non-local interference between lattice-response fields reorganizes ionic correlations within the frustrated ionic subsystem. The resulting collective correlations are directly resolved in reciprocal space through the Li partial static structure factor $S_{\text{Li}^+}(\mathbf{q})$, which is approximated by electron diffraction simulations. In the frozen lattice, $S_{\text{Li}^+}(\mathbf{q})$ is dominated by Bragg diffraction with a weak, isotropic thermal diffuse background (Fig. \ref{fig3}e). In contrast, when full lattice dynamics are included, pronounced diffuse streaks emerge in the vicinity of \{600\}$^*$ reflections (Fig. \ref{fig3}f) (see Supplementary Information S3.1 for further analysis). These features reflect the reciprocal-space signatures of the wave-vector dependence of the renormalization kernel $\mathcal{J}_{\mathbf{q}}^{\mu\mu'}$ in equation \eqref{eq:eq7}. Furthermore, the appearance of satellite peaks around Bragg reflections, together with intensities appearing at systematically absent reflections, i.e. \{150\}$^*$ and \{370\}$^*$, indicates correlated ionic rearrangements that lower the average crystal symmetry. In AgCrSe$_2$, similar local relaxation and reciprocal-space correlation signatures are observed (Supplementary Information S3.2), which demonstrates the same lattice renormalization in chemically and structurally distinct fast ionic conductors.

\section*{Conclusion and outlook}\label{sec_conclusion}
In summary, we have developed a thermodynamic framework that unifies host lattice response and ionic geometric frustration for fast ionic transport. Our central result, the renormalization by lattice relaxation $\mathcal{F}_{\text{LR}}=-\frac{1}{2}\widetilde{F}^\dagger\widetilde{K}^{-1}\widetilde{F}$, reveals that lattice response selectively reshapes the frustrated ionic free-energy landscape by its coupling to configuration-dependent generalized forces. Within this framework, individual transport mechanisms such as lattice softness \cite{Muy2021, Jun2024}, ionic frustration \cite{Wood2021, Adelstein2016, DiStefano2019, Duvel2017}, and concerted ionic migration \cite{He2017} naturally emerge as different aspects of the same renormalized free-energy landscape.

Our present work also raises a fundamental question: what configurational topology, phases, and critical phenomena can emerge from such a renormalized free-energy landscape? For instance, the classical Frenkel-Kontorova model \cite{Kontorova1938, Braun1998} exhibits commensurate–incommensurate structures, collective sliding, and continuous depinning transitions arising from the competition between interparticle elasticity and a periodic substrate \cite{Aubry1983, Fisher1985, Reichhardt2016}. Whether analogous forms of configurational phases and criticality exist in lattice-renormalized geometric frustration systems remains an open question. Addressing it may ultimately reveal rich collective phenomena in solid-state ionics and deepen our understanding of superionic phase transitions \cite{Boyce1979, Hull2004, Hu2026} as well as ionic transport dynamics \cite{Dieterich1980, Tomoyose1991, Dyre2009}.

More broadly, this work reveals a general statistical-mechanical operation: integrating out the response of a deformable host generates a renormalized configurational free-energy landscape for the guest particles it contains. The same operation may therefore apply to other fields of confined mass transport, including diffusion in flexible porous frameworks \cite{Coudert2015, Zhao2019}, intercalation in layered materials \cite{Rajapakse2021, Vanderven2013}, and anomalous transport of confined water \cite{Bocquet2010, Radha2016}. 

\section*{Methods}\label{methods}
\subsection*{Atomistic simulations by machine-learning molecular dynamics}\label{method:MDsim}
Machine-learning molecular dynamics (MLMD) simulations were conducted using the Graphics Processing Units Molecular Dynamics with the neuroevolution potentials \cite{Xuke2025}. The potentials utilized in this work can be obtained from publicly available repositories, as reported in Refs. \cite{Zihan2024, Yang2026}.

Prior to the frozen-lattice simulations, structural optimization was performed using the FIRE algorithm \cite{PhysRevLett.97.170201}, and the atomic forces were converged to below 10$^{-3}$ eV/$\text{\AA}$. Subsequently, framework atoms were fixed at their force-equilibrium positions, whereas the mobile ions evolved exclusively under the static periodic potential and their mutual many-body interactions. This provides an operational realization of the pure GF dynamics, or $\mathcal{H}_{\text{GF}}$, of the ionic subsystem, while the full dynamics of both the ion and framework subsystems ($\mathcal{H}_{\text{GF}}+\mathcal{H}_{\text{lat}}$) are captured by full lattice simulations. The comparison between these two limits therefore directly probes the lattice renormalization. In addition to frozen-lattice and full-lattice simulations, frozen-ion calculations were also performed by fixing the ions at the reference or high-temperature disordered positions, while the host lattice atoms were allowed to evolve. This constrained simulation offers a realization of conditioned lattice dynamics. Table \ref{tab:table1} summarizes the constrained MLMD simulations employed in this work.
\begin{table}[t]
    \caption{\textbf{Summary of (constrained) machine-learning molecular dynamics.} The full-lattice simulation is a normal MLMD simulation, while the host lattice atoms (mobile ions) are kept fixed in the frozen-lattice (frozen-ion) simulations. Frozen-ion simulations are used to evaluate conditional vibrational free energy and lattice stiffness, as reported in Supplementary Information S2.}
    \label{tab:table1}
    \centering
    \renewcommand{\arraystretch}{1.5}
    \begin{tabular}{c c c c}
    \toprule
         & mobile ion & host lattice & quantities\\
    \midrule
    full lattice & free & free & $\mathcal{H}_{\text{GF}} + \mathcal{H}_{\text{lat}}$ \\
    frozen lattice & free & fixed at $\{\mathbf{R}_{l\kappa}^0\}$ & $\mathcal{H}_{\text{GF}}$ \\
    frozen ion & \makecell{fixed at $\{\mathbf{r}_i^{(R)}\}$ or disordered positions taken \\ from full lattice MLMD simulations} & free & $\mathcal{H}_{\text{lat}}$ \\
    \bottomrule
    \end{tabular}
\end{table}

For c-LLZO, a 4$\times$4$\times$4 supercell was constructed using the lattice parameters reported in Ref. \cite{Zihan2024}. MLMD simulations were performed in the temperature range from 700 to 1200 K. Each run comprised a 500 ps equilibrium stage in the canonical (NVT) ensemble using the Langevin thermostat with a timestep of 1 fs, followed by a 2 ns production stage in the microcanonical (NVE) ensemble for collecting atomic trajectories. Both disordered and ordered structures were considered in the frozen-lattice calculations. The disordered structures were built from independent MLMD snapshots at 1200 K to simulate c-LLZO, while the ordered structure was directly constructed from tetragonal LLZO. All calculations reported here were performed using the disordered structural model unless otherwise stated. For AgCrSe$_2$, an orthorhombic supercell was constructed from the conventional hexagonal unit cell using the transformation matrix $\begin{pmatrix} 20 & 0 & 0 \\ 12 & 24 & 0 \\ 0 & 0 & 3\end{pmatrix}$. In this case, both equilibration and production runs were carried out with a timestep of 2 fs.

Diffusion coefficients of Li$^+$ and Ag$^+$ ions were determined from the mean-square displacement as a function of time, as in our previous work \cite{Yang2026},
\begin{equation}
    D = \frac{1}{2N_I d}\lim_{t\to +\infty}\frac{\mathrm{d}}{\mathrm{d}t}\sum_{i=1}^{N_I}\langle\left|\mathbf{r}_i(t_0+t)-\mathbf{r}_i(t_0)\right|^2\rangle_{t_0} \tag{8} \label{eq:eq8}
\end{equation}
where $N_I$ denotes the total number of mobile ions, $d$ is the dimensionality of the system ($d=3$ for c-LLZO, $d=2$ for AgCrSe$_2$), $\mathbf{r}_i(t_0+t)$ and $\mathbf{r}_i(t_0)$ are the positions of the $i$-th ion at times $t_0+t$ and $t_0$, respectively, and $\langle\cdots\rangle$ represents the ensemble average. Activation energies $E_{\text{a}}$ were determined by fitting the temperature-dependent diffusion coefficients $D$ to the Arrhenius relation:
\begin{equation}
    D = D_0\exp\left(-\frac{E_{\text{a}}}{k_B T}\right) \tag{9} \label{eq:eq9}
\end{equation}
where $D_0$ is the diffusion pre-exponential factor.

The probability density distribution of ions $\rho_{\text{ion}}(\mathbf{r})$ is statistically averaged over the MLMD trajectories as
\begin{equation}
    \rho_{\text{ion}}(\mathbf{r}) = \frac{1}{N_I}\langle \sum_i\delta(\mathbf{r}-\mathbf{r}_i)\rangle \tag{10} \label{eq:eq10}
\end{equation}
where $\delta(\cdot)$ is the Dirac delta function, $\mathbf{r}$ and $\mathbf{r}_i$ are ionic coordinates in a unit cell.

The self-part of the van Hove correlation function \cite{Hansen2006} was calculated via:
\begin{equation}
    G_s(r,t) = \frac{1}{N_I}\langle\sum_{i=1}^N\delta(r-|\mathbf{r}_i(t)-\mathbf{r}_i(0)|)\rangle \tag{11} \label{eq:eq11}
\end{equation}
To suppress the trivial $\delta$-function singularity at $r=0$, $t=0$ and improve visual clarity, $4\pi r^2 G_s(r,t)$, instead of $G_s(r,t)$, is plotted.

\subsection*{Static structure factor $S_{\text{ion}}(\mathbf{q})$ of mobile ions}\label{method:sf}
The static structure factor $S(\mathbf{q})$ for an $N$-particle system with density $n(\mathbf{r})=\sum_{i=1}^N\delta(\mathbf{r}-\mathbf{r}_i)$ is defined as the ensemble average of density correlation in reciprocal space as \cite{Hansen2006}:
\begin{equation}
    S(\mathbf{q}) = \frac{1}{N}\langle n_{\mathbf{q}}^* n_{\mathbf{q}}\rangle \tag{12} \label{eq:eq12}
\end{equation}
where $n_{\mathbf{q}}=\int n(\mathbf{r})e^{-i\mathbf{q}\cdot\mathbf{r}}\text{d}\mathbf{r}$.

To approximate experimental observables, $S_{\text{ion}}(\mathbf{q})$ of the mobile ions was calculated using a scattering-weighted proxy of dynamical electron diffraction based on the multislice method implemented in abTEM \cite{abtem2021}. A total of 200 snapshots were extracted from the MLMD trajectories after removal of the framework atoms, such that only the ionic sublattice was retained. Multislice simulations were then performed using 3 MeV electrons, and the atomic potentials were described by the Lobato parametrization \cite{Lobato:mq5024}, with a real-space sampling of approximately $0.03\, \mathrm{\AA}/\text{pixel}$.

\backmatter

\bmhead{Supplementary information}

Details of all the theoretical derivations, numerical estimation of lattice renormalization, and atomistic simulation results are provided in the supplementary file.

\bmhead{Acknowledgements}
L. X. acknowledges the National Natural Science Foundation of China (Grants No. 12174176 and No. 12574045), and the Guangdong Provincial Key Laboratory of Advanced Thermoelectric Materials and Device Physics (Grant No. 2024B1212010001). J. Y. kindly acknowledges support from the Ministry of Education, Singapore, under AcRF Tier 2 (Grant No. MOE-T2EP50122-0016). L. X. further acknowledges Zihan Yan from Westlake University for valuable discussions. Computational resources were supported by the Open Source Supercomputing Center of S-A-I.

The authors acknowledge the use of ChatGPT (OpenAI) and the Overleaf AI Assistant during manuscript preparation. These tools were used to polish language, organize the manuscript, maintain consistent terminology, and enhance the clarity of presentation. All scientific ideas, theoretical formulations, computational results, interpretations, and conclusions were developed, validated, and approved by the authors, who take full responsibility for the contents of this work.

\section*{Declarations}
\begin{itemize}
\item Funding

The author L. X. acknowledges the National Natural Science Foundation of China (Grants No. 12174176 and No. 12574045), and the Guangdong Provincial Key Laboratory of Advanced Thermoelectric Materials and Device Physics (Grant No. 2024B1212010001). J. Y. kindly acknowledges support from the Ministry of Education, Singapore, under AcRF Tier 2 (Grant No. MOE-T2EP50122-0016).

\item Conflict of interest/Competing interests

The authors declare no competing interests.



\item Consent for publication

Not applicable.

\item Data availability

Raw data were generated using the Open Source Supercomputing Center of S-A-I. The machine-learning potentials for LLZO and AgCrSe$_2$ are available at \url{https://github.com/zhyan0603/SourceFiles/tree/main/NEP-ZBL/NEP\_train/NEP802-ZBL\_train} and \url{https://github.com/xielin1984/NEP\_potentials}, respectively. Derived data supporting the findings of this study are available upon request.

\item Code availability

Graphics Processing Units Molecular Dynamics (GPUMD) is available at \url{https://github.com/brucefan1983/GPUMD}. abTEM is available at \url{https://github.com/abTEM/abTEM}. OVITO, which is available at \url{https://ovito.org} \cite{ovito}, is used for data analysis and rendering.

\item Author contribution

The project was conceived by all authors. L. X. performed the atomistic simulations. All authors contributed to the editing of the manuscript.
\end{itemize}

\noindent



\bibliography{sn-bibliography}

\end{document}


\begin{center}
{\Large Supplementary Information}\\[1em]
{\bfseries Lattice-renormalized geometric frustration drives fast ionic transport}\\[1em]
Jianmin Yang$^{1,2}$ and Lin Xie$^{1*}$\\[0.5em]
{\small
$^{1*}$Dongguan Key Laboratory of Solid-State Battery Materials, School of Physical Sciences, Great Bay University, Dongguan, 523000, Guangdong, China.\\
$^{2}$Department of Materials Science and Engineering, National University
of Singapore, 117575, Singapore.\\
$^{*}$Correspondence: xielin@gbu.edu.cn
}
\end{center}

\author[1,2]{\fnm{Jianmin} \sur{Yang}}\email{e0546047@u.nus.edu}
\author*[1]{\fnm{Lin} \sur{Xie}}\email{xielin@gbu.edu.cn}
\affil*[1]{\orgdiv{Dongguan Key Laboratory of Solid-State Battery Materials, School of Physical Sciences}, \orgname{Great Bay University}, \city{Dongguan}, \postcode{523000}, \state{Guangdong}, \country{China}}
\affil[2]{\orgdiv{Department of Materials Science and Engineering}, \orgname{National University of Singapore}, \postcode{117575}, \country{Singapore}}

\section*{S1. Detailed derivations}\label{s1}
\subsection*{S1.1. Microscopic Hamiltonian}\label{s1.1}
In this section, we present the derivation of lattice-renormalized geometric frustration in detail. We first introduce a reference structure $R = \{\mathbf{r}_i^{(R)}, \mathbf{R}_{l\kappa}^0\}$, which consists of a reference ionic configuration $\{\mathbf{r}_i^{(R)}\}$ and a reference lattice framework $\{\mathbf{R}_{l\kappa}^0\}$. The reference structure is in force equilibrium under periodic boundary conditions. All lattice displacements are then defined with respect to this reference lattice framework as $\mathbf{u}_{l\kappa} = \mathbf{R}_{l\kappa} - \mathbf{R}_{l\kappa}^0$. In principle, this reference state can be constructed from any dynamically stable crystal structure in force equilibrium.

The full Hamiltonian of the mobile ions and host lattice can be written as
\begin{equation}
    \mathcal{H}=\sum_i\frac{\mathbf{p}_i^2}{2m_i}+U_{\text{ion}}^{\text{bare}}(\{\mathbf{r}_i\}) + \mathcal{V}_{\text{lat-ion}}(\{\mathbf{r}_i\},\{\mathbf{R}_{l\kappa}\}) + \sum_{l,\kappa}\frac{\mathbf{p}_{l\kappa}^2}{2M_{l\kappa}} + U_{\text{lat}}(\{\mathbf{R}_{l\kappa}\}) \tag{S1} \label{eq:eqS1}
\end{equation}
where $\mathbf{p}_i$, $m_i$, and $\mathbf{r}_i$ are the momentum, mass, and position of the $i$-th mobile ion, while $\mathbf{p}_{l\kappa}$ and $M_{l\kappa}$ are the corresponding quantities for the framework atom $\kappa$ in unit cell $l$. Here $U_{\text{ion}}^{\text{bare}}$ and $U_{\text{lat}}$ represent the interactions within the many-body ionic subsystem and the host lattice, respectively, and $\mathcal{V}_{\text{lat-ion}}$ describes their mutual coupling.

Starting from equation \eqref{eq:eqS1}, we carry out a perturbative expansion of $\mathcal{V}_{\text{lat-ion}}$ and $U_{\text{lat}}$ in powers of displacements $\{\mathbf{u}_{l\kappa}\}$ about the force-equilibrium positions $\{\mathbf{R}_{l\kappa}^0\}$ for a given ionic configuration $\{\mathbf{r}_i\}$
\begin{equation}
    \begin{split}
    \mathcal{V}_{\text{lat-ion}}(\{\mathbf{r}_i\},\{\mathbf{R}_{l\kappa}\}) = & \,\mathcal{V}_{\text{lat-ion}}(\{\mathbf{r}_i\},\{\mathbf{R}_{l\kappa}^0\}) + \sum_{l,\kappa,\alpha}\frac{\partial\mathcal{V}_{\text{lat-ion}}}{\partial R_{l\kappa}^{\alpha}}\bigg|_{\mathbf{R}_{l\kappa}=\mathbf{R}_{l\kappa}^0}u_{l\kappa}^\alpha \, \\ & + \frac{1}{2}\sum_{l,\kappa,\alpha}\sum_{l',\kappa',\beta}\frac{\partial^2\mathcal{V}_{\text{lat-ion}}}{\partial R_{l\kappa}^\alpha \partial R_{l'\kappa'}^\beta}\bigg|_{\mathbf{R}_{l\kappa}=\mathbf{R}_{l\kappa}^0}u_{l\kappa}^\alpha u_{l'\kappa'}^\beta + \cdots
    \end{split} \tag{S2} \label{eq:eqS2}
\end{equation}
and
\begin{equation}
\begin{split}
    U_{\text{lat}}(\{\mathbf{R}_{l\kappa}\}) = & \, U_0 +  \sum_{l,\kappa,\alpha}\frac{\partial U_{\text{lat}}}{\partial R_{l\kappa}^{\alpha}}\bigg|_{\mathbf{R}_{l\kappa}=\mathbf{R}_{l\kappa}^0}u_{l\kappa}^\alpha \\ & + \frac{1}{2}\sum_{l,\kappa,\alpha}\sum_{l',\kappa',\beta}\frac{\partial^2 U_{\text{lat}}}{\partial R_{l\kappa}^\alpha \partial R_{l'\kappa'}^\beta}\bigg|_{\mathbf{R}_{l\kappa}=\mathbf{R}_{l\kappa}^0}u_{l\kappa}^\alpha u_{l'\kappa'}^\beta + U_{\text{anh}}(\{\mathbf{R}_{l\kappa}\})
\end{split} \tag{S3} \label{eq:eqS3}
\end{equation}
where $U_0$ is the potential energy of the static lattice, $\alpha$ and $\beta$ denote the Cartesian indices, and $U_{\text{anh}}(\{\mathbf{R}_{l\kappa}\})$ describes the higher-order anharmonic interactions of the framework. Equation \eqref{eq:eqS3} corresponds to the standard Taylor expansion used in lattice dynamics \cite{PhysRev.128.2589, RevModPhys.40.1, Leibfried1961}. The first term in equation \eqref{eq:eqS2}, $\mathcal{V}_{\text{lat-ion}}(\{\mathbf{r}_i\},\{\mathbf{R}_{l\kappa}^0\})$, represents a static periodic potential generated by the lattice framework and is incorporated into the Hamiltonian of the ionic subsystem, $\mathcal{H}_{\text{GF}}$, as $V_0(\{\mathbf{r}_i\})$.

The remaining terms in equation \eqref{eq:eqS2}, namely $\delta\mathcal{V}_{\text{lat-ion}}\equiv\mathcal{V}_{\text{lat-ion}} - V_0$, are combined with equation \eqref{eq:eqS3}, and $\mathcal{H}$ is rewritten as equation (1) in the main text. Explicitly, the first-order term can be written as
\begin{equation}
    \sum_{l,\kappa,\alpha}\left(\frac{\partial\mathcal{V}_{\text{lat-ion}}}{\partial R_{l\kappa}^{\alpha}} + \frac{\partial U_{\text{lat}}}{\partial R_{l\kappa}^{\alpha}} \right)\bigg|_{\mathbf{R}_{l\kappa}=\mathbf{R}_{l\kappa}^0} u_{l\kappa}^\alpha = -\sum_{l,\kappa,\alpha} F_{l\kappa}^\alpha(\{\mathbf{r}_i\}) u_{l\kappa}^\alpha \tag{S4} \label{eq:eqS4}
\end{equation}
where $F_{l\kappa}^\alpha(\{\mathbf{r}_i\}) \equiv -(\partial\mathcal{V}_{\text{lat-ion}}/\partial R_{l\kappa}^{\alpha} + \partial U_{\text{lat}}/{\partial R_{l\kappa}^{\alpha}})\big|_{\mathbf{R}_{l\kappa}=\mathbf{R}_{l\kappa}^0}$ is the configuration-dependent generalized force acting on the framework atom $(l,\kappa)$. By construction of the reference state, this generalized force vanishes in the reference configuration $\{\mathbf{r}_i^{(R)}\}$, i.e., $\mathbf{F}_{l\kappa}=0$. For arbitrary ionic configuration $\{\mathbf{r}_i\}$, $\mathbf{F}_{l\kappa}\neq 0$ in the general case.

Analogously, the second-order term is given by
\begin{multline}
    \frac{1}{2}\sum_{l,\kappa,\alpha}\sum_{l',\kappa',\beta}\left(\frac{\partial^2\mathcal{V}_{\text{lat-ion}}}{\partial R_{l\kappa}^\alpha \partial R_{l'\kappa'}^\beta} + \frac{\partial^2 U_{\text{lat}}}{\partial R_{l\kappa}^\alpha \partial R_{l'\kappa'}^\beta}\right)\bigg|_{\mathbf{R}_{l\kappa}=\mathbf{R}_{l\kappa}^0} u_{l\kappa}^\alpha u_{l'\kappa'}^\beta = \\ 
    \frac{1}{2}\sum_{l,\kappa,\alpha}\sum_{l',\kappa',\beta}\Phi_{l\kappa,l'\kappa'}^{\alpha\beta}(\{\mathbf{r}_i\}) u_{l\kappa}^\alpha u_{l'\kappa'}^\beta \tag{S5} \label{eq:eqS5}
\end{multline}
where $\Phi_{l\kappa,l'\kappa'}^{\alpha\beta}(\{\mathbf{r}_i\})$ is the generalized second-order interatomic force constant. $\Phi_{l\kappa,l'\kappa'}^{\alpha\beta}(\{\mathbf{r}_i\})$ is symmetric under the simultaneous exchange of the Cartesian components $(\alpha,\beta)$ and the atomic indices $(l\kappa,l'\kappa')$, i.e., $\Phi_{l'\kappa',l\kappa}^{\beta\alpha}(\{\mathbf{r}_i\}) = \Phi_{l\kappa,l'\kappa'}^{\alpha\beta}(\{\mathbf{r}_i\})$. The remaining higher-order expansions in equation \eqref{eq:eqS2} are combined into the anharmonic lattice potential $U_{\text{anh}}(\{\mathbf{R}_{l\kappa}\})$ and are collectively denoted by $\mathcal{U}_{\text{anh}}(\{\mathbf{r}_i\},\{\mathbf{R}_{l\kappa}\})$.

Subsequently, we can rewrite equations \eqref{eq:eqS4} and \eqref{eq:eqS5} by transforming the displacement $\mathbf{u}_{l\kappa}$ into normal-mode coordinates $X_{\mathbf{q}s}$ as \cite{PhysRev.128.2589, RevModPhys.40.1}
\begin{align}
    u_{l\kappa}^\alpha & = \frac{1}{\sqrt{N_c M_{\kappa}}}\sum_{\mathbf{q},s}\epsilon_{\alpha}^{(\kappa,s)}(\mathbf{q})\, e^{i\mathbf{q}\cdot\mathbf{R}_l^0}X_{\mathbf{q}s} \tag{S6a} \label{eq:eqS6a} \\
    X_{\mathbf{q}s} &= \sqrt{\frac{\hbar}{2\omega_{\mathbf{q}s}}}\left(a_{\mathbf{q}s}+a_{-\mathbf{q}s}^\dagger\right) \tag{S6b} \label{eq:eqS6b}
\end{align}
where $N_c$ is the number of unit cells in the crystal, $M_\kappa$ the mass of an atom of type $\kappa$, $\hbar$ the reduced Planck constant, and $\omega_{\mathbf{q}s}$ and $\epsilon_\alpha^{(\kappa,s)}(\mathbf{q})$, respectively, the normal-mode frequency and eigenvector corresponding to wave vector $\mathbf{q}$ and branch index $s$. The normal-mode coordinate $X_{\mathbf{q}s}$ is expressed in terms of the creation and annihilation operators $a_{-\mathbf{q}s}^\dagger$ and $a_{\mathbf{q}s}$. $X_{\mathbf{q}s}$ can be decomposed into an adiabatic lattice displacement $x_{\mathbf{q}s}$ and a thermal/quantum fluctuation $\xi_{\mathbf{q}s}$: $X_{\mathbf{q}s} = x_{\mathbf{q}s} + \xi_{\mathbf{q}s}$, with $\langle X_{\mathbf{q}s}\rangle = x_{\mathbf{q}s}$ and vanishing ensemble average of the fluctuation, $\langle\xi_{\mathbf{q}s}\rangle = 0$. The sum in equation \eqref{eq:eqS6a} is over the full Brillouin zone and the complex coordinates satisfy $X_{-\mathbf{q}s}=X_{\mathbf{q}s}^*$ and $\epsilon_\alpha^{(\kappa,s)}(\mathbf{q}) = \epsilon_\alpha^{(\kappa,s)*}(-\mathbf{q})$.

Upon substituting equation \eqref{eq:eqS6a} into equation \eqref{eq:eqS4}, the generalized force term becomes
\begin{equation}
   -\sum_{l,\kappa,\alpha}F_{l\kappa}^\alpha(\{\mathbf{r}_i\})\, u_{l\kappa}^\alpha = - \sum_{l,\kappa,\alpha}\frac{F_{l\kappa}^\alpha(\{\mathbf{r}_i\})}{\sqrt{N_cM_{\kappa}}}\sum_{\mathbf{q},s}\epsilon_\alpha^{(\kappa,s)}(\mathbf{q})e^{i\mathbf{q}\cdot\mathbf{R}_l^0}X_{\mathbf{q}s} \tag{S7a} \label{eq:eqS7a}
\end{equation}
By interchanging the order of summation, equation \eqref{eq:eqS7a} can be expressed as
\begin{equation}
    - \sum_{l,\kappa,\alpha}\frac{F_{l\kappa}^\alpha(\{\mathbf{r}_i\})}{\sqrt{N_cM_{\kappa}}}\sum_{\mathbf{q},s}\epsilon_\alpha^{(\kappa,s)}(\mathbf{q})e^{i\mathbf{q}\cdot\mathbf{R}_l^0}X_{\mathbf{q}s} = - \sum_{\mathbf{q},s}\left[\sum_{l,\kappa,\alpha}\frac{F_{l\kappa}^\alpha(\{\mathbf{r}_i\})}{\sqrt{N_cM_{\kappa}}}\epsilon_\alpha^{(\kappa,s)}(\mathbf{q})e^{i\mathbf{q}\cdot\mathbf{R}_l^0}\right]X_{\mathbf{q}s} \tag{S7b} \label{eq:eqS7b}
\end{equation}
which naturally motivates the definition of a generalized force in reciprocal space:
\begin{equation}
    F_{\mathbf{q}s}(\{\mathbf{r}_i\}) \equiv \frac{1}{\sqrt{N_c}}\sum_{l,\kappa,\alpha}\frac{F_{l\kappa}^\alpha(\{\mathbf{r}_i\})}{\sqrt{M_{\kappa}}}\epsilon_\alpha^{(\kappa,s)*}(\mathbf{q})\,e^{-i\mathbf{q}\cdot\mathbf{R}_l^0} \tag{S7c} \label{eq:eqS7c}
\end{equation}
The physical interpretation of equation \eqref{eq:eqS7c} is straightforward: $F_{\mathbf{q}s}(\{\mathbf{r}_i\})$ is the Fourier transform of the mass-weighted force $M_{\kappa}^{-1/2}F_{l\kappa}^\alpha$ acting on the framework atom $(l,\kappa)$, projected onto the eigenvector $\epsilon_\alpha^{(\kappa,s)*}(\mathbf{q})$. Equation \eqref{eq:eqS7a} can thus be rewritten as
\begin{equation}
    -\sum_{l,\kappa,\alpha}F_{l\kappa}^\alpha(\{\mathbf{r}_i\})u_{l\kappa}^\alpha = -\sum_{\mathbf{q},s} F_{\mathbf{q}s}^*(\{\mathbf{r}_i\})X_{\mathbf{q}s} = -F^\dagger(\{\mathbf{r}_i\}) X \tag{S7d} \label{eq:eqS7d}
\end{equation}
where $\dagger$ denotes the Hermitian conjugate.

The second-order term [equation \eqref{eq:eqS5}] in normal coordinates can be rewritten as
\begin{equation}
    \frac{1}{2N_c}\sum_{\mathbf{q}',s'}\sum_{\mathbf{q},s}\sum_{l,\kappa,\alpha}\sum_{l',\kappa',\beta}\frac{\Phi_{l\kappa,l'\kappa'}^{\alpha\beta}(\{\mathbf{r}_i\})}{\sqrt{M_\kappa M_{\kappa'}}}\epsilon_\alpha^{(\kappa,s)}(\mathbf{q})\epsilon_\beta^{(\kappa',s')}(\mathbf{q}')\,e^{i(\mathbf{q}\cdot\mathbf{R}_l^0+\mathbf{q}'\cdot\mathbf{R}_{l'}^0)}X_{\mathbf{q}'s'}X_{\mathbf{q}s} \tag{S8a} \label{eq:eqS8a}
\end{equation}
where we introduce a generalized dynamical matrix in reciprocal space, defined as
\begin{equation}
    K_{\mathbf{qq}'}^{ss'}(\{\mathbf{r}_i\}) \equiv \frac{1}{N_c}\sum_{l,\kappa,\alpha}\sum_{l',\kappa',\beta}\frac{\Phi_{l\kappa,l'\kappa'}^{\alpha\beta}(\{\mathbf{r}_i\})}{\sqrt{M_\kappa M_{\kappa'}}}\epsilon_\alpha^{(\kappa,s)*}(\mathbf{q})\epsilon_\beta^{(\kappa',s')}(\mathbf{q}')\,e^{i(-\mathbf{q}\cdot\mathbf{R}_l^0+\mathbf{q}'\cdot\mathbf{R}_{l'}^0)} \tag{S8b} \label{eq:eqS8b}
\end{equation}
Equation \eqref{eq:eqS5} is then recast as
\begin{equation}
    \frac{1}{2}\sum_{l,\kappa,\alpha}\sum_{l',\kappa',\beta}\Phi_{l\kappa,l'\kappa'}^{\alpha\beta}(\{\mathbf{r}_i\}) u_{l\kappa}^\alpha u_{l'\kappa'}^\beta = \frac{1}{2} \sum_{\mathbf{q},s}\sum_{\mathbf{q}',s'}X_{\mathbf{q},s}^* K_{\mathbf{qq}'}^{ss'}(\{\mathbf{r}_i\}) X_{\mathbf{q}'s'} = \frac{1}{2}X^\dagger K(\{\mathbf{r}_i\})X \tag{S8c} \label{eq:eqS8c}
\end{equation}
Given the symmetry of $\Phi_{l\kappa,l'\kappa'}^{\alpha\beta}$, $K$ is Hermitian, i.e., $K^\dagger = K$.

Collecting the above terms, the microscopic Hamiltonian [equation (1) in the main text] can be explicitly expressed as
\begin{align}
    \mathcal{H}_{\text{GF}} & = \sum_i\frac{\mathbf{p}_i^2}{2m_i} + V_0(\{\mathbf{r}_i\}) + U_{\text{ion}}^{\text{bare}}(\{\mathbf{r}_i\}) \tag{S9a} \label{eq:eqS9a} \\
    \mathcal{H}_{\text{lat}} & = \frac{1}{2}\mathbf{P}^\dagger\mathbf{P} - F^\dagger(\{\mathbf{r}_i\}) X + \frac{1}{2}X^\dagger K(\{\mathbf{r}_i\})X + \mathcal{U}_{\text{anh}}(\{\mathbf{r}_i\},X) \tag{S9b} \label{eq:eqS9b}
\end{align}
where $F$, $K(\{\mathbf{r}_i\})$, and $X$ are written in matrix notation, and $\mathbf{P}$ denotes the canonical momentum.

\subsection*{S1.2. Conditional free energy}\label{s1.2}
For a given ionic configuration $\{\mathbf{r}_i\}$ and an adiabatic lattice displacement $x_{\mathbf{q}s}$ that constrains $X_{\mathbf{q}s}$ by $\langle X_{\mathbf{q}s} \rangle = x_{\mathbf{q}s}$, we trace out the momentum and thermal fluctuation $\xi$ \cite{Landau1980, Chandler1987, Pathria2011, Tuckerman2010}. The corresponding partition function is
\begin{align}
    Z(\{\mathbf{r}_i\},x,T) & = \text{Tr}_{\mathbf{p}}\text{Tr}_{\mathbf{P},\xi}\, e^{-\beta\left(\mathcal{H}_{\text{GF}}+\mathcal{H}_{\text{lat}}\right)} \tag{S10a} \label{eq:eqS10a} \\
    & = \text{Tr}_{\mathbf{p}}\, e^{-\beta\left(\sum_i\frac{\mathbf{p}_i^2}{2m_i}+V_0+U_{\text{ion}}^{\text{bare}}\right)} \text{Tr}_{\mathbf{P},\xi}\, e^{-\beta\left(\frac{1}{2}\mathbf{P}^\dagger\mathbf{P}-F^\dagger X+\frac{1}{2}X^\dagger KX+\mathcal{U}_{\text{anh}}\right)} \nonumber \\
    & = Z_{\text{ion}}^{\text{ke}}(T)Z_{\text{GF}}(\{\mathbf{r}_i\})\,\text{Tr}_{\mathbf{P},\xi}\, e^{-\beta\left(\frac{1}{2}\mathbf{P}^\dagger\mathbf{P}-F^\dagger X+\frac{1}{2}X^\dagger KX+\mathcal{U}_{\text{anh}}\right)} \nonumber
\end{align}
where $\beta^{-1} = k_B T$ and
\begin{align}
    Z_{\text{ion}}^{\text{ke}}(T) & \equiv \text{Tr}_{\mathbf{p}}\, e^{-\beta\sum_i\frac{\mathbf{p}_i^2}{2m_i}} = \prod_i\frac{(2\pi m_i k_B T)^{3/2}}{(2\pi\hbar)^3} \tag{S10b} \label{eq:eqS10b} \\
    Z_{\text{GF}}(\{\mathbf{r}_i\}) & \equiv e^{-\beta\left(V_0+U_{\text{ion}}^{\text{bare}}\right)} \tag{S10c} \label{eq:eqS10c}
\end{align}
Here, $Z_{\text{ion}}^{\text{ke}}(T)$ is a prefactor that only depends on temperature.

We now rewrite the framework contribution in equation \eqref{eq:eqS10a} by expanding equation \eqref{eq:eqS9b} with $X = x + \xi$:
\begin{multline}
    \frac{1}{2}\mathbf{P}^\dagger\mathbf{P} - F^\dagger(x+\xi) + \frac{1}{2}(x^\dagger+\xi^\dagger) K (x + \xi) + \mathcal{U}_{\text{anh}} \\ = -F^\dagger x + \frac{1}{2}x^\dagger Kx + \frac{1}{2}\mathbf{P}^\dagger\mathbf{P} - F^\dagger\xi + \frac{1}{2}(x^\dagger K\xi+\xi^\dagger K x) + \frac{1}{2}\xi^\dagger K\xi + \mathcal{U}_{\text{anh}}\tag{S11a} \label{eq:eqS11a}
\end{multline}
Using the property $K=K^\dagger$, we have $\frac{1}{2}(x^\dagger K\xi + \xi^\dagger K x) = \frac{1}{2}(x^\dagger K^\dagger\xi + \xi^\dagger Kx) = x^\dagger K^\dagger\xi$. Therefore, the $\xi$-dependent linear terms combine as $-(F-Kx)^\dagger\xi$, and we can define an uncompensated generalized force acting on the lattice as $\delta F = F - Kx$. Substituting equation \eqref{eq:eqS11a} into equation \eqref{eq:eqS10a} leads to
\begin{multline}
    \text{Tr}_{\mathbf{P},\xi}\, e^{-\beta\left(\frac{1}{2}\mathbf{P}^\dagger\mathbf{P}-F^\dagger X+\frac{1}{2}X^\dagger KX+\mathcal{U}_{\text{anh}}\right)} \\ = e^{-\beta(-F^\dagger x+\frac{1}{2}x^\dagger Kx)} \text{Tr}_{\mathbf{P},\xi}\, e^{-\beta\left[\frac{1}{2}\mathbf{P}^\dagger\mathbf{P} - \delta F^\dagger\xi + \frac{1}{2}\xi^\dagger K\xi + \mathcal{U}_{\text{anh}}\right]} \tag{S11b} \label{eq:eqS11b}
\end{multline}
The operator $\text{Tr}_{\mathbf{P},\xi}$ therefore corresponds to the partition function of a quantum harmonic oscillator subject to an effective external generalized force $\delta F$ and anharmonic interactions $\mathcal{U}_{\text{anh}}(\{\mathbf{r}_i\},x+\xi)$. This motivates the definition of a conditional lattice free energy $\mathcal{F}_{\text{lat}}(\{\mathbf{r}_i\},x,T)$, and an associated conditional lattice vibrational free energy $\mathcal{F}_{\text{vib}}(\{\mathbf{r}_i\},x,T)$, via equation \eqref{eq:eqS11b}
\begin{equation}
\begin{split}
    e^{-\beta\mathcal{F}_{\text{lat}}} & \equiv e^{-\beta(-F^\dagger x+\frac{1}{2}x^\dagger Kx)}\text{Tr}_{\mathbf{P},\xi}\, e^{-\beta\left[\frac{1}{2}\mathbf{P}^\dagger\mathbf{P}-\delta F^\dagger\xi+\frac{1}{2}\xi^\dagger K\xi + \mathcal{U}_{\text{anh}}\right]}\\
    & = e^{-\beta(-F^\dagger x+\frac{1}{2}x^\dagger Kx)}\,e^{-\beta\mathcal{F}_{\text{vib}}}
\end{split} \tag{S11c} \label{eq:eqS11c}
\end{equation}

Combining equations \eqref{eq:eqS10a}-\eqref{eq:eqS11c}, the conditional free energy is determined as
\begin{equation}
\begin{split}
    \mathcal{F}(\{\mathbf{r}_i\},x,T) = & -\frac{1}{\beta}\ln Z(\{\mathbf{r}_i\},x,T) \\
    = & -\frac{1}{\beta}\ln{\left[Z_{\text{ion}}^{\text{ke}}(T)Z_{\text{GF}}(\{\mathbf{r}_i\})\,e^{-\beta\mathcal{F}_{\text{lat}}}\right]}\\
    = & \underbrace{-\frac{1}{\beta}\ln Z_{\text{ion}}^{\text{ke}}(T) + V_0(\{\mathbf{r}_i\}) + U_{\text{ion}}^{\text{bare}}(\{\mathbf{r}_i\})}_{\displaystyle\equiv\mathcal{F}_{\text{GF}}(\{\mathbf{r}_i\},T)} + \mathcal{F}_{\text{lat}}(\{\mathbf{r}_i\},x,T)
\end{split} \tag{S12} \label{eq:eqS12}
\end{equation}
where $\mathcal{F}_{\text{GF}}(\{\mathbf{r}_i\},T)$ is the conditional free energy of the ionic subsystem under the constraints of geometric frustration.

Notably, a conditional free energy inequality holds for arbitrary ionic configurations $\{\mathbf{r}_i\}$ whenever a stable relaxation with $x \neq 0$ exists, namely
\begin{align}
    \min_x\mathcal{F}(\{\mathbf{r}_i\},x,T) & \leqslant \mathcal{F}(\{\mathbf{r}_i\},0,T) \tag{S13a} \label{eq:eqS13a} \\
    \min_x\mathcal{F}_{\text{lat}}(\{\mathbf{r}_i\},x,T) & \leqslant \mathcal{F}_{\text{lat}}(\{\mathbf{r}_i\},0,T) \tag{S13b} \label{eq:eqS13b}
\end{align}
The argument is straightforward: since $x=0$ is always accessible in the space of normal-mode coordinates, any minimum of $\mathcal{F}$ or $\mathcal{F}_{\text{lat}}$, if it exists, cannot exceed its value at $x=0$. Equations \eqref{eq:eqS13a} and \eqref{eq:eqS13b} express a variational principle: lattice relaxation always lowers (or leaves unchanged) the conditional free energy compared to the unrelaxed reference framework. This free-energy lowering is the microscopic origin of the ``barrier flattening'' phenomenon in superionic conductors \cite{Wood2021}.

For analytical convenience, we next perform a Taylor expansion of the conditional lattice free energy $\mathcal{F}_{\text{lat}}$ around $x = 0$. This results in equation (2) of the main text in the more general form:
\begin{equation}
    \mathcal{F}(\{\mathbf{r}_i\},x,T) = \mathcal{F}_{\text{GF}}(\{\mathbf{r}_i\},T) + \mathcal{F}_{\text{lat}}^0(\{\mathbf{r}_i\},T) - \widetilde{F}^\dagger x + \frac{1}{2}x^\dagger\widetilde{K}x + \widetilde{\mathcal{U}}_{\text{anh}}(\{\mathbf{r}_i\},x,T) \tag{S14} \label{eq:eqS14}
\end{equation}
Here, the zeroth-order term $\mathcal{F}_{\text{lat}}^0$ is the conditional lattice vibrational free energy evaluated at the reference position $x=0$, and by definition it coincides with the conditional vibrational free energy $\mathcal{F}_{\text{vib}}(\{\mathbf{r}_i\},0,T)$. Within the self-consistent phonon (SCP) approximation \cite{PhysRevB.89.064302, PhysRevB.96.014111}, it is given by
\begin{equation}
    \mathcal{F}_{\text{lat}}^0(\{\mathbf{r}_i\},T) = \sum_{\mathbf{q},s}\left\{\frac{1}{2}\hbar\Omega_{\mathbf{q}s}(\{\mathbf{r}_i\},T) + k_B T\ln\left[1-e^{-\beta\hbar\Omega_{\mathbf{q}s}(\{\mathbf{r}_i\},T)}\right]\right\} \tag{S15a} \label{eq:eqS15a}
\end{equation}
where $\Omega_{\mathbf{q}s}(\{\mathbf{r}_i\},T)$ is the eigenfrequency of the configuration- and SCP-renormalized dynamical (stiffness) matrix $\widetilde{K}(\{\mathbf{r}_i\},T)$ for the prescribed ionic configuration $\{\mathbf{r}_i\}$ and fixed lattice coordinate ($x=0$). In this approximation, $\mathcal{F}_{\text{lat}}^0$ is a monotonic function of $\Omega_{\mathbf{q}s}$. The conditional lattice vibrational free energy of a configuration $\{\mathbf{r}_i\}$ relative to the reference configuration $\{\mathbf{r}_i^{(R)}\}$ is then
\begin{equation}
\begin{split}
    \Delta\mathcal{F}_{\text{lat}}^0 = \sum_{\mathbf{q},s} & \left\{\frac{1}{2}\hbar\left[\Omega_{\mathbf{q}s}(\{\mathbf{r}_i\},T)-\Omega_{\mathbf{q}s}(\{\mathbf{r}_i^{(R)}\},T)\right]\right. \\ & \left. + k_B T\ln\left[\frac{1-e^{-\beta\hbar\Omega_{\mathbf{q}s}(\{\mathbf{r}_i\},T)}}{1-e^{-\beta\hbar\Omega_{\mathbf{q}s}(\{\mathbf{r}_i^{(R)}\},T)}}\right] \right\}
\end{split} \tag{S15b} \label{eq:eqS15b}
\end{equation}
In the classical (high temperature) limit, $k_BT\gg\hbar\Omega_{\mathbf{q}s}$, equation \eqref{eq:eqS15a} reduces to
\begin{equation}
    \mathcal{F}_{\text{lat}}^0(\{\mathbf{r}_i\},T) \approx k_B T\sum_{\mathbf{q},s}\ln\left(\frac{\hbar\Omega_{\mathbf{q}s}}{k_BT}\right) \tag{S15c} \label{eq:eqS15c}
\end{equation}
Consequently,
\begin{equation}
    \Delta\mathcal{F}_{\text{lat}}^0 \approx k_B T\sum_{\mathbf{q},s}\ln\left[\frac{\Omega_{\mathbf{q}s}(\{\mathbf{r}_i\},T)}{\Omega_{\mathbf{q}s}(\{\mathbf{r}_i^{(R)}\},T)}\right] = k_B T\ln\left[\prod_{\mathbf{q},s}\frac{\Omega_{\mathbf{q}s}(\{\mathbf{r}_i\},T)}{\Omega_{\mathbf{q}s}(\{\mathbf{r}_i^{(R)}\},T)}\right] \tag{S15d} \label{eq:eqS15d}
\end{equation}
It follows that a net softening of the host lattice, i.e., $\prod_{\mathbf{q},s}\Omega_{\mathbf{q}s}(\{\mathbf{r}_i\},T)/\Omega_{\mathbf{q}s}(\{\mathbf{r}_i^{(R)}\},T) \leqslant 1$, implies $\Delta\mathcal{F}_{\text{lat}}^0 \leqslant 0$.

$\widetilde{F}$ and $\widetilde{K}$ in equation \eqref{eq:eqS14} denote the effective generalized force and effective stiffness matrix renormalized by both ionic configuration $\{\mathbf{r}_i\}$ and lattice anharmonicity, respectively. Their components are defined as:
\begin{align}
    \widetilde{F}_{\mathbf{q}s} \equiv & -\frac{\partial\mathcal{F}_{\text{lat}}}{\partial x_{\mathbf{q}s}^*}\bigg|_{x=0} = F_{\mathbf{q}s} - \frac{\partial\mathcal{F}_{\text{vib}}}{\partial x_{\mathbf{q}s}^*}\bigg|_{x=0} \tag{S16a} \label{eq:eqS16a}\\
    \widetilde{K}_{\mathbf{qq}'}^{ss'} \equiv & \frac{\partial^2\mathcal{F}_{\text{lat}}}{\partial x_{\mathbf{q}'s'}\partial x_{\mathbf{q}s}^*}\bigg|_{x=0} = K_{\mathbf{qq}'}^{ss'} + \frac{\partial^2\mathcal{F}_{\text{vib}}}{\partial x_{\mathbf{q}'s'}\partial x_{\mathbf{q}s}^*}\bigg|_{x=0} \tag{S16b} \label{eq:eqS16b}
\end{align}
This construction is analogous to the SCP theory \cite{PhysRevB.89.064302, PhysRevB.96.014111}, in which the bare curvature of the potential-energy surface is replaced by the curvature of a fluctuation-renormalized free-energy surface. It also ensures $\widetilde{K}\geqslant 0$ \cite{Plakida_1978, PhysRevB.108.035155}. The last term $\widetilde{\mathcal{U}}_{\text{anh}}$ represents the effective configuration-dependent higher-order conditional free-energy contributions at finite temperature.

Minimizing equation \eqref{eq:eqS14} adiabatically with respect to $x$, we obtain the full equilibrium condition:
\begin{equation}
    \frac{\partial\mathcal{F}_{\text{lat}}}{\partial x^*}\bigg|_{x^*=x_{\text{eq}}^*} = -\widetilde{F} + \widetilde{K}x_{\text{eq}} + \frac{\partial\widetilde{\mathcal{U}}_\text{anh}}{\partial x^*}\bigg|_{x^*=x_{\text{eq}}^*} = 0 \tag{S17} \label{eq:eqS17}
\end{equation}
Equation \eqref{eq:eqS17} describes the collective lattice relaxation $x_{\text{eq}}$ in response to the configurational force $\widetilde{F}$. 
We now introduce the linear-response approximation, in which the anharmonic contribution $\widetilde{\mathcal{U}}_{\text{anh}}$ is either negligible or its effect on the equilibrium displacement $x_{\text{eq}}$ is absorbed into the renormalized stiffness $\widetilde{K}$. This approximation is justified when the framework displacements remain sufficiently small such that the quadratic term dominates. Under this approximation, Eq. \eqref{eq:eqS17} reduces to the familiar Hooke's law $\widetilde{F}=\widetilde{K}x_{\text{eq}}$, and the conditional free energy becomes
\begin{equation}
    \mathcal{F}(\{\mathbf{r}_i\},T) \approx \mathcal{F}_{\text{GF}}(\{\mathbf{r}_i\},T) + \mathcal{F}_{\text{lat}}^0(\{\mathbf{r}_i\},T) - \frac{1}{2}\widetilde{F}^\dagger\widetilde{K}^{-1}\widetilde{F} \tag{S18a} \label{eq:eqS18a}
\end{equation}
with an explicit expansion of the last term as
\begin{equation}
    -\frac{1}{2}\widetilde{F}^\dagger\widetilde{K}^{-1}\widetilde{F} = -\frac{1}{2} \sum_{\mathbf{q},s}\sum_{\mathbf{q}',s'} \widetilde{F}_{\mathbf{q}s}^*(\{\mathbf{r}_i\},T)(\widetilde{K}^{-1})_{\mathbf{qq}'}^{ss'}(\{\mathbf{r}_i\},T)\widetilde{F}_{\mathbf{q}'s'}(\{\mathbf{r}_i\},T) \tag{S18b} \label{eq:eqS18b}
\end{equation}
Equation \eqref{eq:eqS18a} constitutes a self-consistent equation for the ionic configuration $\{\mathbf{r}_i\}$. On one hand, the geometric frustration of $\{\mathbf{r}_i\}$ is generally penalized by an increase in the corresponding conditional free energy $\mathcal{F}_{\text{GF}}(\{\mathbf{r}_i\})$ due to deviations from its global minima. On the other hand, the generalized forces associated with $\{\mathbf{r}_i\}$ induce a relaxation of the lattice that always compensates the conditional free energy. As a result, the free-energy landscape of the ionic subsystem is effectively reshaped, or ``flattened". A direct implication of this flattening is a reduction in the migration energy barriers for specific ionic configurations \cite{He2017, Wood2021}. It is noteworthy that equation \eqref{eq:eqS18b} is formally analogous to the strain-induced interaction theories developed for concentrated alloys \cite{Khachaturian1983, PhysRevB.90.214108}, in which the elimination of lattice displacements results in a relaxation energy proportional to $-\frac{1}{2}F^TA^{-1}F$ ($F$: Kanzaki force and $A$: configuration-dependent force constant matrix). The key distinction in the present work is that the generalized forces arise from ionic configurations, rather than from substitutional alloy occupancies, thereby extending the same variational principle to the description of ionic transport and geometric frustration. In addition, the present theoretical framework explicitly incorporates finite-temperature anharmonic effects within the SCP formalism.

Alternatively, equation \eqref{eq:eqS18a} can also be expressed as
\begin{equation}
    \mathcal{F}(\{\mathbf{r}_i\},T) \approx \mathcal{F}_{\text{GF}}(\{\mathbf{r}_i\},T) + \mathcal{F}_{\text{lat}}^0(\{\mathbf{r}_i\},T) - \frac{1}{2}x_{\text{eq}}^\dagger\widetilde{K}x_{\text{eq}} \tag{S18c} \label{eq:eqS18c}
\end{equation}
Equations \eqref{eq:eqS18a} and \eqref{eq:eqS18c} are equivalent in the linear-response regime, and both define the lattice-relaxation contribution $\mathcal{F}_{\text{LR}}\equiv -\frac{1}{2}\widetilde{F}^\dagger\widetilde{K}^{-1}\widetilde{F} = - \frac{1}{2}x_{\text{eq}}^\dagger\widetilde{K}x_{\text{eq}}$.

When an eigenfrequency $\Omega_{\mathbf{q}s}$ of the stiffness matrix $\widetilde{K}$ approaches zero (e.g., near a structural phase transition), both representations diverge in the linear-response regime. In the vicinity of such an instability and diverging regime, the anharmonic term $\widetilde{\mathcal{U}}_{\text{anh}}$ must be retained, or the displacement response must be bounded by anharmonicity, as implied by equation \eqref{eq:eqS17}. In this case, the displacement-based form [equation \eqref{eq:eqS18c}] remains useful for order-of-magnitude estimates once a physically bounded value of $x_{\text{eq}}$ is specified. Substituting the equilibrium condition from equation \eqref{eq:eqS17} into equation \eqref{eq:eqS14}, the conditional free energy is
\begin{equation}
\begin{split}
    \mathcal{F}(\{\mathbf{r}_i\},T) = &\, \mathcal{F}_{\text{GF}}(\{\mathbf{r}_i\},T) + \mathcal{F}_{\text{lat}}^0(\{\mathbf{r}_i\},T) - \frac{1}{2}x_{\text{eq}}^\dagger\widetilde{K}x_{\text{eq}} \\ & - \left(\frac{\partial\widetilde{\mathcal{U}}_\text{anh}}{\partial x^*}\right)^\dagger\bigg|_{x^*=x_{\text{eq}}^*}x_{\text{eq}} + \widetilde{\mathcal{U}}_{\text{anh}}(\{\mathbf{r}_i\},x_{\text{eq}},T)    
\end{split} \tag{S19} \label{eq:eqS19}
\end{equation}
Therefore, equation \eqref{eq:eqS18c} is recognized as a truncation of equation \eqref{eq:eqS19} to its leading zeroth-order and quadratic contributions.

\subsection*{S1.3. Coarse-grained lattice renormalization}\label{s1.3}
In this section, we present the formulation of a minimal coarse-grained lattice gas theory for $\mathcal{F}_{\text{lat}}$ or, equivalently, for $\mathcal{F}_{\text{LR}}$ truncated at the level of two-body interactions. In principle, $\mathcal{F}_{\text{GF}}$ can be coarse-grained and expressed in a cluster expansion form as well. Nevertheless, such an extension lies well beyond the present work and it is deferred to future studies.

\subsubsection*{S1.3.1. Fixed stiffness/dynamical matrix approximation}\label{s1.3.1}
At finite temperature, the stiffness, or dynamical matrix of the framework is effectively renormalized by both the instantaneous ionic configuration $\{\mathbf{r}_i\}$ and anharmonicity. To obtain a tractable coarse-grained theory, we employ the \textit{fixed stiffness approximation}, that is, the configuration-dependent changes in the generalized stiffness matrix are neglected compared to the renormalization from anharmonicity. Here, we will briefly discuss the applicability of this approximation.

Without loss of generality, $\widetilde{K}(T)$ can be written as the sum of a configuration-independent effective stiffness $\widetilde{K}_0(T)$ and a configuration-dependent residual stiffness matrix $\delta\widetilde{K}(\{\mathbf{r}_i\},T)$ as
\begin{equation}
    \widetilde{K}(\{\mathbf{r}_i\},T) = \widetilde{K}_0(T) + \delta\widetilde{K}(\{\mathbf{r}_i\},T) \tag{S20a} \label{eq:eqS20a}
\end{equation}
The construction of $\widetilde{K}_0(T)$ is somewhat arbitrary. For example, it can be evaluated at the reference ionic configuration $\{\mathbf{r}_i^{(R)}\}$ or at any accessible configuration $\{\mathbf{r}_i\}$. A sensible selection is the ensemble average of $\widetilde{K}(\{\mathbf{r}_i\},T)$ such that $\widetilde{K}_0(T)\equiv\langle\widetilde{K}(\{\mathbf{r}_i\},T)\rangle$ and $\langle\delta\widetilde{K}\rangle$=0. Substituting equation \eqref{eq:eqS20a} into $\mathcal{F}_{\text{LR}}$ yields a Neumann series expansion
\begin{equation}
\begin{split}
    \mathcal{F}_{\text{LR}} = & -\frac{1}{2}\widetilde{F}^\dagger\widetilde{K}^{-1}\widetilde{F} \\ = & -\frac{1}{2}\widetilde{F}^\dagger\widetilde{K}_0^{-1}\widetilde{F} + \frac{1}{2}\widetilde{F}^\dagger\widetilde{K}_0^{-1}\delta\widetilde{K}\widetilde{K}_0^{-1}\widetilde{F} \\ &-\frac{1}{2}\widetilde{F}^\dagger\widetilde{K}_0^{-1}\delta\widetilde{K}\widetilde{K}_0^{-1}\delta\widetilde{K}\widetilde{K}_0^{-1}\widetilde{F} + \cdots
\end{split} \tag{S20b} \label{eq:eqS20b}
\end{equation}
The explicit form of the first two terms is
\begin{align}
    -\frac{1}{2}\widetilde{F}^\dagger\widetilde{K}_0^{-1}\widetilde{F} = & -\frac{1}{2}\sum_{\mathbf{q},s}\frac{1}{\Omega_{\mathbf{q}s}^2(T)}\left|\widetilde{F}_{\mathbf{q}s}\right|^2 \tag{S20c} \label{eq:eqS20c} \\
    \frac{1}{2}\widetilde{F}^\dagger\widetilde{K}_0^{-1}\delta\widetilde{K}\widetilde{K}_0^{-1}\widetilde{F} = & \frac{1}{2}\sum_{\mathbf{q},s}\sum_{\mathbf{q}',s'}\frac{1}{\Omega_{\mathbf{q}s}^2(T)\Omega_{\mathbf{q}'s'}^2(T)}\widetilde{F}_{\mathbf{q}s}^*\delta\widetilde{K}_{\mathbf{qq}'}^{ss'}(\{\mathbf{r}_i\},T)\widetilde{F}_{\mathbf{q}'s'} \tag{S20d} \label{eq:eqS20d}
\end{align}
where $\Omega_{\mathbf{q}s}(T)$ is the eigenfrequency of $\widetilde{K}_0(T)$. Denoting $x_0 \equiv \widetilde{K}_0^{-1}(T)\widetilde{F}$, equations \eqref{eq:eqS20c} and \eqref{eq:eqS20d} can be rewritten as
\begin{align}
    -\frac{1}{2}\widetilde{F}^\dagger\widetilde{K}_0^{-1}\widetilde{F} = & -\frac{1}{2}x_0^\dagger\widetilde{K}_0(T)x_0 = -\frac{1}{2}\sum_{\mathbf{q}s}\Omega_{\mathbf{q}s}^2(T)x_{\mathbf{q}s,0}^*x_{\mathbf{q}s,0} \tag{S20e} \label{eq:eqS20e} \\
    \frac{1}{2}\widetilde{F}^\dagger\widetilde{K}_0^{-1}\delta\widetilde{K}\widetilde{K}_0^{-1}\widetilde{F} = & \frac{1}{2} x^\dagger_0\delta\widetilde{K}x_0 = \frac{1}{2}\sum_{\mathbf{q},s}\sum_{\mathbf{q}',s'}x_{\mathbf{q}s,0}^*\delta\widetilde{K}_{\mathbf{qq}'}^{ss'}(\{\mathbf{r}_i\},T)x_{\mathbf{q}'s',0} \tag{S20f} \label{eq:eqS20f}
\end{align}
As a consequence, the fixed stiffness approximation should be interpreted as a controlled truncation of $\mathcal{F}_{\text{LR}}$ to its lowest-order contributions [equations \eqref{eq:eqS20c} and \eqref{eq:eqS20e}]. In particular, it leads to the one-body and two-body interactions discussed in the main text, whereas the terms involving $\delta\widetilde{K}$ generate higher-order, namely three-body and four-body interactions. In general, the quantitative relevance of these higher-order terms is material-dependent and must be assessed \textit{a posteriori} by means of numerical simulations. Explicit estimates of the higher-order contributions are presented in Section S2.2. Nonetheless, the fixed stiffness approximation captures the dominant contributions, with higher-order terms providing quantitative corrections but not altering the qualitative physics.

\subsubsection*{S1.3.2. Definition of site-occupation variables $n_{l\mu}$ and coupling amplitude $\lambda_{\mathbf{q}s}^{l\mu}$}\label{s1.3.2}

The ionic configuration $\{\mathbf{r}_i\}$ is coarse-grained in terms of site-occupation variables $n_{l\mu}$, which indicate whether a lattice site $\mu$ in unit cell $l$ is occupied by an ion ($n_{l\mu} = 1$) or not ($n_{l\mu} = 0$). The lattice site can either be a regular reference lattice position (typically corresponding to the global minima of $V_0 + U_{\text{ion}}^{\text{bare}}$) or a frustrated/migration-relevant site (local minima or transition pathways set by $V_0 + U_{\text{ion}}^{\text{bare}}$). In the linear-response approximation, this occupation variable $n_{l\mu}$ deforms the framework atom $(m,\kappa)$ in unit cell $m$ via a generalized force $\mathbf{f}_{l\mu}^{m\kappa}(T)$. A corresponding coupling amplitude $\lambda_{\mathbf{q}s}^{l\mu}(T)$ can then be defined in terms of $\mathbf{f}_{l\mu}^{m\kappa}(T)$ in analogy to equation \eqref{eq:eqS7c} as
\begin{equation}
    \lambda_{\mathbf{q}s}^{l\mu}(T) \equiv \frac{1}{\sqrt{N_c}}\sum_{m,\kappa,\alpha}\frac{f_{l\mu}^{m\kappa,\alpha}(T)}{\sqrt{M_{\kappa}}}\epsilon_{\alpha}^{(\kappa,s)*}(\mathbf{q})\,e^{-i\mathbf{q}\cdot\mathbf{R}_m^0} \tag{S21a} \label{eq:eqS21a}
\end{equation}
Employing translational symmetry, namely, shifting the cell indices such that $l\to 0$ and $m\to n = m-l$, we have $\mathbf{f}_{0\mu}^{n\kappa} = \mathbf{f}_{l\mu}^{m\kappa}$ and equation \eqref{eq:eqS21a} can be rewritten as
\begin{align}\label{eq:eqS21b}
    \lambda_{\mathbf{q}s}^{l\mu}(T) & = \frac{1}{\sqrt{N_c}}\sum_{n,\kappa,\alpha}\frac{f_{0\mu}^{n\kappa,\alpha}(T)}{\sqrt{M_{\kappa}}}\epsilon_{\alpha}^{(\kappa,s)*}(\mathbf{q})\,e^{-i\mathbf{q}\cdot(\mathbf{R}_n^0+\mathbf{R}_l^0)} \notag \\ & = \frac{e^{-i\mathbf{q}\cdot\mathbf{R}_l^0}}{\sqrt{N_c}}\underbrace{\sum_{n,\kappa,\alpha}\frac{f_{0\mu}^{n\kappa,\alpha}(T)}{\sqrt{M_{\kappa}}}\epsilon_{\alpha}^{(\kappa,s)*}(\mathbf{q})\,e^{-i\mathbf{q}\cdot\mathbf{R}_n^0}}_{\displaystyle{\equiv\lambda_{\mathbf{q}s}^\mu}(T)}  \tag{S21b} \\ & = \frac{1}{\sqrt{N_c}}\lambda_{\mathbf{q}s}^\mu(T)\,e^{-i\mathbf{q}\cdot\mathbf{R}_l^0} \notag
\end{align}
where $\lambda_{\mathbf{q}s}^\mu(T)$ is a coupling amplitude that depends solely on the site index $\mu$.

The total effective generalized force acting on the mode $(\mathbf{q},s)$ is the superposition of all couplings to the set of $\{n_{l\mu}\}$
\begin{equation}
    \widetilde{F}_{\mathbf{q}s}(\{n_{l\mu}\}) = \sum_{l,\mu} \lambda_{\mathbf{q}s}^{l\mu}(T) n_{l\mu} = \frac{1}{\sqrt{N_c}}\sum_{l,\mu} \lambda_{\mathbf{q}s}^{\mu}(T)\, n_{l\mu}\,e^{-i\mathbf{q}\cdot\mathbf{R}_l^0} \tag{S22a} \label{eq:eqS22a}
\end{equation} 
Equation \eqref{eq:eqS22a} naturally motivates the definition of the Fourier transform of site-occupation variables $n_{l\mu}$:
\begin{equation}
    n_{\mathbf{q}\mu} \equiv \frac{1}{\sqrt{N_c}}\sum_{l} n_{l\mu}\,e^{-i\mathbf{q}\cdot\mathbf{R}_l^0} \tag{S22b} \label{eq:eqS22b}
\end{equation}
so that equation \eqref{eq:eqS22a} can be expressed in reciprocal space as
\begin{equation}
    \widetilde{F}_{\mathbf{q}s}(\{n_{\mathbf{q}\mu}\}) = \sum_\mu \lambda_{\mathbf{q}s}^\mu(T)\, n_{\mathbf{q}\mu} \tag{S22c} \label{eq:eqS22c}
\end{equation}
Equation \eqref{eq:eqS22c} describes how a collective ionic density modulation $n_{\mathbf{q}\mu}$ at site $\mu$ couples to the lattice degrees of freedom with a site-dependent amplitude $\lambda_{\mathbf{q}s}^\mu(T)$.

Substituting \eqref{eq:eqS22a} into $\mathcal{F}_{\text{LR}}$ and employing the fixed stiffness approximation, we have
\begin{equation}
    \mathcal{F}_{\text{LR}}(\{n_{l\mu}\},T) = - \frac{1}{2N_c}\sum_{\mathbf{q},s}\frac{1}{\Omega_{\mathbf{q}s}^2(T)} \left|\sum_{l,\mu} \lambda_{\mathbf{q}s}^{\mu}(T) n_{l\mu}\,e^{-i\mathbf{q}\cdot\mathbf{R}_l^0}\right|^2 \tag{S23a} \label{eq:eqS23a}
\end{equation}
More compactly, using equation \eqref{eq:eqS22c}, we obtain
\begin{equation}
    \mathcal{F}_{\text{LR}}(\{n_{l\mu}\},T) = - \frac{1}{2}\sum_{\mathbf{q},s}\frac{1}{\Omega_{\mathbf{q}s}^2(T)} \left|\sum_{\mu} \lambda_{\mathbf{q}s}^{\mu}(T) n_{\mathbf{q}\mu}\right|^2 \tag{S23b} \label{eq:eqS23b}
\end{equation}
Both forms explicitly demonstrate that low-frequency soft modes, the key ingredient of all phonon-assisted ion migration mechanisms \cite{Wakamura1997, Muy2021, Zhang2022, Jun2024}, control the non-positive renormalization of the conditional free energy through the weighting factor $\Omega_{\mathbf{q}s}^{-2}$.

From equations \eqref{eq:eqS17} and \eqref{eq:eqS21a}, we can additionally define the mode-resolved ``lattice polarizability" \cite{Kraft2017, Muy2021} as the adiabatic lattice deformation $x_{\mathbf{q}s,\text{eq}}$ induced by a change in ionic occupation $n_{l\mu}$:
\begin{equation}
    \chi_{\mathbf{q}s}^{l\mu}(T) \equiv \frac{\partial x_{\mathbf{q}s,\text{eq}}}{\partial n_{l\mu}} = \frac{1}{\Omega_{\mathbf{q}s}^2(T)}\frac{\partial \sum_{l',\mu'} \lambda_{\mathbf{q}s}^{l'\mu'}(T) n_{l'\mu'}}{\partial n_{l\mu}} = \frac{\lambda_{\mathbf{q}s}^{l\mu}(T)}{\Omega_{\mathbf{q}s}^2(T)} \tag{S24} \label{eq:eqS24}
\end{equation}
A large lattice polarizability is therefore directly associated with both a soft mode frequency $\Omega_{\mathbf{q}s}(T)$ and a strong coupling amplitude $\lambda_{\mathbf{q}s}^{l\mu}(T)$.

Using equation \eqref{eq:eqS24}, equation \eqref{eq:eqS23a} can be expressed as
\begin{equation}
    \mathcal{F}_{\text{LR}}(\{n_{l\mu}\},T) = - \frac{1}{2}\sum_{\mathbf{q},s}\Omega_{\mathbf{q}s}^2(T) \left|\sum_{l,\mu} \chi_{\mathbf{q}s}^{l\mu}(T)\,n_{l\mu}\right|^2 \tag{S25} \label{eq:eqS25}
\end{equation}
Equations \eqref{eq:eqS23a}, \eqref{eq:eqS23b} and \eqref{eq:eqS25} thus elucidate, at lowest order, the microscopic mechanism by which lattice softness and framework polarizability are correlated with fast ionic transport \cite{Muy2021, Krauskopf2018}. Both effects reduce, or flatten the frustrated free-energy landscape and thereby help stabilize ionic occupations at frustration- and migration-relevant sites.

\subsubsection*{S1.3.3. Lattice relaxation in real space} \label{s1.3.3}
In real space, $\mathcal{F}_{\text{LR}}$ can be expanded term by term as
\begin{equation}
\begin{split}
    \mathcal{F}_{\text{LR}}(\{n_{l\mu}\},T) = & -\frac{1}{2N_c}\sum_{\mathbf{q},s}\sum_{l,\mu}\frac{\left|\lambda_{\mathbf{q}s}^\mu(T)\right|^2}{\Omega_{\mathbf{q}s}^2(T)}n_{l\mu} \\ & - \frac{1}{2N_c}\sum_{\mathbf{q},s}\sum_{(l,\mu)\neq(l',\mu')}\frac{\lambda_{\mathbf{q}s}^{\mu*}(T)\lambda_{\mathbf{q}s}^{\mu'}(T)}{\Omega_{\mathbf{q}s}^2(T)}n_{l\mu}n_{l'\mu'}\,e^{-i\mathbf{q}\cdot\mathbf{R}_{l'l}^0} 
\end{split} \tag{S26a} \label{eq:eqS26a}
\end{equation}
where $(l,\mu)\neq(l',\mu')$ denotes $l \neq l'$ or $\mu \neq \mu'$, and $\mathbf{R}_{l'l}^0$ is the difference between $\mathbf{R}_{l}^0$ and $\mathbf{R}_{l'}^0$, i.e., $\mathbf{R}_{l'l}^0 = \mathbf{R}_{l'}^0 - \mathbf{R}_{l}^0$. For a given pair of indices $(l,\mu)$ and $(l',\mu')$ with $(l,\mu)\neq(l',\mu')$, their combined contribution is
\begin{equation}
    -\frac{n_{l\mu}n_{l'\mu'}}{2N_c\Omega_{\mathbf{q}s}^2(T)}\left(\lambda_{\mathbf{q}s}^{\mu*}\lambda_{\mathbf{q}s}^{\mu'}e^{-i\mathbf{q}\cdot\mathbf{R}_{l'l}^0}+\lambda_{\mathbf{q}s}^{\mu'*}\lambda_{\mathbf{q}s}^{\mu}e^{-i\mathbf{q}\cdot\mathbf{R}_{ll'}^0}\right) = -\frac{n_{l\mu}n_{l'\mu'}}{N_c\Omega_{\mathbf{q}s}^2(T)}\text{Re}(\lambda_{\mathbf{q}s}^{\mu*}\lambda_{\mathbf{q}s}^{\mu'}e^{-i\mathbf{q}\cdot\mathbf{R}_{l'l}^0}) \notag
\end{equation}
Consequently, equation \eqref{eq:eqS26a} can be rewritten as
\begin{equation}
\begin{split}
    \mathcal{F}_{\text{LR}}(\{n_{l\mu}\},T) = & -\frac{1}{2N_c}\sum_{\mathbf{q},s}\sum_{l,\mu}\frac{\left|\lambda_{\mathbf{q}s}^\mu(T)\right|^2}{\Omega_{\mathbf{q}s}^2(T)}n_{l\mu} \\ & - \frac{1}{N_c}\sum_{\mathbf{q},s}\sum_{(l,\mu)<(l',\mu')}\frac{\text{Re}[\lambda_{\mathbf{q}s}^{\mu*}(T)\lambda_{\mathbf{q}s}^{\mu'}(T)e^{-i\mathbf{q}\cdot\mathbf{R}_{l'l}^0}]}{\Omega_{\mathbf{q}s}^2(T)}n_{l\mu}n_{l'\mu'}
\end{split} \tag{S26b} \label{eq:eqS26b}
\end{equation}
or, equivalently, in the cluster expansion form,
\begin{equation}
    \mathcal{F}_{\text{LR}}(\{n_{l\mu}\},T) = \sum_{l,\mu}h_{\mu}(T)n_{l\mu} + \sum_{(l,\mu)<(l',\mu')}V_{l\mu,l'\mu'}(T)n_{l\mu}n_{l'\mu'} \tag{S26c} \label{eq:eqS26c}
\end{equation}
where the index $(l,\mu)<(l',\mu')$ is defined by $l<l'$ or ($\mu < \mu'$ and $l' = l$).

The local lattice relaxation (self-trapping) contribution arising from the one-body contribution $h_{\mu}(T)$ has been discussed in the main text. We therefore concentrate on the two-body contributions. For different site occupations ($l\neq l',\mu\neq\mu'$), the quantity $\text{Re}[\lambda_{\mathbf{q}s}^{\mu*}(T)\lambda_{\mathbf{q}s}^{\mu'}(T)\,e^{-i\mathbf{q}\cdot\mathbf{R}_{l'l}^0}]$ is sign-indefinite. As a result, the occupations of two ionic sites can either be positively correlated or negatively correlated, depending on the relative phase and magnitude of the coupling projections $\lambda_{\mathbf{q}s}^{\mu*}(T)\lambda_{\mathbf{q}s}^{\mu'}(T)$ and on the wave vector $\mathbf{q}$ modulation through the factor $e^{-i\mathbf{q}\cdot\mathbf{R}_{l'l}^0}$. Phenomenologically, positive correlations correspond to effectively ``attractive" pair interactions, while negative correlations correspond to effectively ``repulsive" interactions. The coexistence of these effective attractive and repulsive interactions therefore reorganizes the nearly degenerate frustrated configurations, selecting correlations that minimize the free energy.

For occupations of two ions on the same site ($l\neq l',\mu'=\mu$), the sign of $\text{Re}[\lambda_{\mathbf{q}s}^{\mu*}(T)\lambda_{\mathbf{q}s}^{\mu}(T)\,e^{-i\mathbf{q}\cdot\mathbf{R}_{l'l}^0}]$ is determined solely by the wave vector modulation. In the long-wavelength limit ($\mathbf{q}\to 0$), one has $\text{Re}[\lambda_{\mathbf{q}s}^{\mu*}(T)\lambda_{\mathbf{q}s}^{\mu}(T)\,e^{-i\mathbf{q}\cdot\mathbf{R}_{l'l}^0}]\geqslant 0$, which favors a spatially homogeneous, ``ferromagnetic-like" pattern of same site occupations. In contrast, for wave vectors at the zone boundary, an ``anti-ferromagnetic-like" alternation of site occupations is energetically preferred. The wave-vector-dependent ionic modulations and their connection to soft modes will be discussed in the subsequent section.

\subsubsection*{S1.3.4. Lattice relaxation in reciprocal space}\label{s1.3.4}
The lattice relaxation in reciprocal space can be expressed as
\begin{align}\label{eq:eqS27}
    \mathcal{F}_{\text{LR}}(\{n_{\mathbf{q}\mu}\},T) & = -\frac{1}{2}\sum_{\mathbf{q},s}\sum_{\mu,\mu'}\frac{\lambda_{\mathbf{q}s}^{\mu*}(T)\lambda_{\mathbf{q}s}^{\mu'}(T)}{\Omega_{\mathbf{q}s}^2(T)}n_{\mathbf{q}\mu}^*n_{\mathbf{q}\mu'} \notag \\
    & = -\frac{1}{2}\sum_{\mathbf{q}}\sum_{\mu,\mu'}\mathcal{J}_{\mathbf{q}}^{\mu\mu'}(T)n_{\mathbf{q}\mu}^*n_{\mathbf{q}\mu'} \tag{S27} \\
    & = -\sum_{\mathbf{q}}\left\{\frac{1}{2}\sum_{\mu=\mu'}\mathcal{J}_{\mathbf{q}}^{\mu\mu}(T)|n_{\mathbf{q}\mu}|^2 + \sum_{\mu<\mu'}\text{Re}[\mathcal{J}_{\mathbf{q}}^{\mu\mu'}(T)n_{\mathbf{q}\mu}^*n_{\mathbf{q}\mu'}]\right\} \notag
\end{align}
where $\mathcal{J}_{\mathbf{q}}^{\mu\mu'}(T)\equiv\sum_s\Omega_{\mathbf{q}s}^{-2}(T)\lambda_{\mathbf{q}s}^{\mu*}(T)\lambda_{\mathbf{q}s}^{\mu'}(T)$. Note that the diagonal term $\mathcal{J}_{\mathbf{q}}^{\mu\mu}(T)|n_{\mathbf{q}\mu}|^2$ is positive semidefinite at any wave vector $\mathbf{q}$. The energetically favorable ionic modulation $n_{\mathbf{q}\mu}$ is thus determined predominantly by the diagonal couplings $\mathcal{J}_{\mathbf{q}}^{\mu\mu}(T)$, or the leading soft mode $\Omega_{\mathbf{q}s}(T)$. Depending on the soft mode wave vector, three characteristic classes of ionic modulations are anticipated: (i) a ``ferromagnetic-like" homogeneous modulation ($\mathbf{q}=0$), (ii) an ``anti-ferromagnetic-like" modulation ($\mathbf{q}$ at the Brillouin-zone boundary), and (iii) commensurate or incommensurate modulations (general $\mathbf{q}$ in the Brillouin zone). These distinct patterns of ionic modulations can be directly probed by experimental techniques such as X-ray scattering, neutron scattering and electron diffraction \cite{Yang2026}.

``Ferromagnetic-like" ionic correlations are expected in materials exhibiting low-energy optical phonon modes at zone center, including AgI \cite{Buhrer1975}, $\beta$-Ag$_2$S \cite{Alekperov2016}, $\beta$-CuI \cite{Wakamura2002}, Cu$_7$PS$_6$ \cite{Shen2024}, Na$_3$PS$_4$ \cite{Famprikis2021, Brenner2022}, Na$_{2.94}$Sb$_{0.9}$W$_{0.1}$S$_4$ \cite{Maus2023}, Li$_3$N \cite{PhysRevLett.107.118302}, Li$_{0.5}$La$_{0.5}$TiO$_3$ \cite{Sanjuan2005, Pham2026}. For ``anti-ferromagnetic-like" ionic modulations, systems featuring low-frequency modes at the Brillouin-zone boundary, such as Na$_3$PSe$_4$ \cite{Brenner2022}, Na$_3$PS$_4$ \cite{Gupta2021}, Li$_3$YCl$_6$ \cite{Ahammed2024}, Li$_3$OCl \cite{Chen2015}, Li$_2$O \cite{Gupta2012} and CaF$_2$ \cite{Cazorla2014}, are candidate hosts. Finally, systems with commensurate or incommensurate modulated structures are expected in Cu$_{2-\text{x}}$Se \cite{Danilkin2010}, LaNb$_{0.88}$W$_{0.12}$O$_{4.06}$ \cite{Li2019} and LiKSO$_4$ \cite{Zhang1987}.

For different site occupations $\mu'\neq\mu$, the off-diagonal quantity $\text{Re}[\mathcal{J}_{\mathbf{q}}^{\mu\mu'}(T)n_{\mathbf{q}\mu}^*n_{\mathbf{q}\mu'}]$ is sign-indefinite and it describes the softness $\Omega_{\mathbf{q}s}^{-2}$-weighted ``interference" between the occupation modulations $n_{\mathbf{q}\mu}$ and $n_{\mathbf{q}\mu'}$. Specifically, when $\text{Re}[\mathcal{J}_{\mathbf{q}}^{\mu\mu'}(T)n_{\mathbf{q}\mu}^*n_{\mathbf{q}\mu'}] > 0$, the modulations $n_{\mathbf{q}\mu}$ and $n_{\mathbf{q}\mu'}$ jointly ``push" all the framework modes at wave vector $\mathbf{q}$. Conversely, when $\text{Re}[\mathcal{J}_{\mathbf{q}}^{\mu\mu'}(T)n_{\mathbf{q}\mu}^*n_{\mathbf{q}\mu'}] < 0$, the modulations $n_{\mathbf{q}\mu}$ and $n_{\mathbf{q}\mu'}$ are mutually antagonistic: one reinforces while the other opposes the framework, thus partially canceling the net lattice response. In this regime, a nontrivial correlated modulation of $n_{\mathbf{q}\mu}$ and $n_{\mathbf{q}\mu'}$ across different sites emerges.

\subsection*{S1.4. Relations to historical models}\label{s1.4}
\subsubsection*{S1.4.1. Tomoyose's lattice gas model}\label{s1.4.1}
Tomoyose formulated a lattice gas model in 1991 and 1998 to describe phonon-assisted ion hopping \cite{Tomoyose1991} and density fluctuations \cite{Tomoyose1998} in superionic conductors. We briefly recapitulate this theory, which starts from the coupled phonon-lattice gas Hamiltonian:
\begin{equation}
    H=\frac{U}{2}\sum_{j,\delta}\hat{n}_j\hat{n}_{j+\delta} + J\sum_{j,\delta}c_{j+\delta}^\dagger c_j + H_{\text{ph}} + \sum_j V_j\hat{n}_j \tag{S28} \label{eq:eqS28}
\end{equation}
The first term represents the repulsive ion-ion interaction of strength $U$ between the ion at the $j$-th unit cell and its nearest neighbors $j+\delta$. The second term describes ion hopping with the hopping matrix $J$, where $c_j$ ($c_j^\dagger$) are annihilation (creation) operators for the hopping ion at the $j$-th unit cell, and $\hat{n}_j = c_j^\dagger c_j$. In the discrete site occupation and nearest-neighbor approximation, the potential part of our generalized Hamiltonian ($V_0+U_{\text{ion}}^{\text{bare}}$) immediately reduces to these first two terms. More generally, within the discrete multi-site occupation approximation, the Hamiltonian of the ionic subsystem can be expressed in terms of $\{n_{l\mu}\}$ as
\begin{equation}
    \mathcal{H}_{\text{GF}} = \sum_{(l,\mu)\neq(l',\mu')}\frac{U_{ll'}^{\mu\mu'}}{2}n_{l\mu}n_{l'\mu'} + \sum_{(l,\mu)\neq(l',\mu')} J_{ll'}^{\mu\mu'}c_{l'\mu'}^\dagger c_{l\mu} \tag{S29} \label{eq:eqS29}
\end{equation}
Here, $U_{ll'}^{\mu\mu'}$ and $J_{ll'}^{\mu\mu'}$ now represent ion-ion interactions and the hopping between an ion at site $\mu$ in the $l$-th unit cell and that at site $\mu'$ in the $l'$-th unit cell. The summation index $(l,\mu)\neq(l',\mu')$ specifies that either $l'\neq l$ or $\mu'\neq\mu$.

The phonon Hamiltonian $H_{\text{ph}}$ and the last term in equation \eqref{eq:eqS28}, which accounts for the ion-phonon interaction via a local potential $V_j$, are given by \cite{Tomoyose1991, Tomoyose1998}
\begin{align}
    H_{\text{ph}} = & \sum_\mathbf{q}\hbar\omega_\mathbf{q}(a_\mathbf{q}^\dagger a_\mathbf{q} + \frac{1}{2}) \tag{S30a} \label{eq:eqS30a} \\
    V_j = & \sum_\mathbf{q} A_\mathbf{q}e^{i\mathbf{q}\cdot\mathbf{R}_j}(a_\mathbf{q}+a_\mathbf{-q}^\dagger) \tag{S30b} \label{eq:eqS30b}
\end{align}
where $A_{\mathbf{q}}$ is a phonon-ion coupling constant, $\mathbf{R}_j$ the position vector of the mobile ion in the $j$-th unit cell, and $\omega_\mathbf{q}$ the phonon frequency with wave vector $\mathbf{q}$. Comparing these two terms with equation \eqref{eq:eqS9b}, one recognizes that they correspond to the single phonon branch ($s=1$) and single occupation site per unit cell ($\mu'=\mu$) approximation of $\mathcal{H}_{\text{lat}}$ in the harmonic limit. In addition, equation \eqref{eq:eqS30b} is structurally identical to the coarse-grained generalized force of equation \eqref{eq:eqS22a}, except that the latter explicitly resolves the phonon branch $s$ and the site index $\mu$.

The phonon-ion coupling constant $A_{\mathbf{q}}$ can thus be embedded into our present formalism in an analogous manner to that employed in Section S1.2. By integrating out the lattice degrees of freedom in equation \eqref{eq:eqS28}, the resulting lattice-renormalized contribution that corresponds to $\sum_j V_j\hat{n}_j$ is
\begin{equation}
    \mathcal{F}_{\text{LR}}(\{n_j\}) = -\sum_{\mathbf{q}}\frac{|A_{\mathbf{q}}|^2}{\hbar\omega_{\mathbf{q}}}\sum_{j,j'}n_j n_{j'} e^{-i\mathbf{q}\cdot\mathbf{R}_{j'j}^0} \tag{S31a} \label{eq:eqS31a}
\end{equation}
The structure of equation \eqref{eq:eqS31a} is identical to that of equation \eqref{eq:eqS26a}. The diagonal part of $\mathcal{F}_{\text{LR}}$, namely, same site occupation $j'=j$, is $-\sum_\mathbf{q}\frac{|A_{\mathbf{q}}|^2}{\hbar\omega_\mathbf{q}}n_j^2$ and is always non-positive. The off-diagonal contribution is modulated by the phase factor $e^{-i\mathbf{q}\cdot\mathbf{R}_{j'j}^0}$. These are exactly the same self-trapping and interference mechanisms inherent in the lattice gas model. 

Identifying the effective mode frequencies $\Omega_{\mathbf{q}}$ with the phonon frequencies $\omega_{\mathbf{q}}$ under the harmonic approximation and comparing equation \eqref{eq:eqS31a} with equation \eqref{eq:eqS26a}, we obtain
\begin{equation}
    |A_{\mathbf{q}}|^2 = \frac{\hbar}{2N_c\omega_{\mathbf{q}}}|\lambda_{\mathbf{q}}|^2 \tag{S31b} \label{eq:eqS31b}
\end{equation}
which admits the physically reasonable choice
\begin{equation}
    A_{\mathbf{q}} = -\sqrt{\frac{\hbar}{2N_c\omega_{\mathbf{q}}}}\lambda_{\mathbf{q}}^* \tag{S31c} \label{eq:eqS31c}
\end{equation}
Therefore, the lattice gas theory of Tomoyose emerges as a single phonon branch, single-site occupation variable, and harmonic limit of the present framework. It implicitly contains lattice softness, local relaxation and non-local interference terms arising from lattice renormalization. However, these contributions were interpreted primarily in terms of hopping kinetics.

\subsubsection*{S1.4.2. Pardee and Mahan's small polaron theory}\label{s1.4.2}
Pardee and Mahan developed a small polaron theory that accounts for the coupling between mobile ions and the host lattice through longitudinal optical phonons \cite{Pardee1975}. This theory represents another limiting case of the present framework. For a single optical phonon branch approximated by Einstein's model, the basic Hamiltonian is written as
\begin{align}
    H_0 & = \hbar\omega_0\sum_{\mathbf{q}}(a_{\mathbf{q}}^\dagger a_{\mathbf{q}}+\frac{1}{2}) + \sum_{\alpha,\alpha'}V_{\alpha\alpha'}n_{\alpha}n_{\alpha'}+H_I \tag{S32a} \label{eq:eqS32a} \\
    H_I & = \sum_{\mathbf{q},\alpha}(iM_{\mathbf{q}}^*a_{\mathbf{q}}-iM_{\mathbf{q}}a_{\mathbf{q}}^\dagger)\,n_{\alpha}e^{i\mathbf{q}\cdot\mathbf{r}_{\alpha}} \tag{S32b} \label{eq:eqS32b}
\end{align}
where $V_{\alpha\alpha'}n_{\alpha}n_{\alpha'}$ represents the electrostatic repulsion between neighboring cations, and $H_I$ is the interaction between the disordered cations and the optical phonon mode of frequency $\omega_0$. The ion-phonon coupling constant $M_{\mathbf{q}}$ is given by \cite{Pardee1975}
\begin{equation}
    M_{\mathbf{q}} = ZeN|\mathbf{q}|\widetilde{V}(\mathbf{q})\sqrt{\frac{\hbar}{2M\omega_0}} \tag{S32c} \label{eq:eqS32c}
\end{equation}
where $-Ze$ and $M$ are the charge and mass, respectively, of the nonconducting framework, $N$ is the number of unit cells, and $\widetilde{V}(\mathbf{q})=\frac{1}{N\Omega}\frac{4\pi}{q^2}$ is the Fourier transform of the cation's Coulomb potential, with $\Omega$ the crystal volume.

By completing the squares, $H_I$ can be removed from $H_0$, yielding a renormalized effective ion--ion interaction \cite{Pardee1975}:
\begin{equation}
    U_{\alpha\alpha'} = V_{\alpha\alpha'} - \sum_{\mathbf{q}}\frac{|M_{\mathbf{q}}|^2}{\hbar\omega_0}e^{-i\mathbf{q}\cdot(\mathbf{R}_{\alpha'}-\mathbf{R}_\alpha)} \tag{S33a} \label{eq:eqS33a}
\end{equation}
The lattice renormalization in $\sum_{\mathbf{q}}$ has exactly the same decomposition as equations \eqref{eq:eqS26a} and \eqref{eq:eqS31a}. Consequently, Pardee and Mahan’s theory also corresponds to a specific realization of the lattice-renormalized geometric frustration, obtained by adopting a Fr\"{o}hlich-type coupling to a single Einstein optical mode \cite{Frohlich1952}.

The coefficient $|M_{\mathbf{q}}|^2/\hbar\omega_0$ can be written explicitly as
\begin{equation}
    \frac{|M_\mathbf{q}|^2}{\hbar\omega_0} = \frac{4\pi e^2}{\hbar\omega_0\Omega q^2}\left(\frac{1}{\varepsilon_\infty}-\frac{1}{\varepsilon_0}\right) \tag{S33b} \label{eq:eqS33b}
\end{equation}
where $\varepsilon_0$ and $\varepsilon_\infty$ are the static and high-frequency dielectric constants, respectively. Within the notation adopted by Pardee and Mahan, equation \eqref{eq:eqS33b} suggests an enhanced lattice-renormalized interaction for low-frequency optical modes $\omega_0$ at small $\mathbf{q}$, together with the ionic polarizability characterized by $\varepsilon_0-\varepsilon_\infty$. Using equation \eqref{eq:eqS31b}, the coupling amplitude $\lambda_\mathbf{q}$ is obtained as
\begin{equation}
    \lambda_\mathbf{q} \propto \frac{\sqrt{\omega_0}}{q}\sqrt{\frac{1}{\varepsilon_\infty}-\frac{1}{\varepsilon_0}} \tag{S34} \label{eq:eqS34}
\end{equation}
The $q^{-1}\sqrt{\omega_0}$ dependence indicates a strong induction to long-wavelength ($q\to 0$), high-frequency optical phonons, with the interaction strength screened by $\sqrt{\varepsilon_\infty^{-1}-\varepsilon_0^{-1}}$.

As a sectional summary, both the polaron theory of Pardee and Mahan \cite{Pardee1975} and the lattice gas model of Tomoyose \cite{Tomoyose1991, Tomoyose1998} can be viewed as restricted realizations of lattice renormalization. The present framework generalizes these earlier descriptions by explicitly incorporating multi-site frustration, full phonon spectra and the free-energy formulation at finite temperature, thereby connecting phonon-assisted transport, collective ordering and geometric frustration within a unified microscopic picture.

\section*{S2. Numerical evaluations of lattice renormalization and configuration-dependent lattice stiffness} \label{s2}
\subsection*{S2.1. Conditional lattice vibrational free energy $\mathcal{F}_{\text{lat}}^0$}\label{s2.1}
According to equation \eqref{eq:eqS15d}, the change in conditional lattice vibrational free energy $\Delta\mathcal{F}_{\text{lat}}^0$ can be estimated by the net softening/hardening of vibrational frequencies. For instance, a 20\% net softening/hardening per effective vibrational mode gives rise to a correction of $\mathcal{F}_{\text{lat}}^0$ from $-0.22$ to $0.18\, k_BT$ per mode. Note that in the high temperature limit, or when $k_B T \gg \hbar\Omega_{\mathbf{q}s}$, the average energy of a mode is $k_B T$. As a result, moderate softening or hardening of a limited number of vibrational modes gives an entropy contribution to the conditional free energy on the order of a fraction of $k_B T$ per mode.

By definition, an exact calculation of $\mathcal{F}_{\text{lat}}^0(\{\mathbf{r}_i\},T)$ requires constraining the ensemble-averaged lattice displacements $\langle X\rangle$ to zero. Here, for the sake of simplicity, we neglect such constraints and directly evaluate $\mathcal{F}_{\text{vib}}(\{\mathbf{r}_i\},T)$ of the host lattice (conditioned on the ionic configuration $\{\mathbf{r}_i\}$) from the vibrational density of states (VDOS), $D(\omega, \{\mathbf{r}_i\})$, by means of machine-learning molecular dynamics (MLMD) simulations \cite{Dove_1993}:
\begin{equation}
    \mathcal{F}_{\text{vib}}(\{\mathbf{r}_i\},T) = \int_0^{\infty}D(\omega, \{\mathbf{r}_i\})\left[\frac{1}{2}\hbar\omega + k_B T\ln\left(1-e^{-\beta\hbar\omega}\right)\right]\mathrm{d}\omega \tag{S35} \label{eq:eqS35}
\end{equation}
The conditional vibrational free energy difference, $\Delta\mathcal{F}_{\text{vib}}$, is obtained as an ensemble average over five independent simulations: $\Delta\mathcal{F}_{\text{vib}} = \langle\mathcal{F}_{\text{vib}}(\{\mathbf{r}_i\},T)\rangle - \langle\mathcal{F}_{\text{vib}}(\{\mathbf{r}_i^{(R)}\},T)\rangle$.

As summarized in Table \ref{tab:tableS1}, $\Delta\mathcal{F}_{\text{vib}}$ for cubic Li$_7$La$_3$Zr$_2$O$_{12}$ (c-LLZO) is approximately -1.17 eV per c-LLZO unit cell (-2.87 meV per lattice mode) at 700 K and -1.82 eV per c-LLZO unit cell (-4.46 meV per mode) at 1200 K, respectively. For AgCrSe$_2$, $\Delta\mathcal{F}_{\text{vib}}$ is about -9.90 meV per AgCrSe$_2$ primitive cell (-1.10 meV per mode) at 300 K and -22.27 meV per AgCrSe$_2$ primitive cell (-2.47 meV per mode) at 500 K, respectively. These values are consistent with the estimates obtained from equation \eqref{eq:eqS15d}, indicating that the variations of conditional vibrational free energies constitute a small fraction of $k_B T$. Therefore, the conditional lattice vibrational free energy renormalization, $\Delta\mathcal{F}_{\text{vib}}(\{\mathbf{r}_i\},T)$, can either reinforce or partially offset lattice-renormalized stabilization. Nonetheless, the robust negative contribution arises from the adiabatic lattice renormalization $\mathcal{F}_{\text{LR}}$.

\begin{table}[t]
    \centering
    \caption{\textbf{Conditional lattice vibrational free energies of cubic Li$_7$La$_3$Zr$_2$O$_{12}$ and AgCrSe$_2$.} The quantity is ensemble-averaged over ionic configurations $\{\mathbf{r}_i\}$ from full-lattice MLMD simulations and reference ionic configuration $\{\mathbf{r}_i^{(R)}\}$.}
    \label{tab:tableS1}
    \renewcommand{\arraystretch}{1.5}
    \begin{tabular}{c c c c}
        \toprule
        \multicolumn{4}{c}{\textbf{cubic Li$_7$La$_3$Zr$_2$O$_{12}$}} \\
        \midrule
        temperature (K) & $\langle\mathcal{F}_{\text{vib}}(\{\mathbf{r}_i\},T)\rangle$ (eV/cell) & $\langle\mathcal{F}_{\text{vib}}(\{\mathbf{r}_i^{(R)}\},T)\rangle$ (eV/cell) & $\Delta\mathcal{F}_{\text{vib}}$ (eV/cell)\\
        700 & -14.81 $\pm$ 0.28 & -13.64 $\pm$ 0.30 & -1.17 \\
        1200 & -50.42 $\pm$ 0.44 & -48.60 $\pm$ 0.29 & -1.82 \\
        \midrule
        \multicolumn{4}{c}{\textbf{AgCrSe$_2$}} \\
        \midrule
        temperature (K) & $\langle\mathcal{F}_{\text{vib}}(\{\mathbf{r}_i\},T)\rangle$ (meV/cell) & $\langle\mathcal{F}_{\text{vib}}(\{\mathbf{r}_i^{(R)}\},T)\rangle$ (meV/cell) & $\Delta\mathcal{F}_{\text{vib}}$ (meV/cell) \\
        300 & -60.31 $\pm$ 0.08 & -50.41 $\pm$ 0.10 & -9.90 \\
        500 & -318.03 $\pm$ 0.09 & -295.76 $\pm$ 0.06 & -22.27 \\
        \bottomrule
    \end{tabular}
\end{table}

\subsection*{S2.2. Configuration-dependent lattice stiffness}\label{s2.2}
The second moment $M_2$ of the host framework's VDOS is employed to characterize the dependence of lattice stiffness on the ionic configuration \cite{Dove_1993, PhysRevB.53.8993}:
\begin{equation}
    M_2 = \int_0^{\infty}\omega^2D(\omega)\mathrm{d}\omega \tag{S36} \label{eq:eqS36}
\end{equation}
Within the harmonic approximation, it directly quantifies the trace of the interatomic force-constant matrix \cite{Dove_1993, PhysRevB.53.8993}. Accordingly, the overall change in lattice stiffness, characterized by $\widetilde{K}$ and $\widetilde{K}_0$, can be measured by $\Delta M_2=M_2-M_2^{(R)}$, where $M_2^{(R)}$ is the second moment evaluated with the ions constrained at the reference positions. Furthermore, we can also define a spectrally resolved lattice stiffness $M_2(\omega)\equiv\omega^2 D(\omega)$ in terms of the host lattice's VDOS $D(\omega)$.

\begin{figure}[h]
\centering
\includegraphics[width=1.0\textwidth]{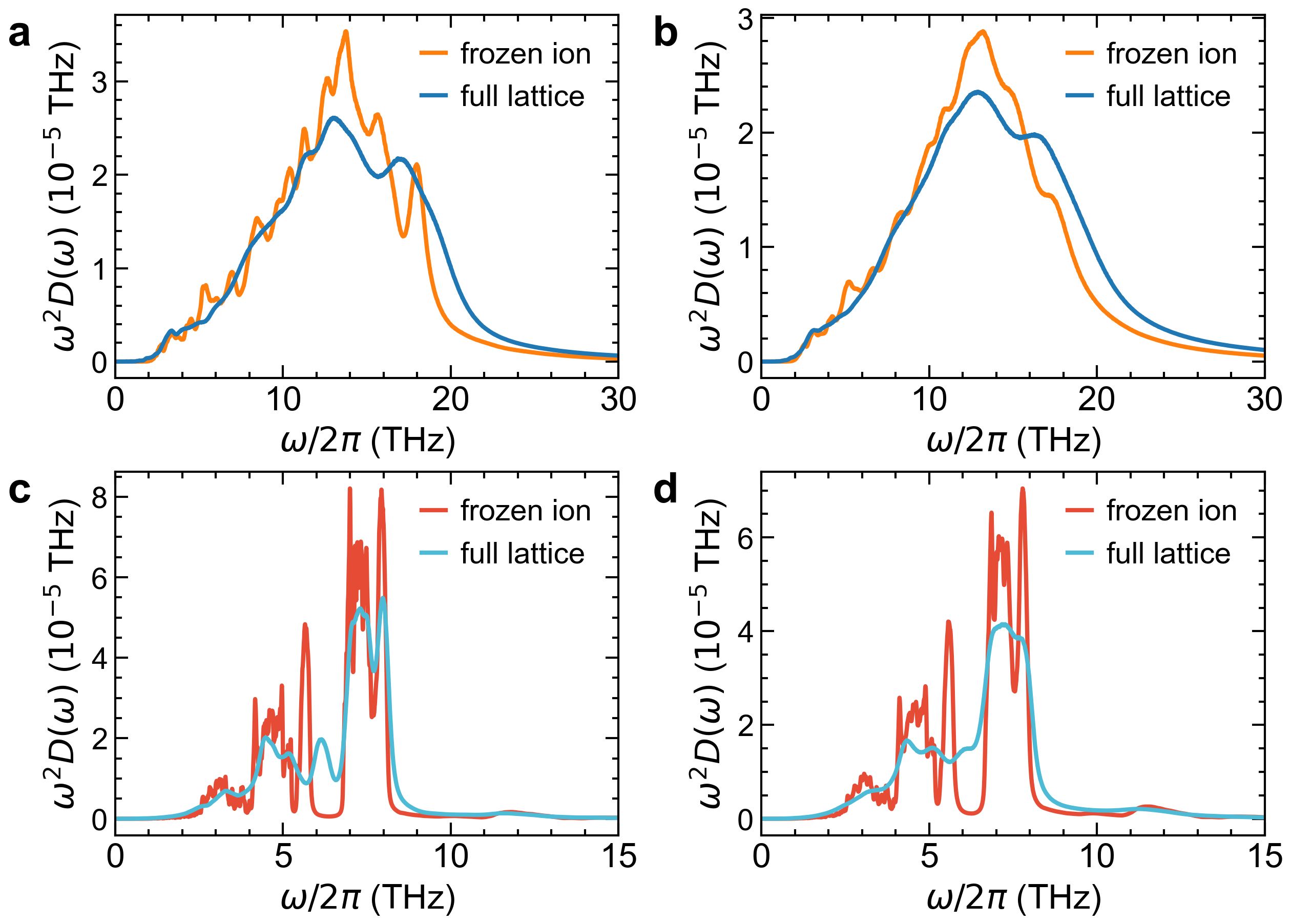}
\caption{\textbf{Spectrally resolved lattice stiffness of cubic Li$_7$La$_3$Zr$_2$O$_{12}$ and AgCrSe$_2$.} \textbf{a,b,} Frozen-ion and full-lattice lattice stiffness $\omega^2 D(\omega)$ of cubic Li$_7$La$_3$Zr$_2$O$_{12}$ simulated at 700 K (\textbf{a}) and 1200 K (\textbf{b}), respectively. \textbf{c,d,} Frozen-ion and full lattice lattice stiffness $\omega^2 D(\omega)$ of AgCrSe$_2$, calculated at 300 K (\textbf{c}) and 500 K (\textbf{d}), respectively.}\label{figS1}
\end{figure}

For c-LLZO, the values of $\Delta M_2$ are approximately 4.35$\times$10$^4$ THz$^2$ at 700 K and 1.00$\times$10$^5$ THz$^2$ at 1200 K for a host framework comprising 8704 atoms. The corresponding relative changes, $\Delta M_2/M_2^{(R)}$, are $\sim$1.54\% and 3.73\%, respectively, indicating a net hardening of the host lattice. As shown in Figs. \ref{figS1}a-b, this enhanced stiffness is predominantly governed by the high-frequency optical modes above 16 THz.

For AgCrSe$_2$, $\Delta M_2$ is approximately 5.76$\times$10$^3$ THz$^2$ at 300 K and -1.49$\times$10$^4$ THz$^2$ at 500 K for a host lattice containing 12960 atoms. The relative changes $\Delta M_2/M_2^{(R)}$ are $\sim$0.47\% and -1.26\%, respectively. In this case, the lattice is stiffened at 300 K, but becomes softened at 500 K. The spectrally resolved lattice softness $M_2(\omega)$ at 300 K and 500 K is presented in Figs. \ref{figS1}c-d. The results reveal that the host lattice is strengthened in the frequency ranges below 2.5 THz, 3.5--4 THz, 6--7 THz, and above 8 THz.

The apparent contradiction between the lattice softening inferred from the vibrational free energy (Table \ref{tab:tableS1}) and the $M_2$-based hardening is resolved by recognizing that these two quantities probe distinct spectral regions. The conditional vibrational free energy $\mathcal{F}_{\text{vib}}$ is dominated by low-frequency modes because of their larger Bose-Einstein weights, whereas $M_2$ is dominated by high-frequency modes through the $\omega^2$ weighting. Consequently, both quantities can exhibit opposite trends when the phonon spectrum undergoes a selective redistribution. Lattice renormalization selectively softens the low-frequency modes that couple strongly to ionic displacements, and their softening directly facilitates ionic transport. In parallel, the high-frequency optical branches, which are associated with framework bond-stretching, may harden as the framework bonds experience increased stress due to the ionic configuration. The conditional free energy primarily captures the former effect (softening), while $M_2$ reflects the concomitant hardening of the host framework.

\subsection*{S2.3. Lattice-relaxation contribution $\mathcal{F}_{\text{LR}}$}\label{s2.3}
\subsubsection*{S2.3.1. Estimation of $\mathcal{F}_{\text{LR}}$ from Debye temperature}\label{s2.3.1}
Regarding $\mathcal{F}_{\text{LR}}$, we start from the expression $-\frac{1}{2}x_{\text{eq}}^\dagger\widetilde{K}x_{\text{eq}}$ and transform it back to real space, which has the form of $-\frac{1}{2} u_{\text{eq}}^{\text{T}}\widetilde{\Phi}\,u_{\text{eq}}$. Here $u_{\text{eq}}$ and $\widetilde{\Phi}$ are the adiabatic lattice relaxation and effective second-order force constant. A reasonable estimate, or upper bound, of $u_{\text{eq}}$ can be determined from the Lindemann criterion \cite{Lindemann}, according to which the square root of mean-squared displacements (MSD) reaches about 0.1 of the interparticle distance at the melting point. For typical lattice periodicity of 2--3 $\rm\AA$, this yields $u_{\text{eq}}\sim$0.2--0.3 $\rm\AA$.

\begin{figure}[t]
\centering
\includegraphics[width=1.0\textwidth]{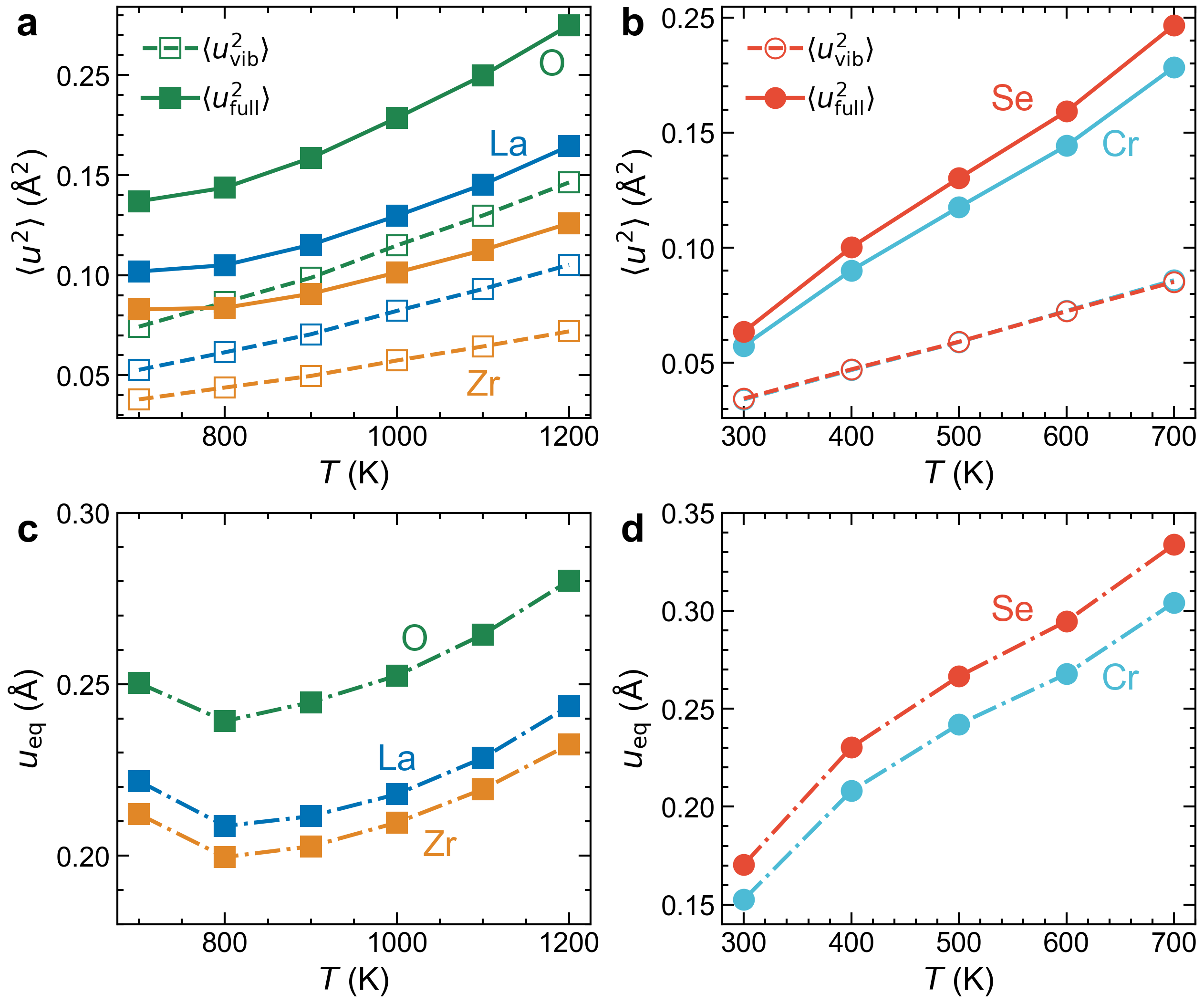}
\caption{\textbf{Mean-square displacements (MSD) of framework atoms calculated by MLMD simulations.} \textbf{a,b,} Frozen-ion MSD $\langle u_{\text{vib}}^2\rangle$ (open symbols) and full-lattice MSD $\langle u_{\text{full}}^2 \rangle$ (filled symbols) of La, Zr and O atoms in cubic Li$_7$La$_3$Zr$_2$O$_{12}$ (\textbf{a}); and Cr and Se atoms in AgCrSe$_2$ (\textbf{b}). \textbf{c,d,} The adiabatic lattice relaxation $u_{\text{eq}}$ of each framework atom in cubic Li$_7$La$_3$Zr$_2$O$_{12}$ (\textbf{c}) and AgCrSe$_2$ (\textbf{d}), estimated as $u_{\text{eq}} = \sqrt{\langle u_{\text{full}}^2\rangle - \langle u_{\text{vib}}^2\rangle}$.}\label{figS2}
\end{figure}

Practically, the framework relaxation can also be evaluated from MLMD simulations. To separate $u_{\text{eq}}$ from thermal fluctuations, we first performed full MLMD simulations and obtained the full MSD $\langle u_{\text{full}}^2\rangle$ of each element. Then frozen-ion MLMD simulations, in which all the ions were fixed at $\{\mathbf{r}_i^{(R)}\}$, were carried out and the MSD corresponding to thermal fluctuations $\langle u_{\text{vib}}^2\rangle$ could be obtained. In this way, we determined $u_{\text{eq}}$ by $u_{\text{eq}}^2 = \langle u_{\text{full}}^2\rangle - \langle u_{\text{vib}}^2\rangle$. The calculation results for c-LLZO and AgCrSe$_2$ are provided in Fig. \ref{figS2}. For both materials, the calculated $u_{\text{eq}}$, especially the framework anions, is in good agreement with the Lindemann criterion in the range of 0.2--0.3 $\mathrm{\AA}$.

The effective force constant of a material can be approximately estimated by its Debye temperature $\Theta_D$ \cite{Schlem2020, Muy2021}
\begin{equation}
    \widetilde{\Phi} \approx M_{\text{eff}}\left(\frac{k_B\Theta_D}{\hbar}\right)^2 \tag{S37} \label{eq:eqS37}
\end{equation}
where $M_{\text{eff}}$ is the effective mass of the framework atoms. For typical $M_{\text{eff}}$ ranging from 16 to 80 atomic mass units (amu) and $\Theta_D$ from 100--300 K, $\widetilde{\Phi}$ is about 0.28--12.82 eV/$\text{\AA}^2$, which leads to $\mathcal{F}_{\text{LR}}$ in the range from -580 to -6 meV per framework atom for $u_{\text{eq}}\sim$ 0.2--0.3 $\mathrm{\AA}$. This value is comparable to $k_B T$ and can give rise to a substantial renormalization of the ionic free-energy landscape.

For c-LLZO, its Debye temperature is reported to be about 498 K from an analysis of the low-temperature specific heat \cite{Wang2025}. Using the sound velocities of c-LLZO, a similar value ($\Theta_D\sim 520$ K) is obtained \cite{Yu2016, Neises2022}. As a result, the effective force constant of c-LLZO is approximately 20.5--22.4 eV/$\text{\AA}^2$ ($M_{\text{eff}}$=46.5 amu) and the local free-energy renormalization ranges from -878 to -410 meV per framework atom. This quantity can be converted to a renormalization of free energy per Li$^+$ ion. For instance, the numbers of framework atoms and Li$^+$ ions in a c-LLZO unit cell are 136 (24 La + 16 Zr + 96 O) and 56, respectively. Therefore, the free-energy renormalization is from -2.13 to -1 eV per Li$^+$ ion. These estimated values are consistent in magnitude with the large difference in activation energies $E_{\text{a}}$, i.e. 0.31 eV for the full simulation and 1.68 eV for the frozen-lattice simulation (Fig. 2a in the main text).

In comparison, due to the soft selenide framework, the Debye temperature of AgCrSe$_2$ is much lower, about 120--137 K \cite{Xie2019, Ding2025}. The corresponding effective force constant is only about 1.79--2.35 eV/$\text{\AA}^2$ ($M_{\text{eff}}$=70 amu), an order of magnitude smaller than that of c-LLZO. The resulting free-energy renormalization is thus from -132 to -20 meV per framework atom, or from -396 to -60 meV per Ag$^+$ ion. Note that this free-energy renormalization is also in accordance with the relatively smaller difference in $E_{\text{a}}$ (Fig. 2b in the main text).

Note that the Debye temperature represents an average over all phonon modes. Since $\mathcal{F}_{\text{LR}}$ is dominated by soft modes with $\Omega_{\mathbf{q}s}(T)\ll\Theta_D$, the actual free-energy renormalization may be smaller than our estimates for a given magnitude of $u_{\text{eq}}$. The results presented here should therefore be understood as an order-of-magnitude estimation.

\subsubsection*{S2.3.2. Estimation of $\mathcal{F}_{\text{LR}}$ from atomistic simulations}\label{s2.3.2}
Besides a rough estimate of $\mathcal{F}_{\text{LR}}$ by Debye temperature, we also carried out further numerical calculations through atomistic simulations. By definition, $\mathcal{F}_{\text{LR}}$ represents the free-energy gain upon adiabatic lattice relaxation from the reference lattice. In principle, it is temperature-dependent and should be evaluated from the full thermodynamic ensemble. In the present work, we evaluate this term in the zero-temperature static limit. Specifically, we obtained multiple snapshots of Li$^+$/Ag$^+$ configurations from full MLMD simulations, and determined the lattice relaxation energy for the given ionic configuration by optimizing the reference lattice. Table \ref{tab:tableS2} summarizes the numerical values of $\mathcal{F}_{\text{LR}}$ for c-LLZO and AgCrSe$_2$. Compared to $\Delta\mathcal{F}_{\text{vib}}$ (Table \ref{tab:tableS1}), $\mathcal{F}_{\text{LR}}$ is larger by an order of magnitude, validating its dominant role in the free-energy landscape renormalization.
\begin{table}[b]
    \centering
    \caption{\textbf{Lattice renormalization in cubic Li$_7$La$_3$Zr$_2$O$_{12}$ and AgCrSe$_2$.}}
    \label{tab:tableS2}
    \renewcommand{\arraystretch}{1.5}
    \begin{tabular}{c c | c c}
        \toprule
        \multicolumn{2}{c}{\textbf{cubic Li$_7$La$_3$Zr$_2$O$_{12}$}} & \multicolumn{2}{c}{\textbf{AgCrSe$_2$}}\\
        \midrule
        temperature (K) & $\mathcal{F}_{\text{LR}}$ (eV/cell) & 
        temperature (K) & $\mathcal{F}_{\text{LR}}$ (meV/cell) \\
        700 & -14.23 $\pm$ 0.26 & 300 & -99.44 $\pm$ 1.73 \\
        800 & -14.68 $\pm$ 0.29 & 400 & -208.22 $\pm$ 14.22\\
        900 & -16.54 $\pm$ 0.52 & 500 & -254.01 $\pm$ 9.62 \\
        1000 & -18.23 $\pm$ 0.27 & 600 & -295.53 $\pm$ 16.48\\
        1100 & -20.24 $\pm$ 0.65 & 700 & -337.07 $\pm$ 14.55 \\
        1200 & -22.80 $\pm$ 0.53 \\
        \bottomrule
    \end{tabular}
\end{table}

Notably, the calculated $\mathcal{F}_{\text{LR}}$ for c-LLZO is from -22.80 to -14.23 eV per unit cell with 192 atoms, which corresponds to -0.41 $\sim$ -0.25 eV per Li$^+$ ion and about a quarter of the values estimated from Debye temperature. In contrast, the lattice-relaxation renormalization obtained for AgCrSe$_2$ (from -337.07 to -99.44 meV per Ag$^+$ ion) is in excellent agreement with the Debye model, and satisfactorily accounts for the reduction in activation energy shown in Fig. 2b.

The large discrepancy observed in c-LLZO is likely attributable to an overestimation of the effective force constants inferred from $\Theta_D$, and/or of the adiabatic lattice-relaxation displacements $u_{\text{eq}}$. Another plausible explanation is that only a subset of Li$^+$ ions within a given snapshot is actively engaged in diffusion. Therefore, only those ions that actually participate in diffusion, i.e. that leave their equilibrium sites and explore transition-state configurations, significantly drive the lattice away from its high-symmetry reference positions. Ions that remain near their equilibrium positions contribute negligibly to the lattice relaxation. As a result, an appropriate normalization of $\mathcal{F}_{\text{LR}}$ from a per-unit-cell quantity to a per-ion quantity is given by the number of Li$^+$ ions that actively participate in diffusion, rather than by the total number of Li$^+$ ions in the simulation cell. In comparison, the Ag sublattice in AgCrSe$_2$ is highly disordered such that nearly all Ag ions contribute to diffusion. In this case, the static relaxation energy can be normalized by the total number of Ag$^+$ ions in the simulation cell, leading to a direct comparison with the calculated reduction in $E_{\text{a}}$.

\section*{S3. Atomistic simulation results}\label{s3}
Beyond reciprocal-space signatures revealed by partial static structure factors in the main text, the lattice renormalization can also be visualized in real space through the evolution of frustrated ionic configurations. Here we adopt an approach similar to the density-of-atomistic-states (DOAS) analysis \cite{Wang2023}. Specifically, we first performed frozen-lattice and full-lattice MLMD simulations at 1200 K for cubic LLZO and 500 K for AgCrSe$_2$ in canonical (NVT) ensembles for an equilibrium period of 500 ps. Subsequently, 1000 configurations were sampled at an interval of 0.1 ps per frame in microcanonical (NVE) ensembles. For each of these sampled configurations, we then conducted frozen-lattice and full-lattice structural relaxation, respectively, with atomic forces converged to below $0.001\; \text{eV/\AA}$. As will be shown below, rather than stabilizing a single preferred arrangement, lattice renormalization transforms a collection of isolated frustrated minima into an interconnected frustrated manifold, providing further real-space and reciprocal-space manifestations of collective ionic behaviors.

\subsection*{S3.1. Cubic Li$_7$La$_3$Zr$_2$O$_{12}$}\label{s3.1}
\subsubsection*{S3.1.1. Lattice-renormalized geometric frustration in real space}\label{s3.1.1}
\begin{figure}[t]
\centering
\includegraphics[width=0.75\textwidth]{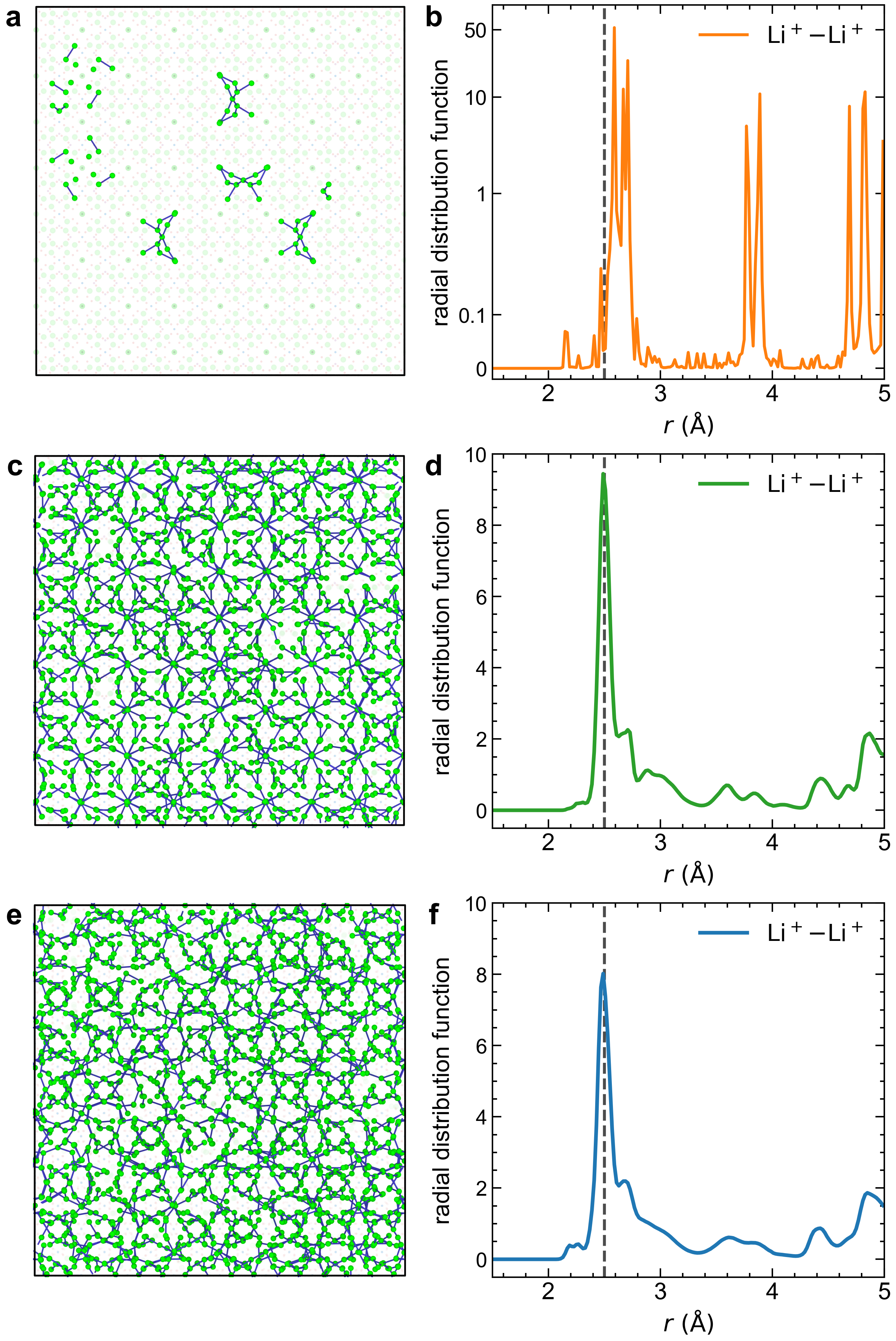}
\caption{\textbf{Real-space analysis of geometric frustration in cubic Li$_7$La$_3$Zr$_2$O$_{12}$.} Li cations are colored in green and Li$^+$-Li$^+$ separations shorter than 2.5 $\rm\AA$ are highlighted in blue. \textbf{a,} Representative snapshot of the frozen-lattice relaxed ordered structure viewed along the $c$-axis. The frustrated Li$^+$ configurations are characterized by butterfly-shaped multimers, V-shaped trimers and correlated chiral rotational motions. \textbf{b,} Radial distribution function of Li$^+$-Li$^+$ obtained from the frozen-lattice relaxation. The dashed vertical line indicates 2.5 $\mathrm\AA$. \textbf{c,d,} Representative snapshot of the full-lattice relaxed ordered structure viewed along the $c$-axis (\textbf{c}), and the corresponding radial distribution function (\textbf{d}). The local ionic configuration is transformed into an interconnected frustrated manifold. The shift of the main peak in the radial distribution function signifies the breakdown of long-range periodicity and the emergence of short-range order. \textbf{e,f,} Snapshot of the full-lattice relaxed disordered structure (viewed along the c-axis) (\textbf{e}), and the corresponding radial distribution function (\textbf{f}). The population of short-distance Li$^+$-Li$^+$ bonds around 2.1--2.3 $\rm\AA$ is further increased.} \label{figS3}
\end{figure}

Figure \ref{figS3}a shows a representative configuration snapshot of ordered c-LLZO viewed along the $c$-axis, extracted from a frozen-lattice 1200 K MLMD trajectory and relaxed under the frozen-lattice constraint. Li$^+$ ions separated by a distance shorter than 2.5 $\mathrm{\AA}$ are highlighted with bonds. The ordered-phase structure was intentionally constructed by placing the tetragonal Li$^+$ configuration into the cubic framework. The most salient feature in this snapshot is a ``multimer" topology with a butterfly-like geometry. This geometrically frustrated motif is found to persist across the ensemble of ionic configurations relaxed in a frozen lattice, despite its relatively low occurrence probability. From the radial distribution function (RDF) of Li$^+$-Li$^+$ pairs, $g_{\text{Li}^+\text{-Li}^+}(r)$ (Fig. \ref{figS3}b), the characteristic inter-Li$^+$ distances $d_{\text{Li}^+\text{-Li}^+}$ associated with this topology are approximately 2.15, 2.41 and 2.47 $\mathrm{\AA}$, respectively. In addition, a V-shaped ``trimer" topology, characterized by an inter-Li$^+$ distance of $\sim$2.27 $\mathrm{\AA}$, is also observed. For comparison, the dominant first nearest-neighbor Li$^+$-Li$^+$ separation is significantly larger, $\sim$2.59 $\mathrm{\AA}$. Beyond these local structural features, we further identify a distinct short-range correlated configuration comprising four Li$^+$-Li$^+$ dimers coupled to chiral Li$^+$ rotational motion. It should be stressed that these structures are intrinsically three-dimensional, and the results presented here should be interpreted as projections along the c-axis. Taken together, these motifs represent local minima belonging to a weakly frustrated manifold of the Li$^+$ subsystem.

Next, we demonstrate how the host lattice can effectively activate the latent geometric frustration within the same ordered Li$^+$ configurations by allowing full lattice relaxation of the configurational snapshots extracted from frozen-lattice simulations. As shown in Fig. \ref{figS3}c, the largely isolated local geometric frustration identified in Fig. \ref{figS3}a now becomes collectively interconnected, resulting in a disordered Li$^+$ network. Within this interconnected network, the characteristic Li$^+$-Li$^+$ separations, quantified by the first nearest-neighbor peak in $g_{\text{Li}^+\text{-Li}^+}(r)$, shift markedly from 2.59 $\mathrm{\AA}$ to 2.49 $\mathrm{\AA}$ (Fig. \ref{figS3}d). This disordered state, stabilized by the flexible La-Zr-O framework, highlights how adiabatic lattice relaxation mediates a redistribution of the populations across frustrated manifolds, thereby enabling distinct local minima to cooperate rather than remain isolated. The resulting shortened Li$^+$-Li$^+$ bonds constitute a real-space manifestation of the effective two-body attractive interaction introduced in equation (6c) of the main text.

For comparison, full-lattice relaxation was also applied to full-lattice MLMD trajectories, in which Li cations are in an intrinsically disordered cubic configuration. As displayed in Fig. \ref{figS3}e, a geometrically frustrated Li$^+$ network similar to Fig. \ref{figS3}c emerges as well. It is thus implied that two initial configurations as different as ordered tetragonal and disordered cubic converge to a common frustrated and correlated structure under adiabatic lattice relaxation. The RDF between Li$^+$ ions, as shown in Fig. \ref{figS3}f, is nearly identical to that in Fig. \ref{figS3}d. The subtle shoulder feature in the range of 2.1--2.3 $\mathrm{\AA}$ indicates the increased population of significantly shorter Li$^+$-Li$^+$ bonds by lattice renormalization.

\begin{figure}[t]
    \centering
    \includegraphics[width=0.6\linewidth]{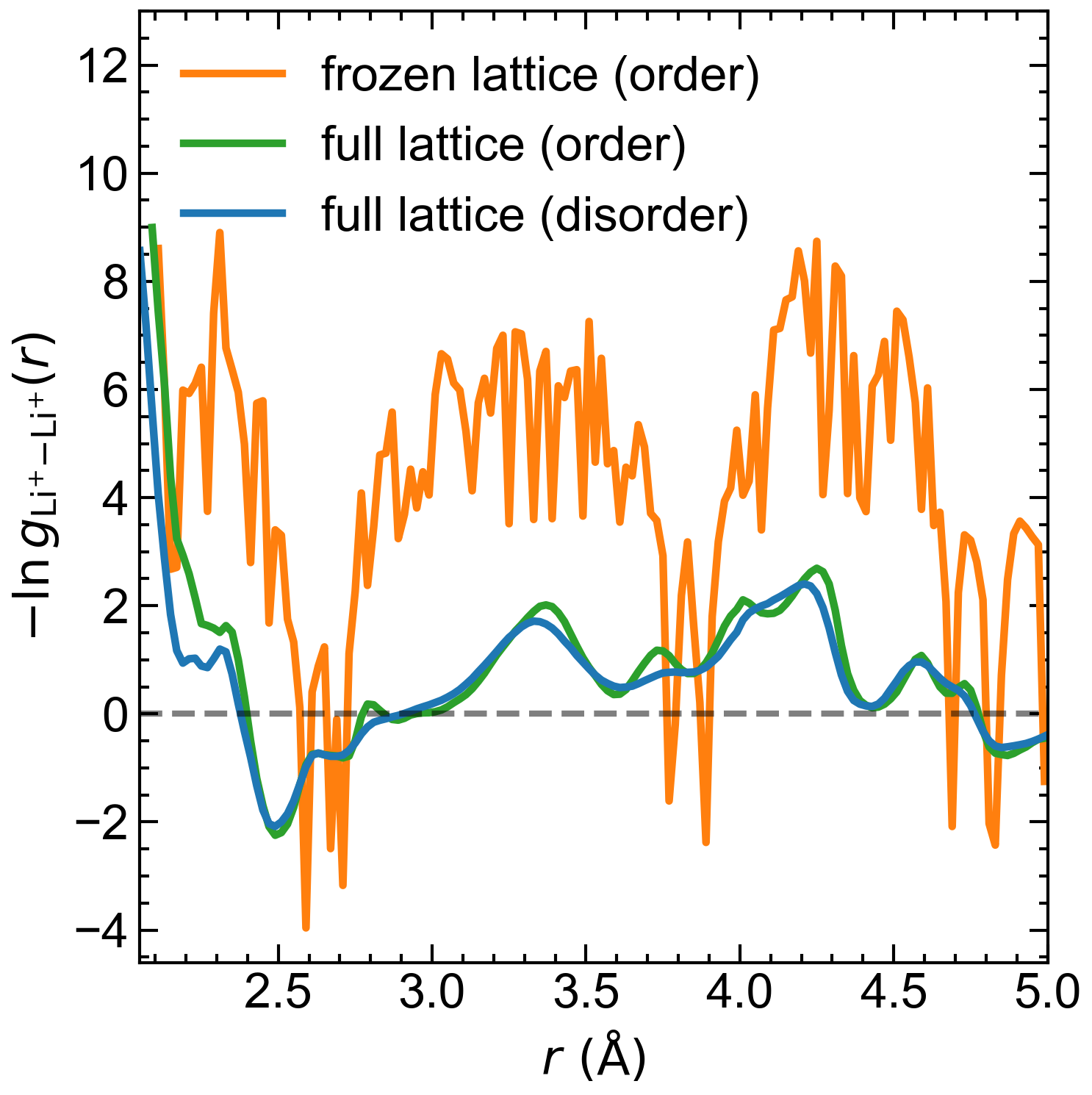}
    \caption{\textbf{The prefactor-reduced potential of mean force between two Li$^+$ cations.} The ordered structures are taken from frozen-lattice MLMD trajectories and relaxed under frozen-lattice constraints or without constraints (full lattice). The disordered structures are extracted from full-lattice MLMD trajectories and fully relaxed. The negative values indicate effectively attractive pair interactions, while positive values are repulsive.}
    \label{figS4}
\end{figure}

The RDFs presented in Figs. \ref{figS3}b, \ref{figS3}d and \ref{figS3}f are further employed to examine the potential of mean force (PMF), $W$, between Li$^+$ ions, which is defined as \cite{FRENKEL2002167}:
\begin{equation}
    W = -k_BT \ln{g_{\text{Li}^+\text{-Li}^+}(r)} \tag{S38} \label{eq:eqS38}
\end{equation}
Since $g_{\text{Li}^+\text{-Li}^+}(r)$ is obtained from static relaxations at $T=0$ rather than from a finite-temperature thermal ensemble, the absolute energy scale set by the thermal prefactor $k_B T$ in $W$ is not meaningful here. We therefore introduce a prefactor-reduced quantity $\tilde{W}\equiv -\ln{g_{\text{Li}^+\text{-Li}^+}(r)}$, whose sign and relative structure serve as a qualitative indicator of the effective Li$^+$-Li$^+$ pair interactions. Figure \ref{figS4} displays the resulting prefactor-reduced PMF between Li$^+$ ions. Negative values, i.e. $\tilde{W}<0$, correspond to effectively attractive pair interactions between Li$^+$ ions, whereas positive values reflect effective repulsion.

For the ordered configurations in which frustration is induced solely by the potential part of $\mathcal{H}_{\text{GF}}$ (Fig. \ref{figS3}a), multiple sharp dips below zero are observed, which indicate strong pinning of Li$^+$ ions by the static periodic potential $V_0$. Over a broad range of separations, the effective Li$^+$-Li$^+$ pair interactions are strongly repulsive, rendering the latent geometric frustration energetically unfavorable under these circumstances. Upon inclusion of lattice renormalization, the effective pair interactions in the range 2.4--2.9 $\mathrm{\AA}$, particularly those below 2.5 $\mathrm{\AA}$, become attractive, thereby stabilizing geometrically frustrated configurations. Simultaneously, the strong repulsive pair interactions in the interval of 2.9--4.75 $\mathrm{\AA}$ are substantially weakened, signifying the crucial role of lattice renormalization. In parallel, the interactions that were previously attractive at $r\approx3.77\;\mathrm{\AA}$ and 3.88 $\mathrm{\AA}$ become repulsive. These features are characteristic of the lattice-renormalized attractive/repulsive interference interactions described by equation (6c) or \eqref{eq:eqS26b}. For the disordered configuration, its prefactor-reduced PMF closely resembles that of the ordered structure relaxed without constraints. The lower values of $\tilde{W}$ in the range 2.1--2.3 $\mathrm{\AA}$ suggest an enhanced stabilization of short-range Li$^+$-Li$^+$ bonds.

\subsubsection*{S3.1.2. Lattice-renormalized geometric frustration in reciprocal space}\label{s3.1.2}
\begin{figure}[t]
\centering
\includegraphics[width=1.0\textwidth]{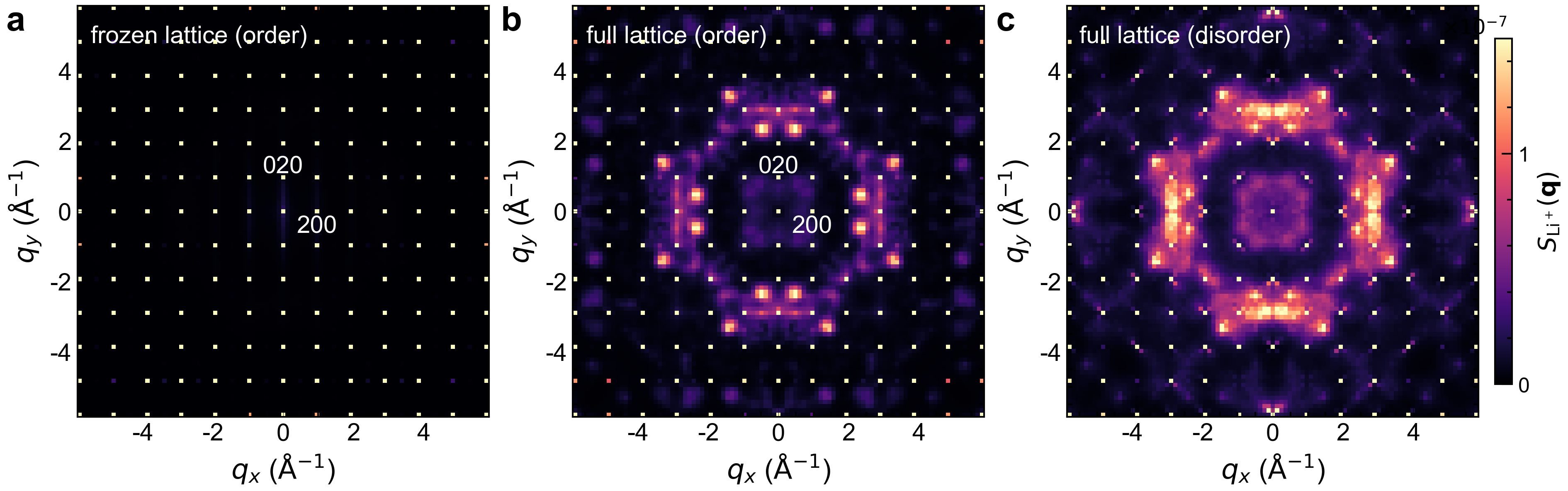}
\caption{\textbf{Analysis of geometric frustration in cubic Li$_7$La$_3$Zr$_2$O$_{12}$ in reciprocal space.} \textbf{a,b,} The simulated partial static structure factors $S_{\text{Li}^+}(\mathbf{q})$ for the ordered structures, relaxed under frozen-lattice constraint (\textbf{a}) and relaxed without constraints (\textbf{b}), respectively. The geometrically frustrated motifs observed in Fig. \ref{figS3}a lead to negligible diffuse scattering, whereas the interconnected frustrated manifolds (Fig. \ref{figS3}c) give rise to enhanced diffuse scattering intensities around \{150\}$^*$ and \{370\}$^*$ Bragg diffraction, as well as diffuse streaks in the vicinity of \{600\}$^*$. \textbf{c,} The simulated partial static structure factor $S_{\text{Li}^+}(\mathbf{q})$ for the full-lattice snapshots optimized without constraints. Similar diffuse scattering pattern with broader intensity distribution and enhanced diffuse streaks is observed.} \label{figS5}
\end{figure}

The partial static structure factors $S_{\text{Li}^+}(\mathbf{q})$ corresponding to the ionic configurations shown in Figs. \ref{figS3}a, \ref{figS3}c and \ref{figS3}e were simulated following the same procedure as described in Methods, and the results are displayed in Fig. \ref{figS5}. For the frozen-lattice snapshots of the ordered Li$^+$ configurations relaxed under the frozen-lattice constraint, $S_{\text{Li}^+}(\mathbf{q})$ is dominated by Bragg diffraction (Fig. \ref{figS5}a). The frustrated butterfly-like topologies, V-shaped trimers, and coupled rotational motions give rise to negligible diffuse scattering due to their low occurrence.

In comparison, the partial static structure factors $S_{\text{Li}^+}(\mathbf{q})$ computed for the same frozen-lattice snapshots (Fig. \ref{figS3}c) and the full-lattice snapshots (Fig. \ref{figS3}e) relaxed without constraints exhibit diffuse streaks around \{600\}$^*$ reflections along with the appearance of spots around the nominally forbidden \{150\}$^*$ and \{370\}$^*$ reflections (Figs. \ref{figS5}b-c). These features are consistent with the diffuse scattering observed in Fig. 3f of the main text. Notably, certain reflections such as \{200\}$^*$ and \{600\}$^*$ are systematically absent in Fig. \ref{figS5}c, but well persist in Figs. \ref{figS5}a-b. These extinct Bragg spots can be well understood from the pure phase cancellation of the partially occupied Li$^+$ sublattice in the space group of Ia$\bar{3}$d. In other words, \{200\}$^*$ and \{600\}$^*$ are allowed reflections in the ordered structure, but become systematically forbidden in the disordered cubic variant.

According to the standard relation between pair correlations and structure factors \cite{Egami2003, Hansen2006}, characteristic inter-ionic length scales define approximate ionic correlation wavevectors $\mathbf{Q}$ via $|\mathbf{Q}|\sim 2\pi/d_{\text{Li}^+\text{-Li}^+}$. For example, an inter-ionic separation $d_{\text{Li}^+\text{-Li}^+}\approx 2.15\,\mathrm{\AA}$ implies $|\mathbf{Q}|\approx 2.92\,\mathrm{\AA}^{-1}$, which is close to the magnitude of the $\{600\}^*$ scattering vector $\mathbf{G}_{{600}^*}\approx 2.86\,\mathrm{\AA}^{-1}$ in c-LLZO. Therefore, Figs. \ref{figS5}b-c not only provide the directional information of Li$^+$-Li$^+$ correlations within the frustrated manifolds, but also the corresponding Li$^+$-Li$^+$ separation distances (projected onto the $a$-$b$ plane). These diffuse-scattering signatures in fast ionic conductors should be experimentally accessible by X-ray, neutron and electron diffraction measurements.

\subsubsection*{S3.1.3. Dynamics of the La-Zr-O host lattice}\label{s3.1.3}
In principle, the effective stiffness matrix $\widetilde{K}(\{\mathbf{r}_i\},T)$ can be obtained by sampling forces and displacements of the host lattice in a self-consistent way as in the stochastic self-consistent harmonic approximation (SSCHA) \cite{PhysRevB.96.014111, PhysRevB.89.064302} or in the temperature-dependent effective potential (TDEP) method \cite{TDEP}. Recent work on the superionic conductor Ag$_2$Te has demonstrated that the full dynamical matrix can also be extracted from MLMD trajectories \cite{Zhou2026}.

Here, for the sake of simplicity, we employed the spectral energy density (SED) method \cite{Thomas2010} implemented in the pySED package \cite{Liang2025} to analyze the phonon dispersion relations of the La-Zr-O host framework. To this end, a 6$\times$6$\times$6 supercell was used. Consistent with the full-lattice simulations, we first performed a 500 ps equilibration in the canonical (NVT) ensemble using the Langevin thermostat. This was followed by an additional 500 ps NVT equilibration, in which the mobile ions were kept fixed. In this second stage, the frozen Li$^+$ positional disorder introduces local structural randomness that renormalizes the stiffness matrix $\widetilde{K}(\{\mathbf{r}_i\},T)$. We refer to this approach as the frozen-ion simulation. The final production stage consisted of a 500 ps trajectory in the NVE ensemble. In addition to the frozen-ion simulation, fully coupled ion-lattice simulations were also carried out for comparison. To ensure sampling over distinct ionic configurations, five independent trajectories were generated.

\begin{figure}[t]
\centering
\includegraphics[width=1.0\linewidth]{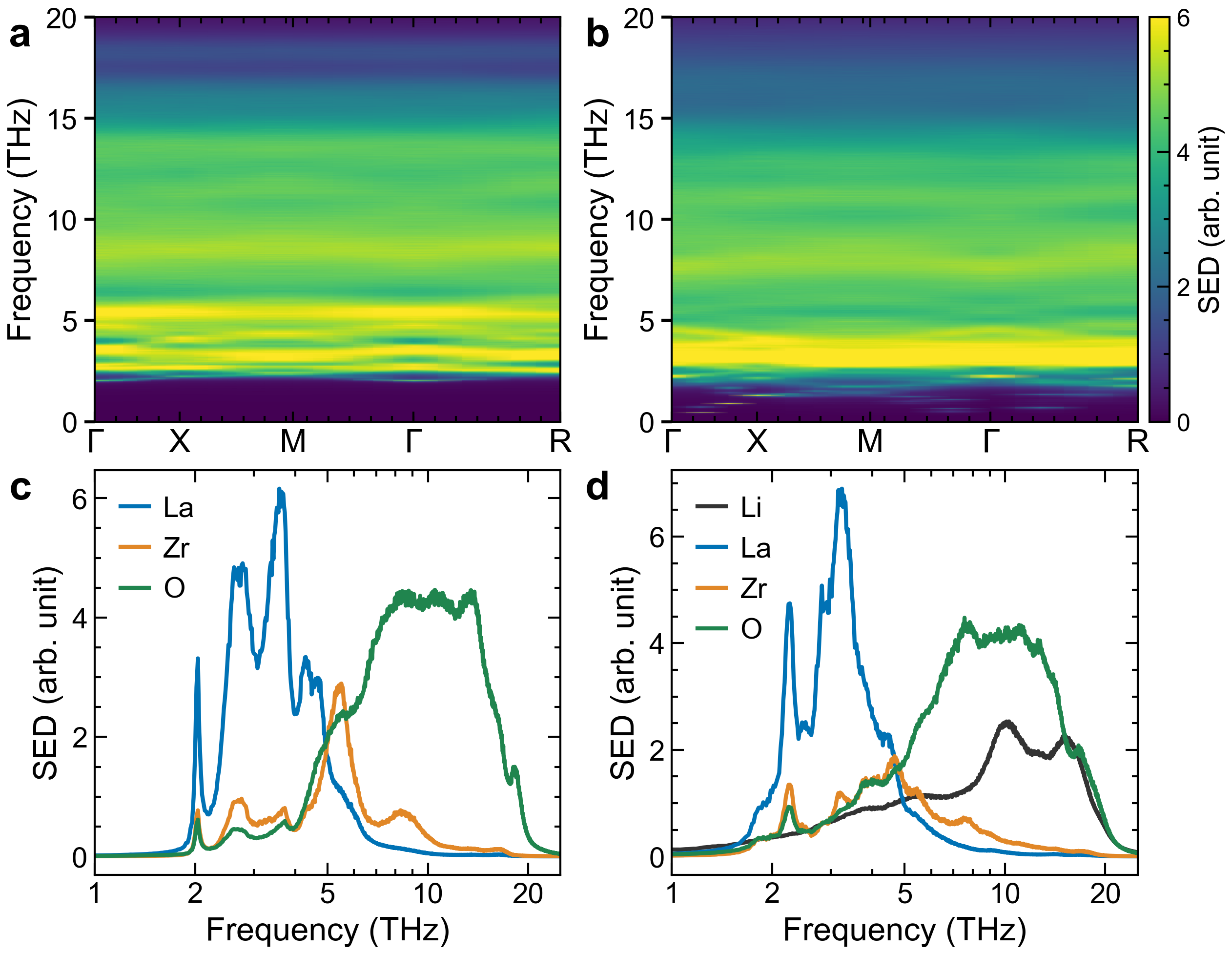}
\caption{\textbf{Spectral energy density (SED) analysis of the La-Zr-O host lattice in cubic Li$_7$La$_3$Zr$_2$O$_{12}$ at 700 K.} \textbf{a,} Spectral energy density of La-Zr-O framework in the frozen Li$^+$ disorder limit. A sharp optical mode at 2 THz can be identified throughout the Brillouin zone. \textbf{b,} Spectral energy density of La-Zr-O framework in the dynamic Li$^+$ disorder regime. \textbf{c,d,} Spectral energy density of La-Zr-O framework at the $\Gamma$ point in the limit of static disorder (\textbf{c}) and dynamic disorder (\textbf{d}). The vibrational modes below 5 THz exhibit frequency shifts and moderate broadening due to dynamic disorder of Li$^+$. The weak broadening of the low-frequency vibrational mode at $\sim$2--2.2 THz indicates a weak coupling between this collective mode and the dynamically disordered Li$^+$ sublattice.} \label{figS6}
\end{figure}

As shown in Fig. \ref{figS6}a, the frozen-ion SED computed at 700 K reveals that the La-Zr-O framework is characterized by weakly dispersive, dense vibrations below 7 THz across the Brillouin zone. By decomposing the frozen-ion SED at the zone center $\Gamma$ point into individual framework elements, we find that the modes below 7 THz are predominantly associated with La atoms, while those of Zr atoms give rise to an intense vibrational density of states around 5 THz (Fig. \ref{figS6}c). The framework O atoms contribute mainly to modes above 4 THz. For comparison, Figs. \ref{figS6}b and \ref{figS6}d illustrate the full SED and its element-resolved decompositions at the $\Gamma$ point at 700 K. The SED associated with La atoms exhibits frequency shifts and phonon bunching at $\sim$3 THz, while the spectral weights of Zr and O atoms broaden. These results are qualitatively consistent with Raman measurements, which assign the modes below 4.5 THz to La-O vibrations along the Li$^+$ hopping pathway \cite{Tietz2013, Lin2025}. The increased peak widths indicate a reduced phonon lifetime arising from the energy exchange between the host lattice and the dynamically disordered Li$^+$ subsystem.

Surprisingly, the low-frequency optical mode at $\sim$2--2.2 THz remains unusually sharp, suggesting weak anharmonicity and a limited dependence on the instantaneous ionic configurations. Such behavior might explain why c-LLZO is a good fast ionic conductor: the low-frequency lattice mode can effectively flatten the ionic free-energy landscape and lower the migration barriers without acting as a strongly dissipative bath \cite{Pollak2023}. A comprehensive understanding of this mechanism would require extending the present theory to explicitly account for ion transport dynamics.

\subsection*{S3.2. AgCrSe$_2$}\label{s3.2}
\subsection*{S3.2.1. Lattice-renormalized geometric frustration in AgCrSe$_2$}\label{s3.2.1}
\begin{figure}[t]
\centering
\includegraphics[width=1.0\textwidth]{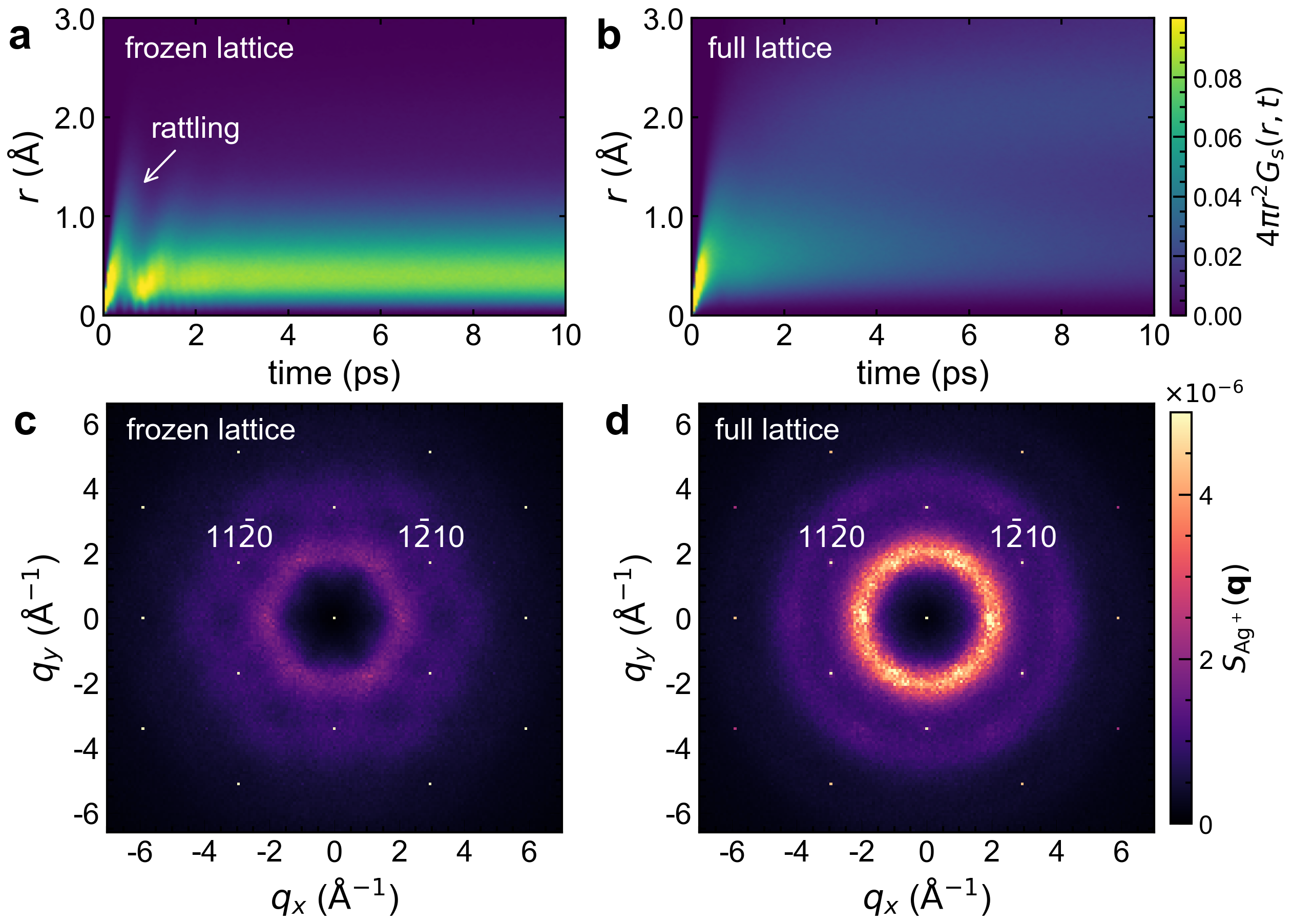}
\caption{\textbf{Lattice renormalization reshapes the frustrated ionic configuration through local lattice relaxation and interference in AgCrSe$_2$ at a simulation temperature of 500 K.} \textbf{a,b,} Single Ag$^+$ dynamics examined by the self-part of the van Hove correlation function $4\pi r^2G_s(r,t)$. In the frozen lattice (\textbf{a}), Ag$^+$ ion rattles with large amplitude and diffuses progressively, while the full lattice enables fast liquid-like diffusion (\textbf{b}). \textbf{c,d,} Simulated frozen-lattice (\textbf{c}) and full-lattice (\textbf{d}) partial static structure factors $S_{\text{Ag}^+}(\mathbf{q})$ viewed along the $c$-axis. The enhanced diffuse scattering in the full lattice (\textbf{d}) compared to the frozen lattice (\textbf{c}) signifies the emergence of collective short-range order \cite{Yang2026}.} \label{figS7}
\end{figure}

The same analysis, i.e. the self-part of the van Hove correlation function $G_s(r,t)$, partial static structure factor $S_{\text{Ag}^+}(\mathbf{q})$ and real-space ionic correlations, was carried out for AgCrSe$_2$ as well. As shown in Figs. \ref{figS7}a-b, the frozen-lattice $G_s(r,t)$ of Ag$^+$ simulated at 500 K reveals strong rattling dynamics but weak diffusion, whereas the full simulation displays the characteristic features of conventional migration. Concurrently, the corresponding $S_{\text{Ag}^+}(\mathbf{q})$ is dominated by an enhanced diffuse ring pattern (Fig. \ref{figS7}d), which has been reported in our previous work \cite{Yang2026}. Despite distinct differences in chemical composition and crystal structures, these commonalities across Li- and Ag-based systems support the universality of the present theoretical framework.

\begin{figure}[h]
\centering
\includegraphics[width=1.0\textwidth]{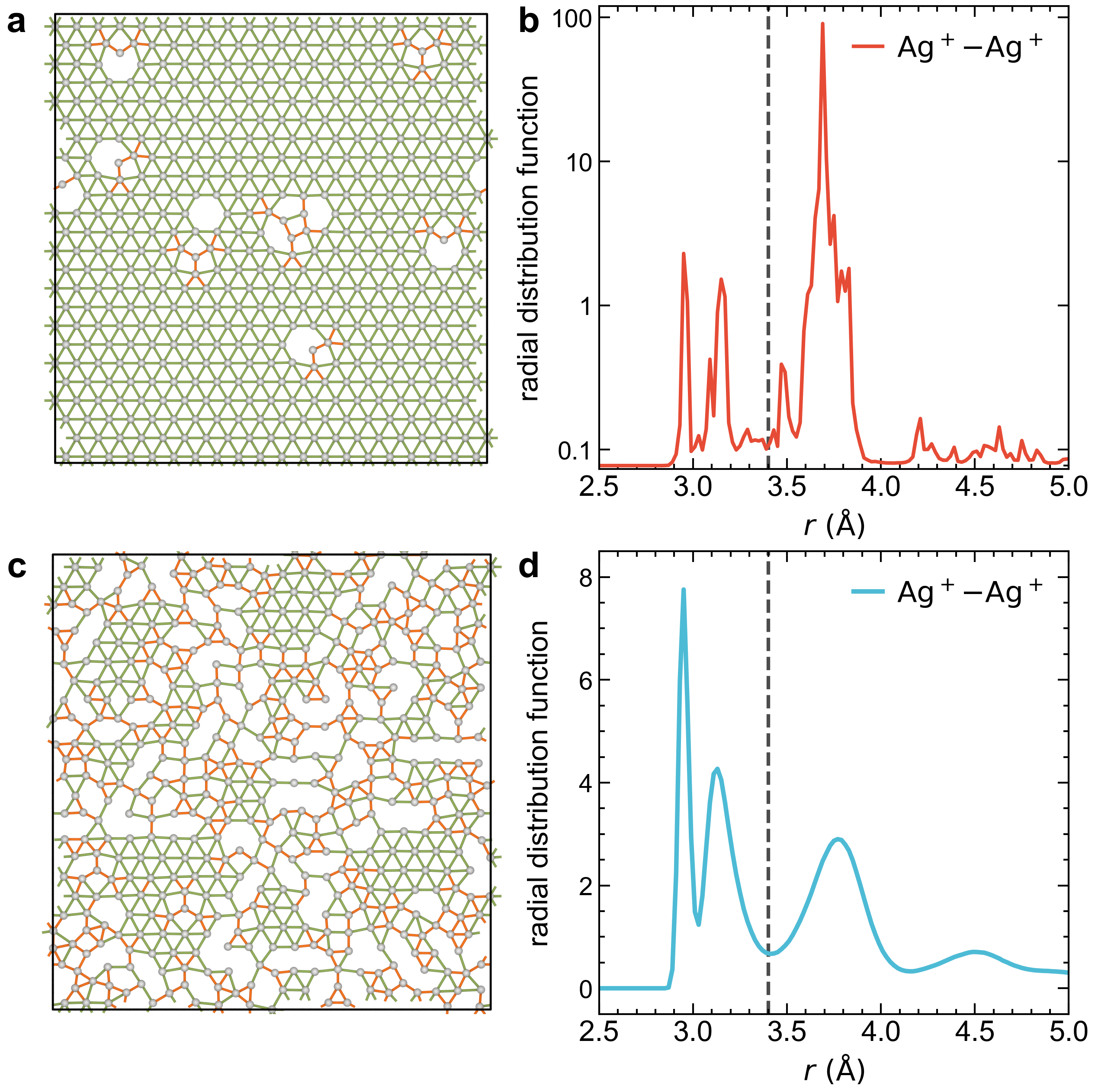}
\caption{\textbf{Analysis of geometric frustration in AgCrSe$_2$ in real space.} \textbf{a,} Representative snapshot of the frozen-lattice relaxed structure viewed along the $c$-axis, where Ag$^+$-Ag$^+$ separations below 3.4 $\rm\AA$ are colored in orange, and in the range of 3.4--4.0 $\rm\AA$ colored in green. \textbf{b,} Radial distribution function of Ag$^+$-Ag$^+$ pairs obtained from the frozen-lattice relaxation. The dashed vertical line indicates 3.4 $\rm\AA$. \textbf{c,} Representative snapshot of the full-lattice relaxed structure (viewed along the $c$-axis) with Ag$^+$-Ag$^+$ distances below 3.4 $\rm\AA$ and in the range of 3.4--4.0 $\rm\AA$ highlighted in orange and green, respectively. \textbf{d,} Radial distribution function of Ag$^+$-Ag$^+$ pairs calculated from the full-lattice relaxation. The local ionic configurations are transformed into an interpenetrating random network and short-range ordered domains by lattice renormalization. The substantial shift of the main peak in the radial distribution function signifies the breakdown of long-range periodicity and the emergence of short-range correlations \cite{Yang2026}.} \label{figS8}
\end{figure}

Figure \ref{figS8}a shows a typical geometrically frustrated AgCrSe$_2$ structure viewed along the $c$-axis, extracted from 500 K frozen-lattice MLMD simulations and optimized with the framework kept fixed. For clarity, only a single layer of Ag$^+$ is illustrated. The first nearest-neighbor inter-Ag$^+$ distances below 3.4 $\text{\AA}$ and in the range of 3.4--4.0 $\text{\AA}$ are color-coded to facilitate the identification of local topologies. The most frequently observed frustrated configuration consists of two interconnected Y-shaped structures, arising from a substantial displacement of a single Ag$^+$ ion from its ideal hexagonal lattice site. The corresponding inter-Ag$^+$ distances, as inferred from Fig. \ref{figS8}b, are 2.95 and 3.15 $\text{\AA}$, respectively. A configuration comprising three interconnected Y-shaped motifs is also observed. The central Ag$^+$, which is topologically shared among these motifs, can be interpreted as an interstitial that likely migrates from a nearby vacant configuration. The remaining complex frustrated structures can similarly be rationalized as variants derived from the fundamental Y-shaped topology.

As shown in Fig. \ref{figS8}c, Ag$^+$ configurations are dramatically frustrated, or reshaped by lattice renormalization. A large number of Y-shaped motifs (identified by a bond length below 3.4 $\text{\AA}$) are created and they form an interconnected two-dimensional random network without long-range periodicity. The inter-Ag$^+$ distances within these disordered, polymer-like networks are identical to those of the Y-shaped structures in the frozen lattice, approximately 2.95 and 3.15 $\text{\AA}$, respectively. Meanwhile, the underlying hexagonal lattice, characterized by a larger separation of 3.68 $\text{\AA}$, becomes fragmented and is confined to short-range ordered domains. These topologically disordered networks and locally ordered crystalline fragments interpenetrate each other, leading to a globally heterogeneous frustrated architecture. In addition, such coexistence of short-bond networks and long-bond domains also generates a large fraction of free volume, which is proposed to be responsible for the liquid-like superionic conduction of AgCrSe$_2$ in our previous work \cite{Yang2026}.

\begin{figure}[ht]
    \centering
    \includegraphics[width=0.6\linewidth]{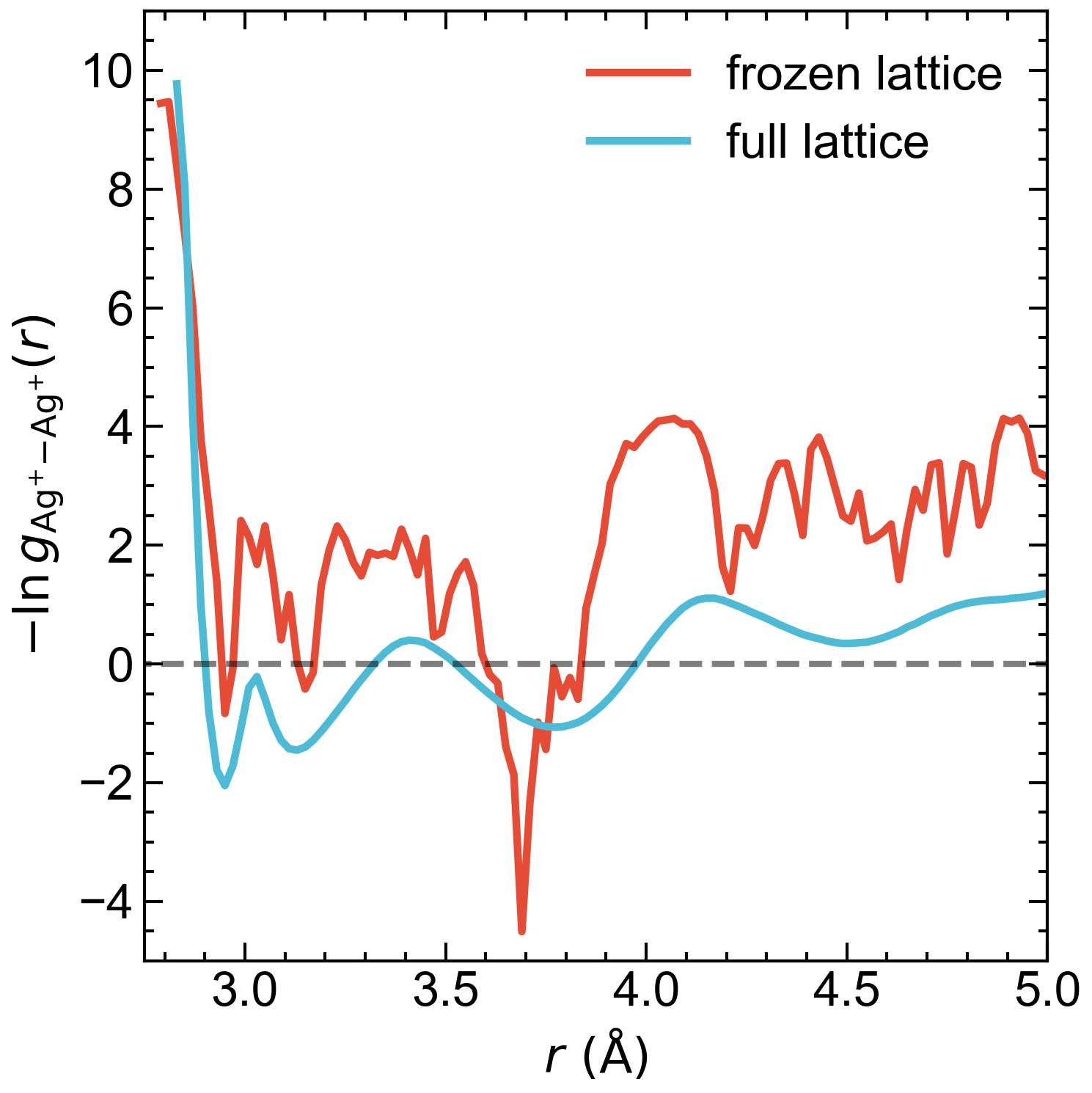}
    \caption{\textbf{The prefactor-reduced potential of mean force between two Ag$^+$ cations.} The frozen-lattice results are taken from frozen-lattice MLMD trajectories and relaxed under frozen-lattice constraints. The full-lattice structures are extracted from full-lattice MLMD trajectories and fully relaxed. The negative values indicate effectively attractive pair interactions, while positive values reflect repulsive interactions.}
    \label{figS9}
\end{figure}

An analogous prefactor-reduced PMF analysis was carried out for AgCrSe$_2$ as well, and the results are presented in Fig. \ref{figS9}. For the ionic configuration obtained under the frozen-lattice constraint, two additional shallow dips ($\tilde{W}<0$) appear at $r\approx2.95\;\mathrm{\AA}$ and 3.15 $\mathrm{\AA}$, in addition to the main dip at $r\approx3.68\;\mathrm{\AA}$ (the periodicity of the host lattice). The Ag$^+$-Ag$^+$ separations associated with these shallow dips deviate substantially from the lattice periodicity, signaling an inherent geometric frustration that is already accessible even in the absence of lattice renormalization. This result is consistent with our recent experimental finding, which demonstrates the existence of static off-center displacements in the low-temperature ground state of AgCrSe$_2$ \cite{Qi2026}. When lattice relaxation is incorporated, the short Ag$^+$-Ag$^+$ separations become more energetically favorable than the longer ones, thus leading to the geometrically frustrated configurations depicted in Fig. \ref{figS8}c. These results illustrate how the lattice renormalizes the effective ion-ion interactions and rearranges the underlying ionic configurations.

\begin{figure}[h]
\centering
\includegraphics[width=1.0\textwidth]{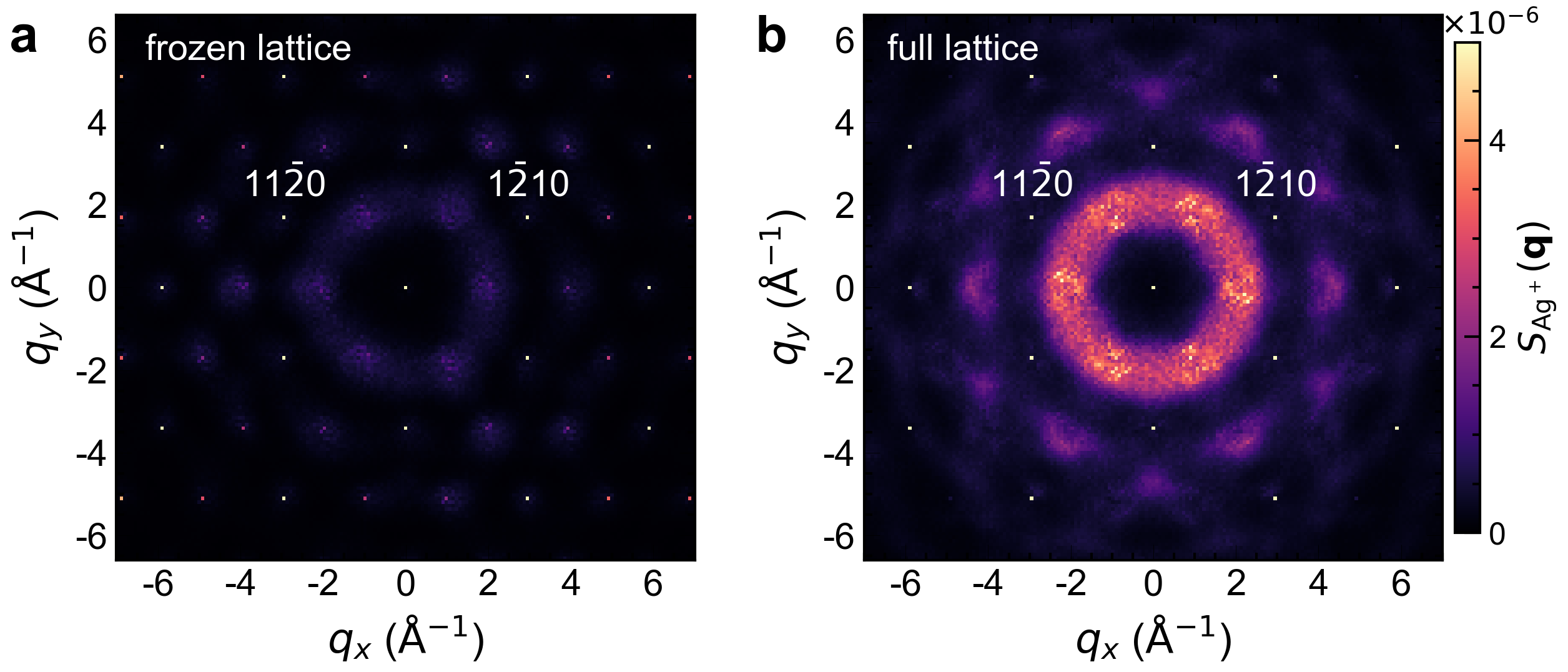}
\caption{\textbf{Analysis of geometric frustration in AgCrSe$_2$ in reciprocal space.}  \textbf{a,} The simulated partial static structure factor $S_{\text{Ag}^+}(\mathbf{q})$ for the frozen-lattice snapshots relaxed under frozen-lattice constraint. The geometrically frustrated motifs (Fig. \ref{figS8}a) lead to weak diffuse intensities. \textbf{b,} The partial static structure factor $S_{\text{Ag}^+}(\mathbf{q})$ for the full-lattice snapshots optimized without constraints. The interpenetrated polymer-like random network and short-range ordered domains (Fig. \ref{figS8}c) give rise to a characteristic diffuse ring around $|\mathbf{Q}|\approx 2.13\,\mathrm{\AA}^{-1}$ \cite{Yang2026}.} \label{figS10}
\end{figure}

The partial static structure factors for the frustrated Ag$^+$ configurations optimized under static conditions are presented in Fig. \ref{figS10}. In contrast to Fig. \ref{figS7}c, the diffuse scattering in $S_{\text{Ag}^+}(\mathbf{q})$ for configurations optimized with a frozen lattice (Fig. \ref{figS8}a) is substantially weaker due to the complete suppression of thermal diffuse scattering. Conversely, $S_{\text{Ag}^+}(\mathbf{q})$ for configurations optimized without constraints exhibits a characteristic diffuse ring pattern. The corresponding scattering vector $\mathbf{Q}$ for this feature is associated with an Ag$^+$-Ag$^+$ separation of $d_{\text{Ag}^+\text{-Ag}^+}\approx$2.95 $\mathrm{\AA}$ via $|\mathbf{Q}|\sim 2\pi/d_{\text{Ag}^+\text{-Ag}^+}\approx 2.13\,\mathrm{\AA}^{-1}$.

\subsection*{S3.2.2. Dynamics of the Cr-Se host lattice}\label{s3.2.2}
\begin{figure}[h]
\centering
\includegraphics[width=1.0\textwidth]{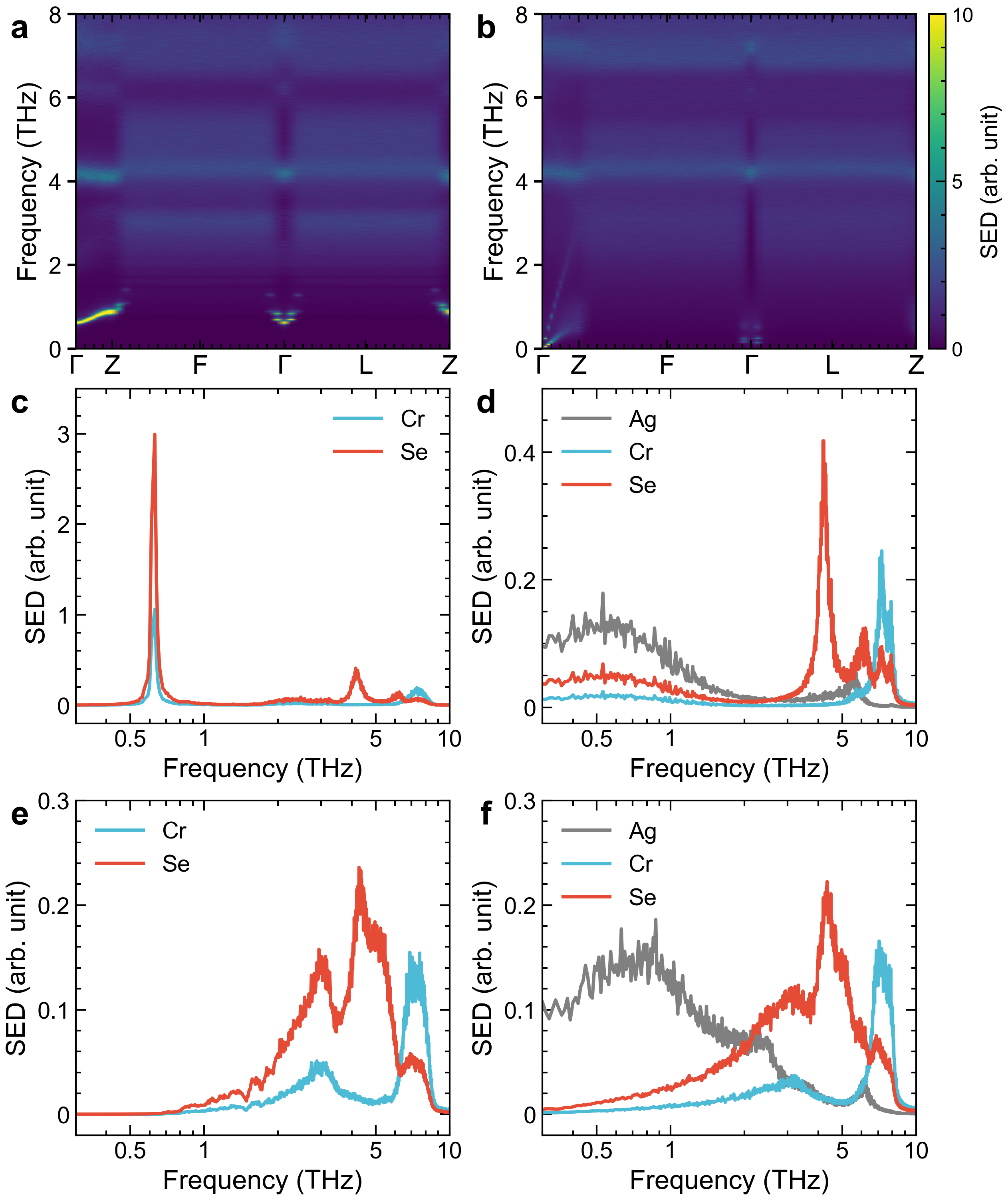}
\caption{\textbf{Spectral energy density (SED) analysis of the Cr-Se host lattice in AgCrSe$_2$  at 500 K.} \textbf{a,} Spectral energy density of the Cr-Se framework in the frozen Ag$^+$ disorder limit. A sharp optical mode in the range of 0.6--0.9 THz along the $\Gamma$--Z path can be identified. \textbf{b,} Spectral energy density of the Cr-Se framework in the dynamic Ag$^+$ disorder regime. \textbf{c,d,} Spectral energy density of the Cr-Se framework at the $\Gamma$ point in the limit of static disorder (\textbf{c}) and dynamic disorder (\textbf{d}). The low-frequency vibrational modes of the Cr-Se framework broaden substantially, indicating a strong coupling between this collective mode and the dynamically disordered Ag$^+$ sublattice. \textbf{e,f,} Spectral energy density of the Cr-Se framework at the F point in the cases of static disorder (\textbf{e}) and dynamic disorder (\textbf{f}). The reduced spectral energy density of the Cr-Se lattice at $\sim$3 THz suggests an energy dissipation channel through the dispersion-less vibrational modes in the full dynamic disorder regime.} \label{figS11}
\end{figure}

The dynamics of the Cr-Se host framework was also examined by the SED method. Figures \ref{figS11}a-b display the SED in the cases of static (frozen) Ag$^+$ disorder and the dynamically disordered Ag$^+$ at 500 K, respectively. In the frozen-disorder limit, the Cr-Se lattice exhibits a flat dispersion relation in the Brillouin zone, except for a sharp optical mode along the $\Gamma$--Z path. However, this low-frequency optical mode vanishes when Ag$^+$ disorder is dynamic (Fig. \ref{figS11}b). A more detailed comparison of this vibrational feature at the $\Gamma$ point is provided in Figs. \ref{figS11}c-d. The low-frequency mode at 0.6 THz evolves from an underdamped vibrational mode into a heavily damped mode. The broadened spectral linewidth indicates a substantially reduced phonon lifetime, which can be attributed to strong coupling between this mode and the dynamically disordered Ag$^+$ sublattice.

In addition to the $\Gamma$ point, the SED at the F point is also shown in Figs. \ref{figS11}e-f. The identical optical branch (at $\sim$3 THz) also broadens, with its spectral weight extending to lower frequencies. Taken together, these results suggest that this low-frequency optical mode is highly susceptible to dynamic Ag$^+$ disorder and acts as an efficient energy-dissipative (i.e., frictional) channel for Ag$^+$ motion. Consequently, the low-frequency lattice vibrations not only reshape the ionic free-energy landscape, but also play an active role in the diffusion dynamics of Ag$^+$ ions. The surprising contrast between Li$^+$ in c-LLZO and Ag$^+$ in AgCrSe$_2$ suggests complex transport dynamics in fast ionic conductors. A light ion can move through a coherent, weakly damped soft-response channel, whereas a heavy ion can generate strong damping if its dynamic disorder decoheres the host-lattice modes.

\medskip
In summary, our atomistic simulation provides direct structural evidence for the lattice-renormalized geometric frustration mechanism. Under frozen-lattice constraints, both ordered c-LLZO and AgCrSe$_2$ exhibit characteristic but isolated frustrated motifs, i.e. butterfly-like multimers and Y-shaped topologies, reflecting the intrinsic local degeneracy of the ionic sublattice. When full lattice relaxation is allowed, these motifs are substantially reorganized: the originally isolated frustrated minima become increasingly connected, giving rise to (\romannumeral1) a heterogeneous but collectively accessible frustrated manifold, (\romannumeral2) systematic shifts in radial distribution functions, and (\romannumeral3) strengthened correlations with shorter inter-ionic distances. These findings corroborate our theoretical prediction that lattice renormalization selectively stabilizes correlated frustrated configurations through two-body interference interactions. Furthermore, these results demonstrate that lattice renormalization simultaneously lowers migration barriers globally, and reorganizes the connectivity of the frustrated manifold itself by transforming isolated local minima into collectively accessible configurational networks. These two important phenomena in superionic conductors are unified within the same lattice-renormalized geometric-frustration landscape. The qualitatively similar behaviors across a three-dimensional oxide electrolyte and a two-dimensional layered chalcogenide underscore the universality of our present framework in fast ionic conductors. Viewed from this perspective, enhanced ionic transport emerges from the collective renormalization of the geometrically frustrated free-energy landscape, thereby providing a direct bridge between the microscopic Hamiltonian and macroscopic transport phenomena.

\bibliography{sn-bibliography}

